\documentclass{emulateapj}

\usepackage{graphicx}
\usepackage{morefloats}
\usepackage{natbib}
\bibpunct{(}{)}{;}{a}{,}{,}

\slugcomment{To be submitted to the Astrophysical Journal}

\shorttitle{Tidal Interactions among Seyfert Galaxies}
\shortauthors{Kuo et al.}

\begin{document}


\title{Prevalence of Tidal Interactions among Local Seyfert Galaxies}


\author{Cheng-Yu Kuo\altaffilmark{1,3}}
\affil{Institute of Astronomy \& Astrophysics, Academia Sinica, PO Box 23-141, Taipei 10617, Taiwan\\
        Department of Astronomy, University of Virginia, Charlottesville, VA 22904}
\email{ck2v@virginia.edu}

\author{Jeremy Lim}
\affil{Institute of Astronomy \& Astrophysics, Academia Sinica, PO Box 23-141, Taipei 10617, Taiwan}
\email{jlim@asiaa.sinica.edu.tw}

\author{Ya-Wen Tang\altaffilmark{1,2}}
\affil{Institute of Astronomy \& Astrophysics, Academia Sinica, PO Box 23-141, Taipei 10617, Taiwan\\
         Department of Physics, National Taiwan University, No. 1, Sec. 4, Roosevelt Rd., Taipei 106, Taiwan}
\email{ywtang@asiaa.sinica.edu.tw}

\and

\author{Paul T. P. Ho\altaffilmark{1,4}}
\affil{Institute of Astronomy \& Astrophysics, Academia Sinica, PO Box 23-141, Taipei 10617, Taiwan\\
       Harvard-Smithsonian Center for Astrophysics, 60 Garden Street, Cambridge, MA 02138}
\email{pho@asiaa.sinica.edu.tw}



\begin{abstract}
No mechanisms have hitherto been conclusively demonstrated to be responsible for initiating optically-luminous nuclear (Seyfert) activity in local disk  galaxies.  Only a small minority of such galaxies are visibly disturbed in optical starlight, with the observed disturbances being at best marginally stronger than those found in matched samples of inactive galaxies.  Here, we report the first systematic study of an optically-selected sample of twenty-three active galaxies in atomic hydrogen (HI) gas, which is the most sensitive and enduring tracer known of tidal interactions. Eighteen of these galaxies are (generally) classified as Seyferts, with over half (and perhaps all) having [OIII] luminosities within two orders of magnitude of Quasi-Stellar Objects. Only $\sim$$28\%$ of these Seyfert galaxies are visibly disturbed in optical DSS2 images.  By contrast, $\sim$$94\%$ of the same galaxies are disturbed in HI, in nearly all cases not just spatially but also kinematically on galactic ($\gtrsim 20$~kpc) scales.  In at least $\sim$$67\%$ and perhaps up to $\sim$$94\%$ of cases, the observed HI disturbances can be traced to tidal interactions with neighboring galaxies detected also in HI.  The majority of these neighboring galaxies have projected separations of $\lesssim 100 {\rm \ kpc}$ and differ in radial velocities by $\lesssim 100 {\rm \ km \ s^{-1}}$ from their respective Seyfert galaxies, and many have optical luminosities ranging from the Small to Large Magellanic Clouds.  In a companion paper, we show that only $\sim$$15\%$ of a matched control sample of inactive galaxies display comparable HI disturbances. Our results suggest that: i) most Seyfert galaxies (with high nuclear luminosities) have experienced tidal interactions in the recent past; ii) in most cases, these tidal interactions are responsible for initiating events that lead to their nuclear activity.
\end{abstract}




\section{INTRODUCTION}
Active Galactic Nuclei (AGNs) are believed to be the luminous visible evidence for the vigorous accretion of gas onto a central SuperMassive Black Hole (SMBH).  In the case of Quasi-Stellar Objects (QSOs), adopting the convenient (but arbitrary) convention that QSOs have absolute $B$-band (nonstellar) nuclear magnitudes $M_B < -23$ \citep{ver98}, the required mass-accretion rate, assuming a $10\%$ efficiency for the conversion of infalling matter into radiation \citep[e.g.,][]{norsil83}, is at least $\sim$$0.1 {\rm \ M_{\odot} \ yr^{-1}}$.  Lower-luminosity AGNs have correspondingly smaller mass-accretion rates, with those described in this paper having mass-accretion rates ranging down to about two orders of magnitude lower ($\S2$).  If the AGNs studied here have lifetimes of $\sim$$10^8 {\rm \ yrs}$, the total amount of matter required to feed their activity is therefore $\sim$$10^5$--$10^7 {\rm \ M_{\odot}}$.  The origin or source of this material, and how this material is brought into the sphere of influence of the central SMBH, is one of the longstanding unresolved problems in AGN research.

Many mechanisms for providing fuel to the central SMBHs of active galaxies have been proposed \citep[e.g., for a recent brief review, see][]{mart04}.  This is particularly true at relatively low accretion rates, where stellar disruptions due to close encounters with the central SMBH, dynamical friction on molecular clouds, stellar winds from massive stars or mass loss from evolved stars \citep[e.g.,][]{norsco88}, or turbulence in the interstellar medium \citep[e.g., review by][]{wad04}, are among the proposed hypotheses.  At higher accretion rates, non-axisymmetric gravitational potentials such as galactic bars \citep[e.g., review by][]{com04} or gravitational interactions with neighboring galaxies \citep[e.g., review by][]{sch04} may be needed to drive sufficient quantities of gas inwards to fuel the AGN.  At even higher accretion rates, corresponding perhaps to that in QSOs, cannibalisms of gas-rich galaxies (minor mergers) or mergers between two gas-rich galaxies (major mergers) may be necessary to carry sufficient fuel to the AGN.


Because of their more accessible spatial scales, most observational studies to date have focussed on whether galactic bars or galaxy-galaxy interactions are responsible for triggering AGNs (in disk galaxies). Nearly all such studies have involved optically-selected samples of Seyfert galaxies, which make up only a very small fraction (roughly $1\%$) of relatively bright field galaxies \citep[e.g.,][]{huc92,ste94}.  Seyfert galaxies are preferentially found among early-tpe spiral galaxies, and exhibit very broad emission lines (from their nuclei) compared with their inactive counterparts.  They are further divided into different classes depending on the relative intensity of their broader ($\sim$1000--$15000 {\rm \ km \ s^{-1}}$) component, which is dominated by permitted lines, to their narrower component ($\sim$200--$2000 {\rm \ km \ s^{-1}}$), which is dominated by forbidden lines.  At the two extremes lie Seyfert~1 galaxies, which have a relatively strong broad component, and Seyfert~2 galaxies, which have a detectable narrow component only (in unpolarized light). 

Searches for an excess of bars in Seyfert compared with inactive galaxies have yielded either null or at best weakly suggestive results.  For example, neither \citet{ho97} nor \citet{mul97} found an excess of large-scale ($\gtrsim 1 {\rm \ kpc}$) bars between Seyfert and inactive galaxies, but \cite{kna00} found a marginal excess of such bars amounting to $79\% \pm 8\%$ in Seyfert compared with $59\% \pm 9\%$ in inactive galaxies.  Both \citet{reg99} and \citet{mart99} found that only a small fraction of Seyfert galaxies in their sample exhibit small-scale ($\lesssim 1 {\rm \ kpc}$) bars, and suggested that such bars cannot therefore be the primary mechanism for fueling Seyfert activity.  On the other hand, \citet{lai02} found that Seyfert galaxies have an excess or at least equal fraction of bars at practically all length scales (from tens of parsecs to tens of kiloparsecs) compared with inactive galaxies; taken together, $73\% \pm 6\%$ of Seyfert galaxies have bars compared with $50\% \pm 7\%$ of inactive galaxies in their sample. \citet{mai03} find bars among $28\%$ of Seyfert compared with $14\%$ of inactive galaxies in their control sample.  By contrast, \citet{hun04} find that Seyfert~1 galaxies exhibit practically no bars, whereas only a small fraction ($15\% \pm 5\%$) of Seyfert~2 and a similar fraction of inactive galaxies in their control sample exhibit bars.

Rather than bars, both \citet{reg99} and \citet{mart99}, as well as \citet{pog02}, found that the majority if not all of the Seyfert galaxies in their sample exhibit nuclear dust spirals.  These dust spirals may trace shocks or turbulence in the central circumnuclear gas disk, and hence correspond to sites of angular momentum dissipation.  \citet{mar03}, however, found that nuclear dust spirals occur with comparable frequency in both active and inactive galaxies.  They concluded that no universal fueling mechanism has yet been demonstrated to operate in disk galaxies at spatial scales larger than $\sim$100~pc from the AGN.

Searches for an excess of Seyfert galaxies among (apparently) interacting galaxies compared with matched samples of normal or apparently isolated spiral galaxies have yielded similarly ambiguous if not contradictory results.  \citet{dah85} find only a marginal excess of Seyfert galaxies among interacting spiral galaxies, but a clear excess when the sample is restricted to just strongly interacting spiral galaxies.  \citet{kee85} also find Seyfert galaxies more frequently in interacting compared with isolated spiral galaxies, but by contrast with \citet{dah85} find a smaller excess in the most disrupted systems.  On the other hand, \citet{bus86} find a smaller fraction of Seyfert galaxies among violently interacting than isolated spiral galaxies.  

Despite the conflicting results, examinations of optical images do reveal that the fraction of Seyfert galaxies that are visibly (or apparently) interacting in the optical is relatively small.  Indeed, even in observations with the {\it Hubble Space Telescope}, the vast majority of local Seyfert galaxies \citep{mal98}, as well as QSO host galaxies at redshifts as high as $z \approx 0.3$ \citep[e.g.,][]{bah97,boy98}, do not appear to be visibly disturbed in the optical.  Apart from their nuclear activity, local Seyfert galaxies appear to otherwise resemble the general population of inactive, preferentially early-type spiral, galaxies.  \citet{der98a} find at best a marginal difference in light asymmetries between Seyfert and inactive galaxies; using the same sample but a more sophisticated analysis, \citet{vir00} find no differences.  On the other hand, \citet{hun04} find that $30\% \pm 9\%$ of Seyfert~1 (and a comparable fraction of LINERs and HII) and $63\% \pm 7\%$ of Seyfert 2 galaxies exhibit isophotal twists compared with just $6\% \pm 4\%$ of a matched sample of inactive galaxies.

With little direct evidence for (more frequent) galaxy-galaxy interactions, studies have turned to whether Seyfert galaxies more frequently exhibit projected neighboring galaxies by comparison with matched samples of inactive galaxies \citep[e.g., summary in][]{sch04}.  Such an excess has been found in some studies \citep[e.g.,][]{sta82,dah84,raf95,dul99,kou06}, but only marginally or not at all in others \citep[e.g.,][]{fue88,der98b}. Selecting Seyfert galaxies from the Markarian catalog (see $\S2$ for a brief description of this catalog), \citet{mac89} find a larger fraction of Seyfert with projected neighboring galaxies compared with inactive galaxies lying in the close vicinity, but not with randomly selected inactive galaxies selected also from the same catalog.  \citet{lau95} find that Seyfert galaxies appear in paired systems no more frequently than inactive galaxies, but have on average more projected neighboring galaxies. \citet{sch01} did not find any differences in the fraction of galaxies with genuine neighboring galaxies (i.e., with known redshifts) among Seyfert, LINERs, HII, transition, and absorption-line galaxies when considering only those with similar morphological types.

In many of the post-1990's \citep[but not all; c.f.][]{raf95} studies, the excess of Seyfert with projected neighboring galaxies (where found) is seen only around Seyfert~2 but not Seyfert~1 galaxies \citep{lau95,dul99,kou06}, or alternatively that Seyfert~2 galaxies have on average a richer environment \citep{der98b}. This supports an earlier study by \cite{per82} that found a difference between the environments of Seyfert~1 and Seyfert~2 galaxies.  
Both \citet{cha02} and \citet{hun04} suggest that Seyfert~2 galaxies are generally at an earlier stage of interactions or dynamical disturbances compared with Seyfert~1 galaxies.  Furthermore, \citet{sto01} find more mergers or close interactions among Seyfert~2 galaxies with recent circumnuclear star formation than those without, and suggest an evolutionary sequence whereby the former are seen earlier after an interaction.

Many of the abovementioned studies are now recognized to contain differences in their selection of Seyfert and/or comparison samples of inactive galaxies, as well as differences in their methodologies, that may have given rise to their disparate results.  Although the more recent studies purport to carefully address these problems, they nevertheless reach entirely different conclusions.  For example, selecting both Seyfert \citep{huc92} and matched inactive galaxies from the CfA Redshift Survey \citep{huc83}, \citet{der98b} find no difference in the fraction of galaxies with projected neighboring galaxies in either samples, but do find (as mentioned earlier) that Seyfert~2 are more likely to have closer projected neighboring galaxies than Seyfert~1 galaxies.  Based on the sample of Seyfert, LINER, transition, absorption-line, and HII galaxies selected from the Revised Shapley-Ames Catalog of Bright Galaxies \citep{san81} as classified spectroscopically by \citet{ho97} (containing mostly low nuclear luminosity galaxies), \citet{sch01} found the same percentage among all the different types of galaxies with projected neighboring galaxies.  Selecting from the catalog of \citet{lip88} containing most if not all of the then known Seyfert galaxies \citep[not, as stated by][based on the Second Byurakan Survey]{dul99}, both \citet{dul99} and \citet{kou06} find an excess of Seyfert~2 but not Seyfert~1 with projected neighboring galaxies compared with matched samples of inactive galaxies.

Failing to find a difference between the fraction of Seyfert and inactive galaxies with projected neighboring galaxies, and with \citet{der98a} finding only a marginal difference in light asymmetries between Seyfert and inactive galaxies, \citet{der98b} suggested that Seyfert galaxies may have cannibalised minor companion galaxies.  To test this idea, \citet{cor00} looked for morphological differences in the light distribution between Seyferts and galaxies having other nuclear spectral types (absorption, H~II, and LINER).  Finding no such differences, nor any qualitative evidence that minor mergers have occurred among any of the Seyfert galaxies or LINERs, \citet{cor00} concluded that either such events occurred $\gtrsim 1 {\rm \ Gyr}$ ago and are essentially complete or not at all.


In summary, most Seyfert galaxies do not appear to be visibly disturbed let alone interacting with neighboring galaxies.  Even if more Seyferts than inactive galaxies have projected neighboring or genuine companion galaxies, it is not clear what fraction are actually involved in tidal interactions.  In this paper, we apply the lessons learnt from studies which demonstrate that neutral atomic hydrogen (HI) gas can reveal gravitational interactions between galaxies not visible in optical starlight. The best known example is the M81 group of galaxies, the closest galaxy group next to our own Local Group.  Optical observations show no visible evidence for interactions between M81 and its nearest neighbours, the starburst galaxy M82 and the inactive galaxy NGC~3077.  HI observations, on the other hand, reveal gaseous filaments connecting all three galaxies that trace tidal interactions between these galaxies \citep{yun94}.  The same situation has also been seen in a small sample of relatively nearby galaxies hosting QSOs \citep{lim99}. Why are tidal interactions more easily seen in HI gas than optical starlight? In normal (i.e., undisturbed) spiral galaxies, the HI gas disk usually extends nearly twice as far out as the stellar optical disk.  The outskirts of the HI disk is not as strongly gravitationally bound and therefore more susceptible to external perturbations than the stellar optical disk; when perturbed, the outer regions of the HI disk also take longer to dynamically relax.  This makes HI a particularly sensitive and enduring tracer of gravitational interactions between galaxies. Comprising simultaneous imaging and spectroscopy, HI observations provide not just measurements of the gas kinematics but also accurate redshifts of surrounding gas-rich galaxies.  These information, together with the spatial structure of the tidal features, can be used to cleanly identify or more confidently pick out possible interacting neighboring galaxies.  

In this paper, we report the complete results of the first systematic imaging study of the HI gas in a sample of Seyfert galaxies, the results for a small subsample of which were reported by \citet{gre04}.  In an accompanying paper \citep{tan07}, we report the corresponding results for a matched control sample of inactive galaxies. Both the Seyfert and control sample are restricted to a narrow range of distances, ensuring that both samples are observed with the same uniform sensitivity and detection threshold. Prior to our study, HI imaging studies of Seyfert galaxies have usually targeted individual objects of special interest.  The available HI images were therefore highly heterogeneous in both spatial resolution and detection threshold \citep[e.g., see review by][]{bri96}.  In particular, not all the HI images have sufficient brightness sensitivity to detect spatially extended emission characteristic of tidal features. Furthermore, it was not clear whether the observed galaxies are representative of the general population of Seyfert galaxies.

This paper is organized as follows.  In $\S2$ we describe our selection of the Seyfert galaxy sample, and in $\S3$ our observations and data reduction.  In $\S4$ we present the results for the individual galaxies, and compile statistical trends for the ensemble sample. We encourage those readers interested in just the statistical results of our study to nevertheless glance at the HI images shown in Figures~1--36, which capture the essence of this work.  In $\S5$ we present our case that tidal interactions initiate events that lead to optically-luminous Seyfert activity in a large fraction of local disk galaxies.  In $\S6$ we discuss the implications of our results for the nature of the interaction, gas infall timescales to fuel the central SMBH, and lifetimes of optically-luminous Seyfert activity. In $\S7$ we provide a concise summary of our results and conclusions.  The catalog from which we draw our sample of active galaxies assumes a Hubble constant of $H_{o} = 50 {\rm \  km \ s^{-1} \ Mpc ^{-1}}$, and we retain this for ease of comparison with other active galaxies (including QSOs) in this catalog.  In the rest of our paper, unless otherwise stated, we assume a Hubble constant $H_{o} = 67 {\rm \ km \ s^{-1} \ Mpc ^{-1}}$ (and $\Omega = 1$) closer to the presently accepted value, and for consistency with the earlier results for a small subsample reported by \citet{gre04}.

\section{SAMPLE SELECTION}
The study reported in this paper is part of a larger program to address the prevalence of galaxy-galaxy interactions among the more optically-luminous AGNs in the Local Universe.  In a separate study, we are conducting a HI imaging survey of all known galaxies hosting QSOs (i.e., with absolute nuclear B-band magnitudes $M_B < -23$ for $H_{o} = 50 {\rm \ km \ s^{-1} \ Mpc ^{-1}}$) at declinations $\delta \geq -40^{\circ}$ and redshifts $z \leq 0.07$ \citep[][]{lim99,lim01}.  Here, we report an extension of that study to lower luminosity Seyfert nuclei, ostensibly (see clarification below) $-19 \geq M_B \geq -23$ (for $H_{o} = 50 {\rm \ km \ s^{-1} \ Mpc ^{-1}}$).  

Prior to the Sloan Digitized Sky Survey (the results from which had not yet been fully published circa 2000), a large fraction (roughly half or more, depending on the redshift interval) of local Seyfert galaxies (excluding QSOs) in the Northern Hemisphere were found in a few major optical surveys.  The first such large-scale survey was the Byurakan Spectral Sky Survey, the results from which are more popularly known as the Markarian catalog \citep{mar89}.  A good fraction of the Seyfert galaxies selected in many of the optical studies mentioned in $\S1$ were drawn (at least in part) from or can be found in this catalog.  A much smaller number of Seyfert galaxies found (or rediscovered) in the (first) CfA Redshift Survey comprise among the most optically-luminous galaxies at their respective redshifts in the Local Universe \citep{huc83}.  About half of these galaxies are contained in the Markarian catalog.  A large number of relatively low-luminosity AGNs (referred to as "dwarf" Seyfert nuclei) in the nearest galaxies (spanning a much lower redshift range than the Markarian or CfA Seyfert catalogs) were selected spectroscopically by \citet{ho97}.  Each of these surveys contain their own peculiar selection effects, and furthermore adopt somewhat different criteria (e.g., line ratios) in classifying galaxies into different activity classes (e.g., Seyfert~1, Seyfert~2, LINER, and inactive galaxies).  Together with those found in other surveys or small-scale studies, the currently known population of optically-selected Seyfert galaxies therefore comprise a heterogeneously selected sample.  Nevertheless, these objects comprise the general population of Seyfert galaxies from which inferences have been drawn about their overall nature.

After looking through the available surveys, it turned out that none individually suited our needs.  For example, the number of objects per unit redshift interval in the Markarian or CfA Seyfert catalogs are not by themselves sufficiently large, whereas the number of objects per unit redshift interval in the \citet{ho97} catalog do not contain enough luminous Seyfert nuclei, for our purpose.  The only catalog that fulfilled our needs (at the time) was that of \citet{ver98}, which collects together all the known AGNs (selected in different ways by the different studies) at the time of compilation (updated approximately every two years since 1991).  By its very nature, this catalog allows us to draw a sample of Seyfert galaxies that are as closely representative of the known (heterogeneously selected) population as is possible based on the most widely accepted classifications for their activity class.  
If the same phenomenon (e.g., tidal interactions) can be widely found in such a heterogeneous sample, this would consitute powerful evidence for its broader significance among the general population.


The \citet{ver98} catalog lists (among other relevant parameters) the absolute B-band magnitude of each galaxy in a relatively small aperture centered on the nucleus.  To compile a reasonably large number of objects with listed absolute B-band (nuclear) magnitudes closely approaching QSOs, we had to reach out to a redshift of $z \approx 0.015$.  To restrict the sample to essentially uniform distances, we selected all twenty-seven disk galaxies listed in the \citet{ver98} catalog at $0.015 \leq z \leq 0.017$ and with $-19 \geq M_B \geq -23$ (for $H_{o} = 50 {\rm \ km \ s^{-1} \ Mpc ^{-1}}$) in the Northern Hemisphere ($\delta \geq 0^{\circ}$).  To fill in the alloted observing time in the year 2000, we added one more galaxy to the sample, UGC~3157, which was selected from \citet{ver00} catalog using the same criteria as described above.  With the selected redshift and absolute magnitude range, the \citet{ver98} catalog contains only two definite elliptical galaxies, Perseus~A  and NGC~315, both of which are well known radio galaxies.  
Imaging the HI emission (if any) of these galaxies is difficult in the presence of their strong radio continuum emission.  The \citet{ver98} catalog also includes objects classified as LINERs (Low Ionization Nuclear Emission Line Regions, which show relatively strong lines of low ionization species such as [OI] and [SII], and which may comprise low-luminosity AGNs), and also those classified as H~II (which show narrow Balmer and [OIII] lines characteristic of H~II regions).  We included these galaxies as they are sometimes classified in reliable studies or other often used catalogs as AGNs; e.g., NGC~1167, which is classified as a LINER in \citet{ver98}, is classified as a Seyfert~2 in \citet{ho97}.  Later on, we will compile the statistical results of our study separately for the entire sample and just those classified as Seyferts in \citet{ver98} for the reader to compare.

Table~1 contains nineteen galaxies with nuclear spectral types classified as Seyfert~1, Seyfert~2, or intermediate Seyfert type (between S1.0 and S2.0), two with uncertain Seyfert classifications, five classified as LINERs, and two as H~II. We retain their names as listed in the \citet{ver98} catalog, except for UGC~3995 which is referred to in that catalog as UGC~3995A (in most literature sources, UGC~3995A is used to refer to the neighboring galaxy of UGC~3995).  Note that NGC~5548 is also known as Markarian~1509, NGC~7469 as Markarian~1514, and NGC 7679 as Markarian~534. All the other listed parameters (coordinates, redshifts, and absolute B-band magnitudes) are also from \citet{ver98} \citep[those for UGC~3157 from][]{ver00}, except for Hubble types and inclinations which are taken from the Hyperleda database \citep[http://leda.univ-lyon1.fr;][]{pat03}.  Note that our selection is entirely blind and hence unlikely to be skewed in any particular way to the HI properties of the selected galaxies.

The absolute B-band (nuclear) magnitudes listed in \citet{ver98} are taken from the literature, and are measured in different ways by different observers (e.g., different aperture sizes).  The listed values may also include an unknown contribution from the host galaxy: this is usually not a problem in the case of QSOs, which (far) outshines their host galaxies, but becomes increasingly problematic with decreasing nuclear luminosities. In addition, the nonstellar nuclear continuum is usually not directly viewed in Seyfert~2 nuclei \citep[e.g.,][]{mul94,mal98}.  Instead of the continuum, \citet{mul94} have shown that the [OIII] emission of Seyfert nuclei is a reliable proxy for their optical luminosities.  The [OIII] emission appears to be isotropic in Seyfert galaxies and well correlated with the nuclear UV continuum in Seyfert~1 galaxies, and therefore a reliable measure too of the nuclear UV continuum in Seyfert~2 galaxies.

We have searched the literature for measurements of the nuclear [OIII] emission from our sample galaxies.  Such measurements are available for a total of fourteen of our sample galaxies in \citet{dah88} and \citet{whi92}, comprising three Seyfert~1, three Seyfert~1.5, six Seyfert~1.9/2, one LINER, and one H~II galaxy; for the seven objects that \citet{dah88} and \citet{whi92} have in common, their quoted [OIII] intensities typically agree to within a factor of $\sim$2 (see Table~1).  The [OIII] luminosities of these fourteen galaxies are quite evenly spread over the range from $\sim$$10^{40.2}$--$10^{42.3} {\rm \ ergs \ s^{-1}}$ (for $H_{o} = 50 {\rm \ km \ s^{-1} \ Mpc ^{-1}}$), even if the LINER and HII galaxies are excluded; their absolute (nuclear) B-band magnitudes as listed in \citet{ver98} span essentially the entire range of our sample, from -19.0 to -21.6.  There is no obvious difference between the range of [OIII] luminosities for the different Seyfert classes, with both the least and most [OIII] luminous objects classified as Seyfert~2.  The [OIII] luminosities for three other LINERs in our sample can be found in \citet{ho97}, including NGC~1167 which is classified in that study as a Seyfert~2; these objects span a lower range in [0III] luminosities of $\sim$$10^{39.5}$--$10^{40.2} {\rm \ ergs \ s^{-1}}$ (for $H_{o} = 50 {\rm \ km \ s^{-1} \ Mpc ^{-1}}$).  Indeed, the bulk of the galaxies (including Seyferts) in \citet{ho97} have [OIII] luminosities below $\sim$$10^{40} {\rm \ ergs \ s^{-1}}$.  Local QSOs \citep[following the definition of][]{ver98} typically have [OIII] luminosities of $\gtrsim 10^{42} {\rm \ ergs \ s^{-1}}$ \citep[e.g., see][scaled to $H_{o} = 50 {\rm \ km \ s^{-1} \ Mpc ^{-1}}$]{kau03}.  At least $\sim$$63\%$ of the Seyfert galaxies in our sample therefore have optical nuclear luminosities spanning a range that is up to about two orders of magnitude below the lower limit for QSOs.  The remainder span the same range of absolute B-band (nuclear) magnitude as listed in \citet{ver98}, and likely so the same range in optical nuclear luminosities.

Our sample contains essentially all of the reliably identified Seyfert (disk) galaxies known in their redshift interval as of 1998.  It includes all seven Seyfert galaxies in the CfA Seyfert catalog \citep{huc83} within our selected redshift interval, and thus comprises a complete subsample from this catalog.  These galaxies are listed separately in Table~2.  Fifteen of the twenty-four galaxies in our sample can be found in the Markarian catalog \citep{mar89}, including twelve of the twenty-one objects classified as Seyferts (under the current classification for their activity classes).  These galaxies also comprise a complete subsample from the Markarian catalog within our selected redshift interval, and again listed separately in Table~2.  As we shall see, our results are the same if restricted to just these two subsamples or for our entire sample, emphasizing the broader significance of our results to the general population.

\section{OBSERVATIONS AND DATA REDUCTION}
We observed our sample galaxies with the Very Large Array (VLA)\altaffilmark{1} of the National Radio Astronomical Observatory (NRAO).  The observations were conducted over four sessions from 2000 to 2004, with a total alloted observing time of $\sim$80~hours, and at night to avoid radio interference from the Sun.  Relevant details of the observations, including the flux, secondary (for amplitude and phase calibration), and bandpass calibrators used for each target object, are summarized in Table~3.  The pointing (phase) center of the telescope was occasionally offset from the position of the target object to better surpress strong continuum sources close to the edge of the field that can otherwise limit the dynamic range of our maps.   We used the most compact configuration of the VLA (the D-array) that in the 21-cm line of HI gas provides a synthesized beam of $\sim$60\arcsec.  The latter corresponds to a spatial resolution of $\sim$20~kpc at the distance to our sample galaxies, which is well matched to the average optical sizes of these galaxies.  Our study is therefore tailored towards the detection of extended emission on galactic scales characteristic of tidal features.  All the galaxies in our sample were observed apart from NGC~266, a LINER, as it had previously been imaged in HI also in D-array by \citet{sim87} \citep[as reported by][]{hib01}. 

At 21~cm, the primary beam of the VLA is $\sim$32\arcmin\ at full-width half-maximum, corresponding to a linear diameter of $\sim$660~kpc.  We configured the correlator to record signals in orthogonal circular polarizations, and in sixty-four spectral channels over a bandwidth of 6.25~MHz and hence a channel separation of 97.7~kHz.  This corresponds to a velocity separation between channels of $\sim$$21.0 {\rm \ km \ s^{-1}}$ (the actual velocity width of each channel is $\sim$$25.2 {\rm \ km \ s^{-1}}$), and a full range in velocity spanned by all channels of $\sim$$1350 {\rm \ km \ s^{-1}}$.  For the assumed Cosmology, the velocity range spanned by the correlator corresponds to a redshift interval of ${\Delta}z \approx 0.005$ (distance range of $\sim$20~Mpc), compared with the redshift interval of ${\Delta}z \approx 0.002$ for our sample galaxies.  The central channel is set to the systemic heliocentric velocity of the HI line if previously measured (usually with a single-dish telescope), or if not then at its reported optical redshift.

\altaffiltext{1}{The VLA is a facility of the National Radio Astronomy Observatory, which is operated by the Associated Universities, Inc. under a cooperative agreement with the National Science Foundation (NSF).}

We targeted a 5$\sigma$ detection threshold in HI column density of $\sim$$3 \times 10^{19} {\rm \ cm^{-2}}$ per synthesized beam in a given channel with a velocity width $\sim$$25.2 {\rm \ km \ s^{-1}}$.  Below roughly this column density, HI gas appears to be ionized by intergalactic UV radiation \citep[][]{bla97}.  We emphasize that this does not necessarily imply that we detect all the HI gas present, as a given parcel of HI gas may have a larger velocity width than that spanned by just one channel (in such a case, the HI column density per channel is lower than that targeted).  The HI column density threshold corresponds to a minimum detectable gas mass of $\sim$$7\times 10^{7} {\rm \ M_{\odot}}$ per synthesized beam in one channel.  The targeted detection threshold corresponds to a rms flux density of $\sim$$0.45 {\rm \ mJy\ beam^{-1}}$ in one channel, requiring an integration time of $\sim$2~hrs (observing time of $\sim$2.5~hrs) per galaxy.

We performed all the data reduction in the standard manner using the NRAO AIPS package.  We inspected the data carefully for radio frequency interference (RFI), which when present were excised from the data.  Occasionally the RFI coincided with part of the HI line from the target galaxy; this did not pose a serious problem when the RFI was intermittent, but when present throughout much if not all of the observation seriously affected the detectability of that portion of the HI line.  As a consequence, we were forced to discard two of our sample galaxies, Markarian 352 and Markarian 1510 (one Seyfert and one H~II galaxy), both of which were badly affected by RFI.  Following bandpass calibration, as well as time-dependent amplitude and phase calibration, we made continuum maps (i.e., combining the data from all the useable channels) for each target galaxy.  To correct for residual amplitude and phase errors, we self-calibrated the data for several iterations first in phase, followed by one or two iterations in both amplitude and phase when the continuum signal detectable over the primary beam was sufficiently strong.  We then subtracted the continuum emission from the visibility data by interpolating between line-free channels (determined through trial and error) on either side of the HI line.  Finally, we made maps of each channel, and corrected for the primary beam response of the antennas.  We also combined the channel maps (without primary beam correction) to make maps in total intensity (zeroth moment) and intensity-weighted mean velocity (first moment), where we tried several different combinations of spectral smoothing and minimum signal cutoff level in each channel so as to suppress the noise and bring out faint diffuse features.

\section{RESULTS}
We omit from further consideration Markarian~352 (Seyfert~1) and Markarian~1510 (H~II), which as mentioned in $\S3$ are badly corrupted by RFI.  Leaving aside these galaxies, only two galaxies, Markarian~359 (Seyfert~1) and 2E~0507+1626 (Seyfert~1.5), were not detected in HI, as described in \citet{gre04}.  In the case of Markarian~359, a strong continuum source detected in the same field limited the accuracy of our passband calibration, and hence resulting map sensitivity. Unfortunately, this leaves only two remaining Seyfert~1s in our sample.  We also leave out NGC~266 (LINER), which although previously observed in D-array as mentioned in $\S3$, has a shallower HI image than the galaxies described here.  The HI image of this galaxy \citep{sim87,hib01} reveals that NGC~266 is interacting with the Seyfert~2 galaxy Markarian~348 \citep[which has a listed redshift of $z=0.014$ in][placing it just outside our selected redshift range]{ver98}.  Including this result would only further strengthen the conclusions we draw from our study.

The results for five galaxies (four Seyferts and one H~II galaxy) have previously been reported by \citet{gre04}, and are not repeated here (except for a short dicussion of Markarian~1157 for clarification).  This leaves a remaining eighteen galaxies in our sample, the results for which are presented below.  These twenty-three galaxies comprise our ensemble sample from which we draw our statistical results.  The measured HI properties of these galaxies  \citep[including all those reported in][]{gre04} are summarized in Table~4, where we list the integrated HI flux density S(HI) and associated uncertainty, the systemic HI velocity $\rm v_{sys}$(HI), the corresponding HI gas mass M(HI), whether they appear disturbed in HI and hence the interacting group to which they belong based on an objective definition of the observed disturbances (see $\S4.1$), and whether they are visibly disturbed based on images from the 2nd Digitized Sky Survey (DSS2).  The HI flux densities were computed from the total intensity (zeroth moment) maps, and converted to HI gas masses in the usual fashion assuming that the HI emission is optically thin.  

We reached our targeted detection threshold for all our sample galaxies except for NGC~1167 (LINER), UGC~3995 (Seyfert~2), and Markarian~341 (unclassified Seyfert type).  These three galaxies were more badly affected by intermittent RFI than the remaining sample, and so more of their data had to be discarded.   This resulted in a rms noise fluctuation that is a factor of $\sim$1.1 higher for Markarian~341, and a factor of $\sim$1.3 higher for both NGC~1167 and UGC~3995; the degradation in sensitivity is not severe, although (perhaps just coincidentally) UGC~3995 and NGC~1167 are the only two galaxies in our sample for which we did not detect HI disturbances.  We also detected in HI a number of galaxies in the same field of view as our target galaxies.  Those neighboring galaxies lying within the primary beam (i.e., within a radius of $\sim$15\farcm8 about phase center) are listed in Table~5, and are referred to by their names from optical catalogs or, when we fail to find them listed in these catalogs, their positional coordinates with the prefix HI.  We also list their projected separation from their corresponding active galaxies, optical redshifts where available, as well as their corrected apparent $B$-band magnitudes and Hubble types as cataloged in the Hyperleda database unless otherwise noted.  Their measured HI properties are summarized in Table~6 \citep[including all those reported in][]{gre04}.

The HI moment and channel maps of all the galaxies in our ensemble sample, apart from those previously reported by \citet{gre04}, are shown in Figures~1--36.  The HI moment maps of neighboring galaxies listed in Table~5 not already shown in Figures~1--36 are shown in Figures~37--43.  For display purposes (i.e., to preserve a uniform rms sensitivity throughout our maps), the maps shown in the next section have not been corrected for the primary beam attenuation (which is in any case negligible over the field of view shown).  The integrated HI flux densities and computed HI gas masses listed in Tables~4 and 6 have, of course, been corrected for the primary beam attenuation.  We have overlaid each integrated HI intensity (zeroth moment) map on an optical (either the red or blue filters) image from the DSS2.  The latter provides a relatively uniform set of optical images for all our sample galaxies, and was used to search for any disturbances in optical starlight.

\subsection{Individual Galaxies}
We present here the results for each galaxy, highlighting features of most relevance to this work.  Based on these HI maps, we can separate the individual galaxies in our ensemble sample into the following three groups.

\begin{itemize}

\item[Group I.] Those that clearly show HI tidal features tracing interactions with neighboring galaxies.  These features are in the form of tidal bridges that connect the two interacting galaxies, or an extension from one or both galaxies in the direction of the other (i.e., an incomplete tidal bridge).  In addition, features in the form of tidal tails comprising a protrusion or curved extension from one galaxy on the side away from the other interacting galaxy are usually seen.  Thirteen of our sample galaxies fall into this group.


\item[Group II.] Those that clearly show both spatial and kinematic disturbances in HI, but which cannot be directly (but are probably linked) to interactions with neighboring galaxies where detectable.  These disturbances are in general less prominent than the tidal features observed in the first group.  Four of our sample galaxies fall into this group.
                               
\item[Group III.] Those that show marginal or no detectable HI disturbances.  Six of our sample galaxies fall into this group, with at least two exhibiting possible disturbances.

\end{itemize}

\subsubsection{Group I: HI tidal features tracing interactions with neighboring galaxies}  Twelve of our sample galaxies exhibit HI tidal features that clearly trace interactions with neighboring galaxies.  In nearly all cases, counterparts of these HI tidal features also are seen in the interacting neighboring galaxies.  By contrast, only two of these galaxies exhibit optical disturbances in DSS2 images that can be traced to interactions with neighboring galaxies.  

Four galaxies --- Markarian 1, NGC~7679, NGC~7682, and ARK~539, all Seyfert~2s --- exhibit a complete HI tidal bridge with a neighboring galaxy.  NGC~7679 and NGC~7682 are interacting with each other, and are the only interacting Seyfert pair in our sample.  Of these, only NGC~7679 appears disturbed in the optical.  Another eight galaxies --- NGC~5548, Markarian~461, Markarian~1157 \citep[reported in][]{gre04}, NGC~841, Markarian~573, NGC~7591, NGC~7469 \citep[reported in][]{gre04}, Markarian~1040, arranged in descending order of the observed prominence of their tidal features --- exhibit either incomplete or relatively short tidal bridges, and usually also tidal tails.  One is a Seyfert 1 (Markarian 1040), one a Seyfert~1.5 (NGC~5548), four Seyfert~2s (Markarian~461, Markarian~1157, Markarian~573, and NGC~7469), one LINER (NGC~841), and one an unclassified Seyfert (NGC~7591).  Of these, only one (Markarian~1040) exhibits optical tidal features.  NGC~5548 does not appear to be optically disturbed in DSS2 images, but in a deep ground-based optical image was found to exhibit a linear feature that we confirm to be a tidal feature in HI.  NGC~7469 does not exhibit any appreciable optical disturbances, but its neighboring galaxy IC~5238 exhibits prominent tidal features produced presumably by interactions with NGC~7469.

To this group we add Markarian~341, which shows a tidal tail in the optical, and forms an optically-overlapping pair with its larger and even more optically-disturbed neighbour NGC~169.  We cannot spatially separate these two galaxies in our HI observations, but detect extended tidal features from NGC~169 that we attribute to interactions with Markarian~341.  Thus, altogether, Group~I contains thirteen galaxies, of which only three exhibit detectable tidal features or disturbances in optical starlight based on DSS2 images.

Apart from extended tidal features, the HI disks of the majority of these galaxies have kinematics that deviate strongly from normal rotating disks.  Detected here on spatial scales of $\sim$20~kpc or larger, the observed kinematic perturbations correspond to disturbances on a global scale.\newline

\hspace{3.0 cm}
Markarian~1
\newline

Markarian~1 (Mrk~1) is a Sb galaxy hosting a Seyfert~2 nucleus.  A neighboring galaxy with comparable size and brightness, NGC~451, is visible about 1\farcm9 (40~kpc) to the south-east of Mrk~1, differing in optical systemic velocity by only $\sim$$100 {\rm \ km \ s^{-1}}$ (all optical systemic velocities from the NASA Extragalactic Database, NED).  Neither galaxies, however, exhibit any visible disturbances in the optical, nor does Mrk~1 exhibit any detectable bar even at the angular resolution of the HST \citep{mal98}.

Mrk~1 has previously been imaged in HI with the Giant Meterwave Radio Telescope (GMRT) by \citet{oma02} at a factor of $\sim$2 higher angular resolution than that attained here.  It is instructive to compare their results (in particular, their Figs.~5 and 6) with ours (Fig.~1).  \citet{oma02} find that the HI morphology and kinematics of Mrk~1 are disturbed, whereas those of the neighboring galaxy NGC~451 appear normal.  No direct link, however, can be made between the observed disturbances in Mrk~1 and tidal interactions with NGC~451.

On the other hand, our HI image (Fig. 1) shows that Mrk~1 and NGC~451 shares a common gaseous envelope.  An examination of the channel maps (Fig.~2) reveals a tidal bridge connecting the two galaxies, as well as tidal tails on opposing sides of the two galaxies.  By contrast, the gaseous disturbances detected by \citet{oma02} cannot be clearly associated with tidal features.  Our HI image is a factor of $\sim$5.6 times more sensitive in flux density than that of \citet{oma02}; furthermore, our synthesized beam is a factor of $\sim$4 larger in area, resulting in a factor of $\sim$22 higher brightness sensitivity than the HI image of \citet{oma02}.  The higher brightness sensitivity of our HI map allows us to detect more extended and dimmer emission characteristic of tidal features.

Consistent with the higher angular-resolution but lower sensitivity image of \citet{oma02}, our image reveals that the spatial distribution and kinematics of the HI disk in Mrk~1 is strongly perturbed.  Although the largest variation in velocity is along the optical major axis, the velocity does not change from redshift to blueshift along this axis as would be expected of a normal rotating disk.  Like \citet{oma02} we detect relatively weak HI absorption at velocities lower than that spanned by the HI emission (Fig. 2). These absorption features are most likely due to the HI clouds close to the central region of the galaxy which absorbed the continuum photons from the AGN. Note that these absorptions do not affect the moment maps, and are not responsible for the irregular kinematics observed.  By contrast, the interacting neighboring galaxy NGC~451 exhibits what appears to be a normal rotating disk, as confirmed in the higher angular-resolution image of \citet{oma02}.  This trend towards more disturbed HI kinematics in the active galaxy compared with its interacting neighboring inactive galaxy is seen in a number of interacting pairs in our sample.\newline

\hspace{1.7 cm}
NGC~7679 and NGC~7682
\newline

NGC~7679 and NGC~7682 are a closely-separated pair of Seyfert~2 galaxies with virtually identical optical systemic velocities.  Pairs of Seyfert galaxies are comparatively rare, and this is the only such example in our sample.  NGC~7679 is a barred S0-a galaxy and NGC~7682 a barred Sab galaxy, with NGC~7679 lying $\sim$4\farcm5 ($\sim$99~kpc) to the south-west of NGC~7682.  NGC~7679 is clearly disturbed in the optical, exhibiting a particularly prominent extension to the east of the galaxy; i.e., in a direction closely towards NGC~7682.  On the other hand, NGC~7682 appears to be normal in the optical, even at the angular resolution of the HST \citep{mal98}.

Like Mrk~1, our HI image of NGC~7679 and NGC~7682 (Fig.~3) reveals that the two galaxies share a common gaseous envelope.  The channel maps (Fig. 4) clearly show a tidal bridge connecting the two galaxies, as well as tidal tails on roughly opposing sides of these galaxies.  The HI kinematics of both galaxies also are disturbed, with the isovelocity contours exhibiting a twist along their HI major axes.\newline

\hspace{3 cm}
Ark 539
\newline

ARK~539 is a spiral galaxy with an unclassified Hubble type (in HyperLeda) hosting a Seyfert~2 nucleus.  A relatively faint galaxy, PGC~2370360, with no previously reported redshift lies $\sim$$3\farcm4$ (74~kpc) to the south-east of ARK~539.  There is no evidence for optical disturbances in either galaxy.  

Our HI image (Fig.~5) and channel maps (Fig.~6) reveal a curved tidal bridge connecting the southern or south-western side of ARK~539 with the south-western side of PGC~2370360.  In addition, the HI gas on the eastern side of ARK~539 curls towards the south in the direction of PGC~2370360.  The HI kinematic axis of ARK~539 twists with radial distance from the center of the galaxy, following the same curl as the HI spatial distribution.  PGC~2370360 has a narrow HI width suggestive of a face-on galaxy in accord with its optical image.  Although therefore more difficult to judge whether the HI gas in this galaxy also is disturbed kinematically, its HI gas does not appear to be as strongly disturbed spatially as ARK~539.

We also detected HI gas in two other galaxies in the same field, PGC~2369294 and HI~1828$+$503 (Fig.~41).  PGC~2369294 lies $\sim$1\farcm5 ($\sim$33~kpc) to the south-east of HI~1828$+$503.  The systemic velocities of PGC~2369294 and HI~1828$+$503 differ by about $\sim$$250 {\rm \ km \ s^{-1}}$ from those of ARK~539 and PGC~2370360, and these two pairs of galaxies do not appear to be interacting with each other.  PGC~2369294 and HI~1828$+$503 share a common HI envelope, with at least PGC~2369294 exhibiting disturbed HI kinematics indicative of interactions with HI~1828$+$503.\newline


\hspace{3 cm}
NGC~5548
\newline

NGC~5548 is a S0-a ring galaxy hosting a Seyfert 1.5 nucleus.  In the DSS2 image, the optical morphology of this galaxy appears to be asymmetric, with an extended spiral arm emerging from the western side of the galaxy towards the south before winding north.  \citet{mal98} noted that NGC~5548 appears normal in their HST image.  In the deep ground-based optical image, however, \citet{tys98} detected a linear feature (length $\sim$35~kpc) with a low surface brightness emerging from NGC~5548 towards the south-east.  \citet{tys98} suggest that this feature is a remnant tidal tail from a past interaction involving NGC~5548.  NGC~5548 does not have a detectable bar even at the angular resolution of the HST \citep{pog02}.  A relatively dim inclined disk galaxy, PGC~1725892, which has a comparable optical systemic velocity as NGC~5548, lies $6\farcm0$ ($\sim$132~kpc) to the south-west of NGC~5548.  This galaxy does not appear to be visibly disturbed in the optical.

We detected in HI both NGC~5548 and PGC~1725892, as well as the faint galaxy SDSS~J141824.74+250650.7 lying $\sim$$5\farcm9$ ($\sim$130~kpc) to the south-east of NGC~5548 (Figs.~7 and 8).  The galaxy SDSS~J141824.74+250650.7 differs in systemic velocity by $\sim$$200 {\rm \ km \ s^{-1}}$ from NGC~5548, and does not appear to be optically disturbed.  Of these three galaxies, the HI gas in NGC~5548 appears to be the most strongly disturbed.  NGC~5548 exhibits a prominent HI extension on the southern side of the galaxy that points east towards SDSS~J141824.74+250650.7.  We identify this feature as an incomplete tidal bridge produced by interactions with SDSS~J141824.74+250650.7.  This feature lies close to, but is spatially displaced from, the low-surface-brightness linear optical feature found by \citet{tys98} as mentioned above.  SDSS~J141824.74+250650.7 exhibits a protuberance in HI on its north-western side, which we identify as the counterpart of the same incomplete tidal bridge from NGC~5548.  NGC~5548 also exhibits a less prominent extended HI feature on the north-western side of the galaxy that may be a tidal tail produced by interactions with SDSS~J141824.74+250650.7.  More likely, this feature is an incomplete tidal bridge produced by interactions with PGC~1725892, as PGC~1725892 exhibits a prominent HI extension on its north-eastern side that we identify as an incomplete tidal bridge produced by interactions with NGC~5548.

\citet{tys98} found that NGC~5548 has a de Vaucouleurs (i.e., elliptical galaxy like) radial surface brightness profile, and suggest that this galaxy may comprise a recent merger between two gas-rich galaxies.  Our HI image, however, suggests that NGC~5548 is more likely a disk galaxy that has been strongly disturbed by recent tidal interactions with PGC~1725892 and SDSS~J141824.74+250650.7.\newline


\hspace{2.6 cm}
Markarian~461
\newline

Markarian~461 (Mrk~461) is a Sab galaxy with a Seyfert~2 nucleus.  No optical disturbances nor bars are apparent in this galaxy even at the angular resolution of the HST \citep{pog02}.  A relatively dim galaxy, PGC~2044413, differing in optical systemic velocity by $\sim$$150 {\rm \ km \ s^{-1}}$, lies just $\sim$1\farcm1 ($\sim$22~kpc) to the south-west of Mrk~461.  A brighter inclined disk galaxy, PGC~48870, can be seen at a larger angular separation of $\sim$3\farcm9 ($\sim$82~kpc) also to the south-west of Mrk~461; there are no reported redshift measurements for this galaxy.  Neither PGC~2044413 nor PGC~48870 appear to be disturbed in the optical.

We detected both Mrk~461 and PGC~48870 in HI (Fig.~9), but apparently not PGC~2044413 as can be seen mostly clearly in the channel maps (Fig.~10).  (With a different systemic velocity, PGC~2044413 should appear in a different set of channels than Mrk~461.)  The HI kinematic axis of Mrk~461 is oriented along its optical major axis.  The HI major axis, however, is elongated approximately orthogonal to the HI kinematic axis, intersecting PGC~2044413 and also in a direction closely intersecting PGC~48870.  The HI gas in PGC~48870 also is disturbed, with the HI major and kinematic axis twisting away from the optical major axis on the western side of PGC~48870 towards Mrk~461.  We therefore attribute the elongated HI feature from Mrk~461 towards PGC~48870 to an incomplete tidal bridge, and the elongated HI feature from Mrk~461 away from PGC~48870 to a tidal tail, produced by tidal interactions between these two galaxies.  We cannot rule out, however, that these features also may have been produced by interactions between Mrk~461 and PGC~2044413, which lies closer in projection than PGC~48870.\newline



\hspace{2.6 cm}
Markarian~1157
\newline

A detailed description of Markarian~1157 (Mrk~1157), which does not appear to be disturbed in the optical, can be found in \citet{gre04}.  Not immediately obvious in the HI images of both Mrk~1157 and its neighboring galaxy PGC~5848 shown in \citet{gre04} is that the north-west to south-east HI extensions seen in both galaxies lie along a line joining these galaxies; i.e., PGC~5848 lies $\sim$10\farcm4 ($\sim$202~kpc) to the south-east of Mrk~1157.  These extensions are closely aligned with the major HI kinematic axes but not with the corresponding optical major axes of both galaxies, and trace an incomplete tidal bridge and opposing tidal tails.\newline


\hspace{3 cm}
NGC~841
\newline

NGC~841 is barred Sab galaxy hosting a LINER.  A comparably bright galaxy, NGC~834, differing in optical systemic velocity by only $\sim$$50 {\rm \ km \ s^{-1}}$, lies $\sim$10\farcm7 ($\sim$209~kpc) to the north-west of NGC~841.  Neither NGC~841 nor NGC~834 appear to be optically disturbed in the DSS2 image.

Our HI image of NGC~841 (Fig.~11) reveals a rotating gas disk that appears to be normal on the south-eastern side, but which is disturbed both spatially and kinematically on the north-western side; i.e., on the side facing NGC~841.  The HI disk of NGC~834 is also disturbed, with the HI kinematic axis having a different orientation than the optical major axis of the galaxy.  In addition, the HI disk of NGC~834 appears to be disturbed (elongated) on the side facing towards and away from NGC~841.  All these HI disturbances can also be seen in the channel maps (Figs.~12 and 13).  We attribute the disturbed HI features in both NGC~841 and NGC~834 on their sides facing each other as an incomplete tidal bridge, and the disturbed HI feature in NGC~834 on the side away from NGC~841 as a tidal tail.\newline



\hspace{3.0 cm}
Markarian~573
\newline

Markarian~573 (Mrk~573) is a double-barred S0-a galaxy hosting a Seyfert~2 nucleus.  The nuclear bar is clearly defined in the HST image of \citet{pog02}.  A relatively dim galaxy, PGC~1226059, with no previously reported redshift, lies $\sim$3\farcm8 ($\sim$83~kpc) to the north-east of Mrk~573.  Neither galaxies appear to be optically disturbed in the DSS2 image, nor Mrk~573 even at the angular resolution of the HST \citep{mal98,pog02}.

We detected both Mrk~573 and PGC~1226059 in HI (Figs.~14 and 15).  The HI emission of PGC~1226059 is not spatially resolved, but reveals that this galaxy has a systemic velocity that differs by only $\sim$$100 {\rm \ km \ s^{-1}}$ from that of Mrk~573.  The HI gas in Mrk~573 is clearly disturbed spatially and kinematically.  The HI kinematic axis is better aligned with the optical major axis on the north-eastern side of center, but twists away from this axis on the south-western side.  In addition, a prominent extension can be seen towards the south-eastern side of Mrk~573.  These disturbances are located on the side of Mrk~573 opposite to PGC~1226059, and likely traces a tidal tail produced by interactions with its northern neighbour.\newline



\hspace{3 cm}
NGC~7591
\newline

NGC~7591 is a barred Sb galaxy with an uncertain Seyfert classification in \citet{ver98}.  This galaxy does not appear to be disturbed in the optical.  \citet{kim95} classify NGC~7591 as a LINER.  On the other hand, \citet{ver97} show that NGC~7591 has a composite Seyfert and H~II nuclear spectrum.  As Seyfert galaxies quite commonly exhibit enchanced circumnuclear star formation, we will assume here that NGC~7591 is indeed a Seyfert galaxy.  An inclined disk galaxy, PGC~214933, with no previously reported redshift, lies $\sim$1\farcm8 ($\sim$39~kpc) to the south-west of NGC~7591.  Amram et al. (1994) suggested that this galaxy is warped based on their optical image.

\citet{tho04} have presented a relatively shallow HI image of NGC~7591, taken at a factor of $\sim$2 high angular resolution in C-array with the VLA.  In their image, PGC~214933 was not detected in HI.  By comparison, we detected not just PGC~214933, but also a relatively dim galaxy lying $\sim$3\farcm6 ($\sim$77~kpc) to the east of NGC~7591, hereafter referred to as HI~2318$+$065 (Figs.~16 and 17).

The HI disk of NGC~7591 is clearly asymmetric (Fig.~16).  The main HI body is much more extended on the north-western side than on the south-eastern side.  In addition, the HI gas exhibits a prominent extension on the south-eastern side reminiscent of a tidal feature.  NGC~7591 and its south-western neighbour PGC~214933 share a common HI envelope, which in the channel maps (Fig.~17) is more clearly revealed as a tidal bridge.  NGC~7591 and its eastern neighbour HI~2318$+$065 also are linked in HI, which again in the channel maps is more clearly revealed as a tidal bridge.  This is one of the few active galaxies in our sample, together with NGC~5548, which are interacting with multiple neighboring galaxies.\newline



\hspace{3 cm}
NGC~7469
\newline

A detailed description of NGC~7469 can be found in \citet{gre04}.  NGC~7469 is disturbed in HI by tidal interactions with IC~5283, which is disturbed both optically and in HI.\newline


\hspace{2.6 cm}
Markarian~1040
\newline

Markarian~1040 (Mrk~1040) is a highly inclined Sbc galaxy hosting a Seyfert~1 nucleus.  It forms an optically-overlapping pair with PGC~212995 (UGC~1935), which lies just $\sim$19\arcsec\ ($\sim$7~kpc) to the north of Mrk~1040.  PGC~212995 has a comparable optical redshift to, but is both smaller and dimmer than, Mrk~1040.

It is not clear whether Mrk~1040 is disturbed in the optical.  Although not reportedly disturbed in the HST image of \citet{mal98}, in the DSS2 image Mrk~1040 exhibits relatively dim and highly extended spiral-like arms on both the north-eastern and south-western sides.  These features extend approximately equidistance from center and appear to be symmetric, and hence may just be ordinary spiral arms.  On the other hand, these spiral-like arms may actually be tidal features dragged outwards in the plane of the galactic disk.  \citet{war78} find that PGC~212995 also exhibits strong but narrower emission lines than Mrk~1040, and suggest that the activity in PGC~212995 is caused by interactions with Mrk~1040.  In addition, \citet{amr92} find that the optical velocity field of PGC~212995 is disturbed, and suggest that this disturbance is caused by interactions with Mrk~1040.  By contrast, \citet{amr92} attribute the asymmetric optical velocity field in Mrk~1040 to a large-scale and massive bar.

Mrk~1040 and PGC~212995 cannot be spatially separated at the angular resolution of our HI observation.  In our HI map (Fig.~18), the centroid in integrated HI intensity as well as body of HI as a whole is centered on Mrk~1040, the larger galaxy.  There is no evidence from our channel maps (Fig.~19) for any detectable emission from PGC~212995.  The HI disk of Mrk~1040 is clearly disturbed both spatially and kinematically, although these disturbances can be much more clearly seen in the channel rather than moment maps.  A prominent HI feature can be seen on the eastern side of the galaxy, located beyond the extended sprial-like optical arm on this side of the galaxy.  This HI feature curls north at its easternmost periphery; i.e., to the side where PGC~212995 is located.  A similarly extended HI feature can be seen on the western side of the galaxy, again located beyond the extended sprial-like optical arm on this side of the galaxy.  We attribute these features to tidal disturbances caused by interactions with PGC~212995.\newline

\hspace{3.0 cm}
Markarian~341
\newline

Markarian~341 (Mrk~341) is a S0-a galaxy that forms an optically-overlapping pair with NGC~169, a bigger and brighter galaxy located just $\sim$23\arcsec\ ($\sim$7~kpc) to the north.  These two galaxies have nearly identical optical redshiftes. NGC~169 is a Sab galaxy, and is clearly disturbed in the optical.  It exhibits a prominent extension to the east that was presumably produced by tidal interactions with Mrk~341.  Mrk~341 itself also exhibits a relatively dim extension to the south, which is on the side opposite to NGC~169 and hence a tidal tail.  Mrk~341 has an unclassified nuclear spectral type in \citet{ver98}.  \citet{kee85} suggest that Mrk~341 hosts a LINER, whereas \citet{ver97} find a composite H~II and Seyfert nuclear spectra type, as in the case of NGC~7591.  We shall henceforth assume that Mrk~341, just like NGC~7591, is indeed a Seyfert galaxy.

Mrk~341 and NGC~169 cannot be spatially separated at the angular resolution of our HI observation.  In our HI map (Fig.~20), the centroid in integrated HI intensity as well as body of HI as a whole is centered on NGC~169, the larger galaxy.  There is no evidence from our channel maps (Fig.~21) for any detectable emission from Mrk~341.  The HI disk of NGC~169 is clearly disturbed both spatially and kinematically.  It shows a prominent extension to the east coinciding with the optical extension mentioned above but extending to much larger distances.  In addition, the HI kinematic axis twists away from the HI and optical major axes on both the eastern and western sides of the galaxy, and on both sides curl in a direction towards Mrk~341.  All these disturbances can be most obviously linked to tidal interactions with Mrk~341.



\subsubsection{Group II: HI disturbances likely produced by interactions}
Four of our sample galaxies clearly exhibit both spatial and kinematic perturbations in HI, but unlike the situation in Group~I their observed perturbations cannot be immediately linked to interactions with neighboring galaxies (if any).  By contrast with their appearences in HI, only one appears to be disturbed optically in DSS2 images.  Two --- MS~04595$+$0327 (Seyfert~1) and NGC~513 (Seyfert~1.9), both reported in \citet{gre04} --- have neighboring galaxies with projection separations of only about 50~kpc detected also in HI, providing circumstantial evidence that their disturbances are indeed caused by tidal interactions with these galaxies.  Neither, however, appear disturbed in the optical.  One --- UGC~1395 (Seyfert~1.9) --- is connected to a separate HI feature that may be related to a projected neighboring galaxy.  The optical disk of this galaxy is asymmetric, suggesting a disturbance.  The last --- Markarian~1158 (H~II galaxy), reported in \citet{gre04} --- does not have any neighboring galaxies detected in HI, and does not appear to be disturbed in the optical although clearly disturbed in HI.  Although the observed perturbations in all four galaxies are in general less prominent than the tidal features observed in Group~I, these perturbations nevertheless correspond to disturbances on a global scale. 

Detailed descriptions of MS~04595$+$0327, NGC~513, and Markarian~1158 can be found in \citet{gre04}.  Here, we present only the results for UGC~1395.\newline


\hspace{3 cm}
UGC~1395
\newline

UGC~1395 is a Sb ring galaxy hosting a Seyfert~1.9 nucleus.  Although not classified as a barred galaxy in Hyperleda, \citet{mlr95} find a strong bar in the near-IR.  UGC~1395 exhibits a relatively dim but highly extended spiral arm on its western side that gives this galaxy an asymmetric appearence in the optical.  A relatively faint galaxy, PGC~1304037, with an unknown optical redshift lies $\sim$2\farcm9 ($\sim$65~kpc) west of UGC~1395; i.e., on the same side as the extended western spiral arm.

Our HI image of UGC~1395 (Fig.~22) reveals that this galaxy is disturbed both spatially and kinematically.  First, the HI kinematic axis exhibits a prominent twist on the northern side of the galaxy.  Second, UGC~1395 is linked to a HI condensation lying $\sim$2\farcm1 ($\sim$48~kpc) to the west.  This HI condensation spans a different velocity range than UGC~1395; furthermore, an examination of the channel maps (Fig.~23) reveals that this condensation is indeed a separate object.  This HI condensation lies in close proximity to, but its centroid does not coincide with, PGC~1304037; we believe this HI condensation is most likely associated with PGC~1304037, but to be conservative we consider the possibility that it is not and refer to this condensation as HI~0155$+$066.  Both the moment and channel (Fig.~23) maps reveal a tidal bridge between UGC~1395 and HI~0155$+$066.

\subsubsection{Group III: Weak or no HI disturbances}
Three of the galaxies in our sample --- Markarian~993 (Seyfert~1.5), UGC~3157 (Seyfert~2) and Markarian~1419 (LINER) --- exhibit relatively weak HI disturbances.  Two of these (UGC~3157 and Markarian~1419) appear to be disturbed optically in DSS2 images, with the possible HI disturbances apparently coinciding with the optical disturbances.  Two (Markarian~993 and UGC~3157) have neighboring galaxies detected in HI that are excellent candidates for having produced the observed disturbances in the active galaxies.  Another galaxy in our sample --- IRAS~14082$+$1347 (LINER) --- also appears to exhibit HI disturbances, but this galaxy was detected at a relatively low signal-to-noise ratio.  It does not appear to be optically disturbed in DSS2 images.  The remaining two galaxies --- UGC~3995 (Seyfert~2) and NGC~1167 (LINER)  --- do not exhibit detectable HI disturbances. In the optical, UGC~3995 (Seyfert~2) forms an overlapping pair with UGC~3995A, and may therefore comprise an interacting system.  NGC~1167 (LINER) is the only galaxy in our sample where we find no evidence whatsoever for disturbances in either the optical or HI.\newline


\hspace{3.0 cm}
Markarian~993
\newline

Markarian~993 (Mrk~993) is a Sa galaxy hosting a Seyfert~1.5 nucleus.  No optical disturbances nor bars are visible in this galaxy even at the angular resolution of the HST \citep{mal98,pog02}. 

In our HI image (Fig.~24), we detected Mrk~993 as well as a comparatively optically-dim galaxy lying $\sim$$8\farcm7$ ($\sim$175~kpc) to the east, hereafter referred to as HI~0126$+$321.  These galaxies have systemic velocities that differ by $\sim$$450 {\rm \ km \ s^{-1}}$, the largest of any confirmed or candidate interacting galaxy pairs in our sample.  In the moment maps (Fig.~24), the HI disk of Mrk~993 appears to be quite normal except for a weak protuberance on its northern-eastern side.  This protuberance can also be seen in the channel maps (Fig.~25), and points east in the direction of HI~0126$+$321.  Weak protuberences pointing west, in a direction opposite to HI~0126$+$321, also can be seen on the northern-eastern side of the HI disk.  We tentatively associate these weak disturbances to tidal interactions with HI~0126$+$321.\newline



\hspace{3 cm}
UGC~3157
\newline

UGC~3157 is a barred Sbc galaxy hosting a Seyfert~2 nucleus.  It has a highly asymmetric appearence in the DSS2 image caused by a relatively dim and diffuse swirl on the eastern side of the galaxy.  There are, however, no previously cataloged optical galaxies nor are there any obvious galaxies in the DSS2 image within several arcminutes of UGC~3157.  

Our HI image of UGC~3157 (Fig.~27) reveals that this galaxy has an asymmetric HI disk.  The centroid in integrated HI intensity is shifted east of the geometrical and kinematic center of the HI disk, as well as the center of the optical galaxy.  An examination of the channel maps (Fig.~28) reveal that the HI emission is stronger on the eastern side of the disk, tracing the diffuse optical swirl mentioned above.  The disturbance seen in the optical is therefore also apparent in HI, albeit less clearly presumably because of our lower angular resolution.  The HI kinematics of the galaxy, however, appears normal.

We also detected in HI a relatively dim object located $\sim$4\farcm8 ($\sim$95~kpc) to the north-west of UGC~3157, differing in systemic velocity by $\sim$$150 {\rm \ km \ s^{-1}}$.  In the DSS2 image, this object appears to be diffuse, apart from a relatively bright star-like object at its north-western periphery, and hence is a genuine galaxy.  It is henceforth referred to as HI~0446$+$185.  The HI kinematic axis of this galaxy coincides with its optical major axis, providing additional confidence that we have correctly identified the optical counterpart of HI~0446$+$185.  It is possible that the optical and HI disturbances seen in UGC~3157 is caused by tidal interactions with HI~0446$+$185.\newline

\hspace{3.0 cm}
Markarian~1419
\newline

Markarian~1419 (Mrk~1419) is a Sa galaxy hosting a LINER nucleus.  It has an asymmetric appearence in the DSS2 images, caused by a relatively dim and diffuse swirl on the north-eastern side of the galaxy.  There are, however, no cataloged galaxies in the vicinity of Mrk~1419 at a comparable redshift.

Our HI image of Mrk~1419 (Fig.~29) reveals that this galaxy has an asymmetric HI disk.  Analogous to the case of UGC~3157, in Mrk~1419 the centroid in integrated HI intensity is shifted north-east of the geometrical and kinematic center of the HI disk, as well as center of the optical galaxy.  In the channel maps (Fig.~30), the HI emission is stronger and clearly disturbed on the north-eastern side of the disk, coincident with the diffuse optical swirl mentioned above.  The disturbance seen in the optical is therefore also apparent in HI, albeit less clearly presumably because of our lower angular resolution.  The HI kinematics of the galaxy, however, appears normal, just as in the case UGC~3157.  Compared with UGC~3157, the only difference is that we did not detect in HI any neighboring galaxy (nor any other galaxies in the same field of view) that could have caused the disturbances seen in Mrk~1419.\newline


\hspace{2.6 cm}
IRAS~14082$+$1347
\newline

IRAS~14082$+$1347 has an uncertain Hubble type in Hyperleda.  This galaxy hosts a LINER nucleus.  There are no cataloged galaxies at a comparable redshift, nor are there any obvious galaxies in the DSS2 images, in the vicinity of IRAS~14082$+$1347.

We detected IRAS~14082$+$1347 at a relatively low signal to noise (Fig.~31).  This galaxy has a HI gas mass of only $\sim$$3.5 \times 10^7 {\rm \ M_{\odot}}$, which is at least an order of magnitude smaller compared with the other galaxies in our sample.  The HI body of this galaxy is clearly displaced from the optical center of the galaxy.  In addition, the HI major and apparently also kinematic axes of this galaxy is clearly different from its optical major axis.

\citet{hei94} reported shells around the amorphous body of IRAS~14082$+$1347.  Given its appearence in the DSS2 image, and the HI properties of this galaxy, it is possible that IRAS~14082$+$1347 is an early-type (elliptical) galaxy that has cannibalized a gas-rich dwarf galaxy.\newline


\hspace{3 cm}
UGC~3995
\newline

UGC~3995, a Sb galaxy, is one of an optically-overlapping pair of spiral galaxies separated by just $\sim$0\farcm5 ($\sim$11~kpc) lying at the same redshift.  UGC~3995 is the brighter eastern galaxy, and hosts a Seyfert~2 nucleus.  The western galaxy is UGC~3995A, a Sbc galaxy.  (We take the names of the galaxies from the NED.  In some literature sources, UGC~3995 is referred to as UGC~3995A \citep[as in][]{ver98}, and UGC~3995A referred to as UGC~3995B.)  Despite their overlapping disks, \citet{marz99} found no signs of interactions between the two galaxies from their HST image.

This pair of galaxies cannot be separated in our HI image (Fig.~33).  An examination of the channel maps (Fig.~34) indicate that we have detected HI emission from both galaxies.  In both cases, the HI traces normal rotating disks with no obvious signs for disturbances.\newline


\hspace{3 cm}
NGC~1167
\newline

NGC~1167 is a S0 galaxy that hosts a LINER nucleus.  The galaxy does not appear to be disturbed in the optical.

Our HI image of NGC~1167 (Figs.~35 and 36) reveals an enormous gas disk around this galaxy, the largest seen in our sample.  The HI kinematics are consistent with a normal rotating disk with no signs of disturbances.  In addition to HI in emission, we also detected spatially-unresolved HI in absorption (spanning a velocity width of $\sim$$250 {\rm \ km\ s^{-1}}$) towards the center of the galaxy. Apart from NGC~1167, we detected in HI a relatively dim galaxy henceforth referred to as HI 0302+352 lying $\sim$6\farcm9 (147~kpc) to the north-east of NGC~1167.  This galaxy also is not obviously disturbed in the optical nor HI.

\subsection{Other Galaxies detected in Target Fields}
As described in \S~4.1, we detected a number of galaxies in HI around our target objects, many of which we show to be definitely (in Group~I), probably (in Group~II), or possibly (in Group~III) interacting with the target galaxies.  In addition, we detected a number of other galaxies mostly at larger projected separations from the target galaxies than the (candidate) interacting neighboring galaxies.  The optical and HI images (moment maps) of these other galaxies are shown in Figures~38--43.  Many of these galaxies, like the (candidate) interacting neighboring galaxies, are relatively faint (and small) in the optical, and are not obvious candidates for neighboring galaxies.  The properties of all the neighboring galaxies are listed in Tables~4 and 5.  We separate, in our discussion, those neighboring galaxies lying within the $\sim$$80\%$ response (radius of $\sim$$9\farcm1$ or $\sim$190~kpc) of the primary beam and hence observed at approximately the same sensitivity as the active galaxies, and those lying further out up to the $\sim$$50\%$ (radius of $\sim$$15\farcm8$ or $\sim$320~kpc) response of the primary beam and hence observed at a significantly lower sensitivity.

Altogether, we detected nineteen surrounding galaxies within $\sim$190~kpc of our ensemble sample.  Ten of these are confirmed interacting neighboring galaxies in Group~I, three are probable interacting neighboring galaxies in Group~II, and two are possible interacting neighboring galaxy in Group~III (i.e., totaling  fifteen of the nineteen).  
Thus, the vast majority of galaxies detected in HI within $\sim$190~kpc of the target galaxies are confirmed or likely to be interacting with the target galaxies.  An additional eight galaxies lie between $\sim$190~kpc and $\sim$320~kpc of the target galaxies, two of which (in the range 200--210~kpc) we show to be definitely (Group~I) interacting with the target galaxies.  

In Figure~44 (upper panel), we show the number as well as cumulative fraction of Seyfert galaxies in our ensemble sample having (candidate) interacting neighboring galaxies (i.e., in Groups~I, II, or III), plotted as a function of their projected separations (for the few with multiple interacting neighboring galaxies, their nearest such neighbor).  Within 210~kpc, there are only two neighboring galaxies not identified to be (possibly) interacting with the same Seyfert galaxy, comprising two of the three neighboring galaxies (the other being identified as the interacting neighbor) around ARK~539; including these galaxies do not therefore change the statistics.  A dashed line is drawn at a projected separation of 190~kpc, within which our HI detection threshold is relatively uniform.  As can be seen, the fraction of Seyferts with neighboring galaxies increases steeply with projected separation up to $\sim$100~kpc; by this separation, $\sim$$80\%$ of the Seyferts already have neighboring galaxies.  Within a projected separation of $\sim$180~kpc, nearly all ($\sim$$95\%$) have neighboring galaxies.

In Figure~44 (lower panel), we plot the difference in radial velocity between the Seyfert and all their neighboring galaxies galaxies detected in HI within 210~kpc.  We use different symbols for those in the different interacting classes (e.g., our defined Group~I, II, and III), as well as those not apparently interacting with the Seyfert galaxy.  The vast majority of the confirmed or probable interacting companion galaxies lie within $\pm 100 {\rm \ km \ s^{-1}}$, and all but one within about $\pm 200 {\rm \ km \ s^{-1}}$.

\subsection{Ensemble Statistics}
We now compile statistics relevant to this study from the results for the ensemble sample of twenty-three galaxies that we mapped in HI.  Two are Seyfert~1s, three are Seyfert~1.5s, two are Seyfert~1.9s, nine are Seyfert~2s, two are unclassified Seyfert types (i.e., total of eighteen Seyferts in all), four are LINERs, and one is a H~II galaxy.

\subsubsection{Comparison with optical starlight}
Based on DSS2 images, six of the twenty-three galaxies ($\sim$$26\%$) --- comprising five Seyferts and one LINER --- are disturbed in the optical.  Among these six, only three (NGC~7679, Mrk~1040, and Mrk~341, all Seyferts) exhibit optical tidal features that can be confidently (in the case of NGC~7679, with the aid of our HI image) traced to interactions with neighboring galaxies.  Two of these galaxies (Mrk~1040 and Mrk~341) comprise the closest interacting pairs, and overlap in the optical with their interacting neighbour.  Two (NGC~7679 and Mrk~1040) exhibit corresponding HI tidal features, and one (Mrk~341, itself not detected in HI) corresponding HI tidal features in its interacting companion.  The remaining three galaxies that appear to be disturbed in the optical (UGC~1395, UGC~3157, and Mrk~1419) all exhibit asymmetric disks in the form of an extended spiral arm.  Two (UGC~1395 and UGC~3157) are Seyferts, and one (Markarian~1419) is a LINER.  All three also exhibit corresponding HI disturbances.

The remaining seventeen of the twenty-three galaxies ($\sim$$74\%$) are not visibly disturbed in DSS2 images.  One (NGC~7469) lies in an interacting system where only its neighbour exhibits tidal features.  Another (NGC~5548) exibits an extended feature detectable only in a deep ground-based optical image that we confirm from our HI image to be a tidal feature.  In dramatic contrast with their optical appearences, fifteen of these seventeen galaxies are disturbed in HI.  Ten (Mrk~1, NGC~7682, ARK~539, NGC~5548, Mrk~461, Mrk~1157, NGC~841, Mrk~573, NGC~7591, and NGC~7469), comprising nine Seyferfs and one LINER, exhibit HI tidal features that can be directly traced to interactions with neighboring galaxies (i.e., those in Group~I).  The remaining five (MS~04595+0327, NGC~513, Mrk~1158, Mrk~993, and IRAS~14082+1347) all exhibit spatial and kinematic disturbances in HI, and which in four cases is likely produced by interactions with neighboring galaxies detected also in HI.  Three are Seyferts, one a Liner, and one a H~II galaxy.  Only two of the twenty-three galaxies in our ensemble sample do not exhibit any visible disturbaces in (optical and) HI.  One is a Seyfert, and the other a LINER.  

In summary, only six of the twenty-three galaxies ($\sim$$26\%$) in our ensemble sample are visibly disturbed in DSS2 images.  In dramatic contrast, twenty-one of the same twenty-three galaxies ($\sim$$91\%$) are disturbed in HI, thirteen ($\sim$$62\%$) of which are interacting with neighboring galaxies (Group~I), another four ($\sim$$19\%$) probably interacting with neighboring galaxies (Group~II), and the remaining four ($\sim$$19\%$) weakly disturbed (Group~III).  In nearly all these cases, the HI is disturbed not just spatially but also kinematically on global scales.  Of the eighteen galaxies in our ensemble sample classified as Seyferts, only five ($\sim$$28\%$) are visibly disturbed in DSS2 images.  By contrast, seventeen ($\sim$$94\%$) exhibit HI disturbances, twelve ($\sim$$67\%$) of which are interacting with neighboring galaxies (Group~I), another three ($\sim$$17\%$) probably interacting with neighboring galaxies (Group~II), and the remaining two ($\sim$$12\%$) weakly disturbed (Group~III).

Our parent sample contains all the Seyfert galaxies in the CfA Seyfert and Markarian catalogs within our selected redshift range (see \S2).  The trends seen in our ensemble sample are also seen in the complete subsamples from these catalogs.  Of the seven Seyfert galaxies in our sample present also in the CfA Seyfert catalog \citep{huc83}, only one ($\sim$$14\%$) of these galaxies is visibly disturbed in DSS2 images.  By contrast, in HI, five ($\sim$$71\%$) are found to be interacting with neighboring galaxies (Group~I), and the other two ($\sim$$29\%$) disturbed (Group~III).  Fifteen of the galaxies in our sample are present in the Markarian catalog \citep{mar89}, twelve of which (including two classified here as LINER and H~II respectively) are in our ensemble sample.  Four of these twelve galaxies ($\sim$$33\%$) are visibly disturbed in DSS2 images, whereas nine ($\sim$$75\%$) are interacting (Group~1) and three ($\sim$$25\%$) probably or weak disturbed (Group~II or III) in HI.  Among just the ten galaxies classified here as Seyferts, three ($\sim$$30\%$) are visibly disturbed in DSS2 images, whereas nine ($\sim$$90\%$) are interacting with neighboring galaxies (Group~1) and one ($\sim$$10\%$) weakly disturbed (Group~III) in HI.

\subsubsection{Comparison with other galaxies in target fields}
The other galaxies detected in our target fields, lying within a projected separation of $\sim$320~kpc from the target objects, may provide a natural sample that is best matched in mean environment to our ensemble sample.  These galaxies, however, are clearly not matched in optical luminosities.  Where cataloged and have apparent $B$-band magnitudes listed in Hyperleda, many have corresponding absolute $B$-band magnitudes close to or spanning the range between the Small and Large Magellanic Clouds.   (If placed at the same distance, the SMC would have a corrected apparent $B$-band magnitude $m_B \approx 18.5$, and the LMC $m_B \approx 16.5$.)  Those not cataloged or do not have apparent $B$-band magnitudes listed in Hyperleda likely have even smaller optical luminosities.  These galaxies also are not matched in other optical properties such as size or Hubble type; where cataloged in Hyperleda, many are later-type Sb-c galaxies.

Eleven of the nineteen neighboring galaxies ($\sim$$58\%$) detected at approximately the same sensitivity in HI as the ensemble sample (i.e., those lying within $\sim$190~kpc of the target galaxies) are disturbed in HI.  All eleven of these disturbed neighboring galaxies are among the fifteen we identify to be definitely (Group~I), probably (Group~II), or possibly (Group~III) interacting with the target galaxies.  The trend therefore is towards a higher incidence of HI disturbances among the target compared with their neighboring galaxies.  We caution, however, that this may be an observational selection effect, as the HI gas masses and extents of the neighboring galaxies are usually much smaller than that of the target galaxies, and hence any comparable level of disturbances more difficult to detect.



\subsubsection{Tidal interactions versus mergers}
Our HI images allow us to directly determine the cause of the gaseous disturbances in many of the target galaxies.  Thirteen of the galaxies in Group~I (twelve Seyferts and one LINER) out of the ensemble sample of twenty-three galaxies ($\sim$$57\%$) have recently experienced tidal interactions with neighboring galaxies.  Adding to this the three galaxies in Group~II and the two galaxies in Group~III (all Seyferts) that have likely been disturbed by neighbouring galaxies also detected in HI, eighteen of the twenty-three galaxies in our ensemble sample ($\sim$$78\%$) have likely experienced recent tidal interactions with neighboring galaxies.  Among those classifed as Seyferts, twelve of eighteen ($\sim$$67\%$) have experienced recent tidal interactions with neighboring galaxies (Group~I), and another five ($\sim$$28\%$) likely or probably have experienced such interactions (Group~II or III); i.e., in total, seventeen of the eighteen ($\sim$$94\%$) Seyfert galaxies.

Only three of the twenty-one galaxies that are disturbed in HI do not have neighboring galaxies cataloged in the optical or detected in HI.  One is a H~II galaxy, and the remaining two LINERS.  Thus, only three of the twenty-one galaxies in our ensemble sample ($\sim$$14\%$) are good candidates for recent (minor) mergers, none of which are Seyferts.  Note that another two galaxies (one Seyfert and one LINER) are not disturbed in either the optical or HI (and for which we cannot immediately rule out the possibility of ancient minor mergers).  Of course, in all these cases, the disturbances or interacting neighboring galaxies may simply not have been detectable in our HI observations.

\section{INTERPRETATION}
\subsection{Bias towards Interacting Systems?}
Given the prevalence of tidal interactions or disturbances seen in our HI images, is it possible that our sample is somehow biased towards interacting systems?  This could possibly happen if we had imposed in our selection criteria a threshold in HI gas mass based on single-dish observations, thus ensuring that all the galaxies are detected.  The more gas-rich systems could preferentially comprise interacting or merging systems; i.e., contain two or more galaxies in the single-dish beam, or constitute merger remnants.   We remind the reader, however, that our selection is entirely blind to the HI content of the galaxies; indeed, the HI gas mass of the detected Seyfert galaxies span the broad range $\sim$$10^8$--$10^{10} {\rm \ M_\sun}$, with a median of $\sim$$6 \times 10^9 {\rm \ M_\sun}$.  For the same reason, if only a relatively small number of the more gas-rich galaxies are detected, this could skew the results towards interacting systems.  As mentioned at the start of $\S4$, only two of our sample galaxies were not detected (with two others omitted because of RFI); in one case (Mrk~359) the presence of a strong continuum source resulted in a higher noise detection threshold than targeted, and in the other case (2E~0507+1626) the redshift is not accurately known \citep[see][]{gre04}.  Because we detected the vast majority of the galaxies in our sample, our results are unlikely to be biased towards the more gas-rich and therefore perhaps preferentially interacting systems.

Is there any evidence from their optical morphology that we had somehow, perhaps just because of small number statistics, preferentially selected interacting systems?  Only five of the eighteen galaxies ($\sim$$28\%$) in our ensemble sample classified as Seyferts are visibly disturbed in DSS2 images.  Including the remaining four galaxies classified as LINERS and one galaxy classified as H~II, six of the twenty-three galaxies ($\sim$$26\%$) are visibly disturbed in DSS2 images.  The relatively small fraction of optically disturbed --- let alone (apparently) optically interacting --- galaxies in our sample is typical of that found in other optical studies utilizing different samples and images.

For instance, \citet{der98a} have studied thirty-four Seyfert galaxies selected from the CfA Seyfert catalog (nearly half of which can also be found in the Markarian catalog) using much deeper optical images than DSS2 images.  Only about $26\%$ (9/34) of the Seyferts in their sample were classified as disturbed (a fraction comparable with that found in our sample).  The majority (7/9) of these disturbed Seyferts and their interacting neighbors form overlapping pairs, with some perhaps comprising late mergers.  Most (28/34) of the Seyferts in that sample have absolute $B$-band magnitudes \citep[as listed in][]{ver98} spanning the same range as our sample.  Considering only this subsample, about $29\%$ (8/28) of the Seyferts are disturbed.  These galaxies, however, occupy a much broader redshift range than our sample, spanning $0.0033 \leq z \leq 0.0422$ with an average redshift of $z=0.0189$; i.e., reaching both lower and higher redshifts than our sample.  Disturbances are of course more difficult to detect in more distant galaxies, although \citet{der98a} found no apparent distance bias in their ability to detect disturbances with the disturbed galaxies spread quite evenly over the entire redshift interval of their sample.

Although a direct comparison between the incidence of optical disturbances we and \citet{der98a} find in our respective samples cannot be made, we note that the disturbances and/or tidal features \citet{der98a} see in all their disturbed galaxies can also be seen in DSS2 images.  Thus, a study using just DSS2 images would reach the same conclusion as \citet{der98a} on the incidence of optical disturbances in their sample.  There are only two galaxies in our sample in common with that of \citet{der98a}, Mrk~461 and NGC~5548.  \citet{der98a} find neither galaxies to be optically disturbed, whereas we find both to be interacting with neighboring galaxes detected also in HI (i.e., Group~I).

\subsection{Bias towards Higher Density Environments?}
We also checked to see whether our sample is, perhaps just by chance, somehow biased towards higher density environments compared with the general Seyfert population.  Eleven of the twenty-three Seyfert galaxies in our parent sample were studied by \citet{dah84}, about half of whose sample comprise Markarian galaxies.  This study remains to date that which has found the largest disparity between the fraction of Seyfert galaxies likely to have physical companions ($\sim$$15\%$) compared with a control sample of inactive galaxies ($\sim$$3\%$).  Only two of the eleven ($\sim$$18\%$) Seyfert galaxies in our parent sample were listed by \citet{dah84} to have projected neighboring galaxies, similar to that found in their overall sample of Seyfert galaxies.  
Nine (including the two with projected neighboring galaxies) of these eleven galaxies are in our ensemble sample, all of which we find in our HI images to be definitely or probably interacting with neighboring galaxies; i.e., seven ($\sim$$78\%$) are in Group~I, and two ($\sim$$22\%$) in Group~II.

Why did \citet{dah84} not find many of the interacting neighboring galaxies that we detected in HI?  The limiting magnitude used in his work was 18.3, comparable with that of the SMC at the distance of our sample and hence sufficient to detect most of the neighboring galaxies that we detected.  \citet{dah84}, however, searched for projected neighboring galaxies only at angular separations within three times the optical diameters of the Seyfert (or control inactive) galaxies (the search radius therefore differs for Seyfert/inactive galaxies with different optical diameters).  Projected neighboring galaxies (genuine neighboring galaxies, as it turns out) were found for NGC~513 and NGC~7469, but not for Mrk~1, 352, 359, 573, 1157, NGC~5548, 7682, ARK~539, and UGC~1395.  In all the latter cases (excluding Mrk~352 and Mrk~359, which are not in our ensemble sample) apart from UGC~1395, the actual interacting companion galaxy is located beyond the search radius used; the possible interacting companion galaxy of UGC~1395 (Group~II) is unrecognizable even in modern (DSS2) optical images as a galaxy.  

\citet{raf95} have searched for an excess of projected neighboring galaxies around all (261) spectroscopically-classified Seyfert galaxies listed in the \citet{ver91} catalog at $z \lesssim 0.05$ (mostly at significantly lower redshifts than the upper limit) and $\delta \geq -23^{\circ}$.  No significant differences were found between the frequency of Seyfert~1 and 2 galaxies with projected neighboring galaxies, but this frequency was found to be much higher than in matched samples of inactive field spiral galaxies.  Twelve of the Seyfert galaxies in our parent sample are included in that study.  Only four of these twelve ($33\%$) Seyfert galaxies are listed in \citet{raf95} as having projected neighboring galaxies, a fraction comparable with that found for their entire sample ($\sim$$26\%$). 
Ten of these twelve galaxies are in our ensemble sample, seven of which ($70\%$) are interacting with neighboring galaxies detected also in HI (Group~I), two others ($20\%$) probably interacting with neighboring galaxies detected also in HI (Group~II), and the remaining galaxy ($10\%$) probably disturbed in HI (Group~III).

\citet{raf95} used the same search criteria as \citet{dah84}, but required the Seyfert and projected neighboring galaxy to have a difference in apparent magnitude not exceeding 3.  They identified projected neighboring galaxies for Mrk~1, 1040, NGC~7469, IRAS~1408+1347 (almost certainly the genuine neighboring galaxies for the first three galaxies; we do not detect an interacting neighboring galaxy for IRAS~1408+1347, classified here as a LINER), but not for Mrk~352, 359, 573, 993, 1157, NGC~513, 5548, and 7682.  Once again, in the latter case (apart from NGC~513) the interacting neighboring galaxies detected in HI lie beyond the search radius used.  \citet{raf95}, unlike \citet{dah84}, failed to identify a projected neighboring galaxy for NGC~513; its interacting neighboring galaxy falls outside the magnitude difference criteria used by \citet{raf95}.

There is therefore no independent evidence from optical studies that our Seyfert sample is somehow biased towards higher density environments than the general Seyfert population.  On the contrary, when we can find a large number of the Seyfert galaxies in our sample included in optical studies, the fraction with projected neighboring galaxies as identified in those studies is identical with the fraction found in their overall sample.

\subsection{Prevalence of Tidal Interactions in HI Gas}
The results from our HI study, by contrast with those from optical studies, dramatically alter our perception of the frequency (and strength) of large-scale disturbances in local Seyfert galaxies.  Of the eighteen galaxies in our ensemble sample of twenty-three galaxies classified as Seyferts, seventeen ($\sim$$94\%$) exhibit HI disturbances, compared with only five ($\sim$$28\%$) exhibiting optical disturbances in DSS2 images.  In twelve (including Mrk~341) of these eighteen cases ($67\%$), the HI disturbances can be directly traced to tidal interations with neighboring galaxies detected also in HI.  In the other five of six cases, the observed HI disturbances are likely produced by tidal interactions with neighboring galaxies detected also in HI.  Thus, nearly all ($\sim$$94\%$) of the Seyfert galaxies in our ensemble sample can be directly or indirectly shown to be involved in recent or ongoing tidal interactions with neighboring galaxies.


Our parent sample comprises all known (circa 1998) Seyfert (disk) galaxies within a given redshift interval in the northern hemisphere.  These galaxies were compiled from the same surveys that found a large fraction of the known population, and are therefore likely representative of this heterogeneously selected general population.  The only intended selection bias is towards Seyfert nuclei with relatively high optical luminosities; specifically, within about two orders of magnitude of the lower limit for QSOs.  Our results therefore suggest that a large fraction (if not the vast majority) of local Seyfert galaxies with high nuclear luminosities are involved in or have recently experienced tidal interactions with neighboring galaxies. Much more often than not, these interactions are visible only in HI gas.  The prevalence of such tidal interactions constitutes a shift in our understanding of local Seyfert galaxies, and is the first of the two important results of our study.

We note that there are only two Seyfert~1 and three Seyfert~1.5, compared with eleven Seyfert~2, in our ensemble sample (the remaining two being unclassified Seyfert types).  Given that many optical studies find an excess of projected neighboring galaxies around Seyfert~2 but not Seyfert~1 galaxies, it may be that these two classes of galaxies have genuinely different environments and that our results can be confidently extrapolated only to Seyfert~2 galaxies.  On the other hand, all the Seyfert~1 and Seyfert~1.5 galaxies in our sample also are interacting or disturbed in HI (three classified in our Group~I, one in Group~II, and one in Group~III), and so (with the caveat of small number statistics) we do not see a dichotomy in the frequency of tidal interactions between Seyfert~1 and Seyfert~2 galaxies in HI.

\subsection{Implicating Tidal Interactions in Triggering AGNs}
The observed prevalence of tidal interactions in Seyfert galaxies does not, by itself, necessarily imply that such interactions are responsible for initating events that lead to their AGN activity.  It may be that the HI disks of many galaxies --- both active and inactive --- are moderately to strongly disturbed at their outskirts by tidal interactions with neighboring galaxies.  To test whether this is the case, we have conducted a HI imaging survey of a matched control sample of inactive galaxies at an identical spatial resolution and detection threshold to those attained here.  The results of that study are reported in a companion paper by \citet{tan07}.  Below, we provide just a synopsis of that work.

\citet{tan07} selected a sample of twenty-seven inactive galaxies from the CfA Redshift Survey \citep{huc83,huc95} that were closely matched on a one-to-one basis in Hubble type, to the degree possible galaxy optical luminosity, and in range of size and inclination to our parent active galaxy sample.  Other than that, the inactive galaxies were essentially selected at random from the field, and in this way we hoped representative of the environment of inactive field galaxies.  To reduce the required observing time, \citet{tan07} selected inactive galaxies at redshifts of $0.0075 \leq z \leq 0.0085$, half that of our active galaxy sample.  All the galaxies in the control sample were imaged in HI at the same spatial resolution ($\sim$20~kpc) and detection threshold as our active galaxy sample.

Twenty-one of the twenty-seven inactive galaxies detected in HI comprise the ensemble control sample.  These galaxies have approximately the same distribution in HI gas masses as our ensemble sample of active galaxies.  \citet{tan07} found that only four of these twenty-one galaxies ($\sim$$19\%$) exhibit spatial and usually also kinematic disturbances on galactic ($\gtrsim 20$~kpc) scales.  One of these disturbed galaxies has since been found to be a Seyfert~2 galaxy, and so only three of the twenty ($15\%$) inactive galaxies in our ensemble control sample are actually disturbed in HI.  By contrast, seventeen of the twenty-one ($\sim$$81\%$) galaxies show no HI disturbances whatsover on galactic ($\gtrsim 20$~kpc) scales.  Excluding again the one disturbed galaxy since found to be a Seyfert~2 galaxy, seventeen of the twenty ($85\%$) inactive galaxies in our ensemble control sample are not disturbed in HI.

The contrast between the incidence of HI disturbances in the active (or just Seyfert) and inactive galaxy samples is dramatic, and difficult to explain away as selection or other spurious effects.  In \citet{tan07}, we compare in greater detail the contrasting results found for the Seyfert and control samples in our HI study with the results found for such samples in optical studies, and show that all these results need not be incompatible given the different criteria used to find candidate neighboring galaxies in optical studies. The prevalence of HI disturbances in our sample of active galaxies, but the low incidence of disturbances at a comparable level in our control sample of inactive galaxies, directly implicates tidal interactions in initiating events that lead to optically-luminous Seyfert activity in a large fraction of local disk galaxies.  This constitutes another shift in our understanding of local Seyfert galaxies, and is the second of the two important results of our study. As before, our results can be more confidently extrapolated to Seyfert~2 galaxies, although we find no difference in our results for Seyfert~1 galaxies.

\section{DISCUSSION}
If our sample is indeed representative of the general population, our results have interesting and important implications for the nature of gravitational interactions involving Seyfert galaxies, the timescale for gas to be transported inwards to fuel their central SMBHs, and the lifetime of optically-luminous Seyfert activity.

\subsection{Bound Interactions or Flybys?}
Do the observed tidal interactions constitute gravitationally bound interactions or flybys?  Seyfert galaxies preferentially comprise luminous early-type disk galaxies, as is the case also in our sample as can be seen in Table~1.  To esimate whether the Seyfert galaxies in our ensemble sample are gravitationally bound to their confirmed or likely interacting neighboring galaxies, let us assume that these Seyfert galaxies have a typical mass (including dark matter) of $\sim$$10^{12} {\rm \ M_\sun}$, comparable with that of our own Milky Way galaxy.  The projected spatial separation between the Seyferts and their (candidate) interacting neighbors ranges from a minimum of $\sim$6~kpc to a maximum of $\sim$210~kpc, with the majority ($\sim$$85\%$) lying within $\sim$100~kpc.  The projected escape velocity (i.e., velocity along the plane of the sky) therefore ranges from $\sim$$1200 {\rm \ km \ s^{-1}}$ at the closest separation to $\sim$$200 {\rm \ km \ s^{-1}}$ at the largest separation, and corresponds to $\sim$$800 {\rm \ km \ s^{-1}}$ at $\sim$100~kpc.  For comparison, most of the interacting neighboring galaxies differ in radial velocities by $\lesssim 100 {\rm \ km \ s^{-1}}$, and nearly all by $\lesssim 200 {\rm \ km \ s^{-1}}$, from their respective Seyfert galaxies.  Scaling as the square root of the Seyfert galaxy mass, the escape velocity remains in most cases larger than the observed difference in radial velocities even if the Seyfert galaxy mass is an order of magnitude smaller than has been assumed.   Because the overall difference in radial velocities should be statistically comparable with the overall difference in velocities along the plane of the sky, most of the interacting neighboring galaxies are likely to be gravitationally bound to their respective Seyfert galaxies.

\subsection{Gas Infall Timescale}
The relative simplicity of many of the tidal features observed --- clearly identifiable tidal bridges and tails --- suggests that most of the systems studied are caught at a relatively early phase of an encounter; i.e., seen just after their first close approach.  If this is indeed the case, then our results place a stringent upper limit on the timescale for gas to be transported inwards to fuel the central SMBH.  For our ensemble sample of Seyfert galaxies at a mean redshift of $z \approx 0.016$ (distance $\sim$71~Mpc), the time, t, since closest approach (when tidal forces are strongest, and during which theoretical simulations suggest tidal distortions first become appreciable) in an interacting pair is:
\begin{equation}
t \approx 2 \times 10^8 \ {\rm \theta_{kpc} \over (20~kpc)} \ \left({v \over 100 \rm \ km/s}\right)^{-1} \ {\rm yrs},
\end{equation}
where $\theta_{\rm kpc}$ is the spatial separation (in kpc) between the two galaxies, and $v$ is their relative velocity in the plane of the sky.  As mentioned above, the projected spatial separation between the (candidate) interacting neighboring galaxies and the Seyfert galaxies in our ensemble sample ranges from a minimum of $\sim$6~kpc to a maximum of $\sim$210~kpc, with most ($\sim$$85\%$) within $\sim$100~kpc.  If we assume that their relative velocities along the plane of the sky is comparable with their relative radial velocities of usually $\lesssim 100 {\rm \ km \ s^{-1}}$, then the time since closest approach is $\sim$$10^8$--$10^9 {\rm \ yrs}$.  This is comparable with the orbital period in the outer regions of normal disk galaxies, allowing perhaps sufficient time for infalling gas even at quite large radial distances from center to reach the nucleus.

Numerical simulations show that mergers between two gas-rich disk galaxies with comparably large masses (i.e., major mergers) can be highly effective at transporting large amounts of gas into the centers of the merger remnants \citep[e.g.,][]{bar92,bar96,mih96b}.  The same also can be true for minor mergers, where a relatively massive galaxy cannibalizes a lower mass but gas-rich galaxy \citep[e.g.,][]{mih96a}. On the other hand, while tidal interactions can strongly perturb the outskirts of galaxies, it is not immediately obvious how such interactions promote gas infall in the inner regions of the galaxy.  More theoretical simulations of such tidal interactions are needed to address the possible gas infall mechanisms involved.

\subsection{Longevity of Seyfert Activity}
If indeed most systems have just experienced their first close encounter, then the inferred timescale since closest approach of $10^8$--$10^9 {\rm \ yrs}$ also places a stringent upper limit on the longevity of Seyfert activity produced during a given interaction episode.  If many Seyfert nuclei shine continuously or sporadically for much longer timescales, then we should see a high fraction of systems with relatively complex tidal features or in (near) mergers.  Instead, there are very few good candidates for mergers, with most of the interacting systems having well separated neighboring galaxies and quite simple tidal features.

\section{SUMMARY AND CONCLUSIONS}
We reported in this paper the first systematic imaging program in atomic hydrogen (HI) gas of a representative sample of Seyfert galaxies.  Our parent sample comprised all twenty-seven disk galaxies in the \citet{ver98} catalog, plus another in the \citet{ver00} catalog, classified as either Seyfert, LINER, or H~II galaxies (the LINER and H~II sometimes classified as Seyferts in other catalogs or studies) at redshifts $0.015 \leq z \leq 0.017$ (i.e., essentially uniform distances) and with absolute $B$-band magnitudes $-19 \geq M_B \geq -23$ (i.e., within about two orders of magnitude of the minimum optical luminosities for QSOs, as confirmed for over half with measured [OIII] luminosities).  We successfully imaged twenty-three of these galaxies in HI at essentially uniform sensitivities and spatial resolutions ($\sim$20~kpc), comprising eighteen Seyferts, four LINERs, and one H~II galaxy.  These twenty-three galaxies comprised our ensemble sample from which we drew the follo
 wing statistical results.

\begin{itemize}

\item[1.]  In DSS2 images, only $\sim$$26\%$ (6/23) of the galaxies are visibly disturbed in the optical.  Among those classified as Seyferts, only $\sim$$28\%$ (5/18) are visibly disturbed in the optical.  These results are comparable with other studies involving different Seyfert samples and using different optical images, all of which show that only a small minority of such galaxies are visibly disturbed in optical starlight.

\item[2.]  In dramatic contrast with their appearences in optical starlight, $\sim$$91\%$ (21/23) of the galaxies are disturbed in HI gas.  The HI gas is disturbed not just spatially, but usually also kinematically, on global (i.e., galactic) scales.  Among those classified as Seyferts, $\sim$$94\%$ (17/18) exhibit HI disturbances.

\item[3.]  The disturbances in $\sim$$57\%$ (13/23) of the galaxies can be directly traced to tidal interactions with neighboring galaxies detected also in HI.  Among those classified as Seyferts, the fraction is $\sim$$67\%$ (12/18).  In another five galaxies (all Seyferts), the observed HI disturbances are most likely caused by tidal interactions with neighboring galaxies detected also in HI.  Including the latter, $\sim$$78\%$ (18/23) of the galaxies in our ensemble sample, and $\sim$$94\%$ (17/18) of those classified as Seyferts, can be directly or indirectly shown to have recently experienced tidal interactions with neighboring galaxies.

\item[4.]  The confirmed or probable interacting neighboring galaxies typically have projected separations of $\lesssim 100 {\rm \ kpc}$ and radial velocities differing by $\lesssim 100 {\rm \ km \ s^{-1}}$ from their respective Seyfert galaxies.  Many have optical $B$-band luminosities ranging from that of the Small to Large Magellanic Clouds.

\item[5.]  Three galaxies that are disturbed in HI (two LINERs and one H~II galaxy) do not have neighboring galaxies apparent in the optical or detected in HI.  These galaxies are the only good candidates for (minor) mergers.

\item[6.]  Only $\sim$$9\%$ (2/23) of the galaxies exhibit no detectable disturbances in HI, nor in the optical.  One is a Seyfert, and the other a LINER.  The Seyfert galaxy with no HI disturbances forms an optically-overlapping pair with a disk galaxy detected also in HI.

\end{itemize}

Our results, if found to prevail in studies incorporating larger samples, have three important implications.

\begin{itemize}

\item[1.]  The majority of local Seyfert galaxies with relatively high nuclear luminosities are involved in tidal interactions with gravitationally bound neighboring galaxies.  Much more often than not, the resulting tidal features are visible in HI gas but not in optical starlight (in DSS2 images).

\item[2.]  The incidence of HI disturbances in a matched control sample of inactive galaxies is only $\sim$$15\%$ \citep{tan07}.  The prevalence of HI disturbances in our Seyfert sample but low incidence of comparable HI disturbances in the control sample implicates tidal interactions in initiating events that lead to optically-luminous Seyfert activity in a large fraction of local field disk galaxies.

\item[3.]  The relative simplicity of the tidal features observed suggest that many Seyfert galaxies are in an early phase of an encounter with neighboring galaxies; i.e., seen just after the first close approach.  If this is indeed the case, then both the timescale for gas infall to fuel the central supermassive black hole, and the lifetime of Seyfert nuclei in their optically luminous phase, is $\lesssim 10^8$--$10^9 {\rm \ yrs}$.

\end{itemize}

Our work demonstrates in a systematic way the power of HI imaging observations to reveal tidal interactions where none are otherwise visible, and in most cases to positively identify and in many of the remaining cases pick out the most likely interacting neighboring galaxies.  The application of this technique to Seyfert galaxies has provided the first strong link between tidal interactions and luminous Seyfert activity.


\acknowledgments
\hspace{2.5 cm}
Acknowledgemetns
\newline

The National Radio Astronomy Observatory is a facility of the National Science Foundation operated under cooperative agreement by Associated Universities, Inc.  A part of the work presented in this paper was the basis of a Masters thesis by Liau Wen-Shuo, who sadly decided to leave astronomy.  We are grateful to Richard Chou for his help in putting together many of the images shown in this paper.  Both C.-Y. Kuo and Y.-W. Tang acknowledge the support of a Research Assistantship at the ASIAA where the bulk of this work was done.  J. Lim acknowledges the National Science Council of Taiwan for providing a grant in support of this work, and in support of Liau Wen-Shuo during his Masters thesis.  This research has made use of NASA's Astrophysics Data System Bibliographic Services, and the NASA/IPAC Extragalactic Database (NED) which is operated by the Jet Propulsion Laboratory, California Institute of Technology, under contract with the National Aeronautics and Space Administration.  We acknowledge the usage of the HyperLeda database (http://leda.univ-lyon1.fr).



{\it Facilities:} \facility{VLA}.

\clearpage
\clearpage
\begin{deluxetable}{lcccccccll}
\tablewidth{0 pt}
\tablecaption{Active Galaxy Sample}
\tablehead{
\colhead{}               & \colhead{R.A.}         &
\colhead{Decl.}          & \colhead{}             &
\colhead{Spectral}        & \colhead{}             &
\colhead{Hubble}   & \colhead{Inclination}  &
\colhead{log L$_{[OIII]}$}  & \colhead{Refs.}\\
\colhead{Name}           & \colhead{(J2000)}      &
\colhead{(J2000)}        & \colhead{z}            &
\colhead{Type}  & \colhead{$M_{B}$}      &
\colhead{Type}              & \colhead{(degrees)}  &
\colhead{(erg s$^{-1}$)}                     & \colhead{L$_{[OIII]}$}}
\startdata
Markarian 341   & 00 36 52.2 & 23 59 06 & 0.017 & S     & $-21.9$ & S0-a & 70.5 &...& \\
NGC 266         & 00 49 48.2 & 32 16 43 & 0.015 & LINER & $-20.6$ & Sab & 14.5&...& \\
Markarian 352   & 00 59 53.3 & 31 49 37 & 0.015 & S1.0  & $-19.5$ & S0   & 33.9&40.89&2\\
Markarian 1     & 01 16 07.2 & 33 05 22 & 0.016 & S2    & $-19.0$ & Sb   & 58.7&41.56, 41.85&1,2\\
NGC 513         & 01 24 26.8 & 33 47 58 & 0.016 & S1.9  & $-21.5$ & Sc   &61.9&...&\\
Markarian 993   & 01 25 31.4 & 32 08 11 & 0.017 & S1.5  & $-20.0$ & Sa   & 90.0&40.49&1\\
Markarian 359   & 01 27 32.5 & 19 10 44 & 0.017 & S1    & $-20.2$ & Sb   & 39.1&41.05, 41.13&1,2\\
Markarian 1157  & 01 33 31.2 & 35 40 05 & 0.015 & S2    & $-20.1$ & S0-a &40.2&41.39, 41.39&1,2\\
Markarian 1158  & 01 34 59.5 & 35 02 22 & 0.015 & H~II  & $-19.3$ & S?   &58.7&40.95&1\\
Markarian 573   & 01 43 57.8 & 02 20 59 & 0.017 & S2    & $-20.1$ & S0-a &28.0&41.92, 42.30&1,2\\
UGC 1395        & 01 55 22.1 & 06 36 42 & 0.017 & S1.9  & $-20.5$ & Sb   &54.8&40.89&2\\
NGC 841         & 02 11 17.3 & 37 29 49 & 0.015 & LINER & $-21.4$ & Sab  &64.8&...&\\
Markarian 1040  & 02 28 14.4 & 31 18 41 & 0.016 & S1.0  & $-19.2$ & Sbc  &81.1&40.98, 41.19&1,2\\
NGC 1167        & 03 01 42.4 & 35 12 21 & 0.016 & LINER & $-21.1$ & S0   &49.0&...&\\
UGC 3157        & 04 46 29.7 & 18 27 40 & 0.016 & S2    & $-19.9$ & Sbc & 35.2&...& \\
MS 04595$+$0327 & 05 02 09.0 & 03 31 51 & 0.016 & S1    & $-19.5$ & S?   & 50.2&...&\\
2E 0507+1626    & 05 10 45.5 & 16 29 56 & 0.017 & S1.5  & $-19.4$ & E?  & 39.7&...&\\
UGC 3995        & 07 44 09.2 & 29 14 50 & 0.015 & S2    & $-21.5$ & Sb   & 67.6&...&\\
Markarian 1419  & 09 40 36.5 & 03 34 38 & 0.015 & LINER & $-19.8$ & Sa   & 41.9&40.72&1\\
Markarian 461   & 13 47 17.7 & 34 08 56 & 0.016 & S2    & $-20.4$ & Sab  &43.6&...&\\
IRAS 14082+1347 & 14 10 41.7 & 13 33 27 & 0.017 & LINER & $-19.5$ & S?   & 53.5&...&\\
Markarian 1510  & 14 17 58.6 & 26 24 47 & 0.015 & H~II  & $-20.7$ & Sbc  & 48.2&...&\\
NGC 5548        & 14 17 59.6 & 25 08 13 & 0.017 & S1.5  & $-20.7$ & S0-a & 41.0&41.81, 41.91&1,2\\
ARK 539         & 18 28 48.2 & 50 22 22 & 0.016 & S2    & $-19.9$ & S?   &31.9&40.22&1\\
NGC 7469        & 23 03 15.6 & 08 52 26 & 0.017 & S1.5  & $-21.6$ & Sa   &30.2&42.00, 41.84&1,2\\
NGC 7591        & 23 18 16.3 & 06 35 09 & 0.017 & S     & $-21.6$ & SBb  & 67.1&...&\\
NGC 7679        & 23 28 46.6 & 03 30 42 & 0.016 & S2    & $-21.4$ & S0-a & 59.1&...&\\
NGC 7682        & 23 29 03.9 & 03 32 00 & 0.017 & S2    & $-20.7$ & Sab  & 22.7&41.46&2\\
\enddata
\tablecomments{Units of right ascension are hours, minutes, and seconds, and units of declination are degrees, arcminutes, and arcseconds.  Spectral type S denotes unclassified Seyfert galaxy.  All data apart from Hubble types and inclinations are from the \citet{ver98} catalog, except for UGC~3157 from the \citet{ver00} catalog.  UGC~3995 is referred to in \citet{ver98} as UGC~3995A.  Hubble types and inclinations are from the Hyperleda database, where E? and S? denote unclassified elliptical and spiral galaxies respectively. The [OIII] luminosities are computed from measurements by \citet{dah88} and/or \citet{whi92} assuming $H_{o} = 50 {\rm \ km \ s^{-1} \ Mpc ^{-1}}$ and q$_{0}$ $=$ 0.  In the last column, reference 1 refers to \citet{dah88} and refrence 2 to \citet{whi92}.}

\end{deluxetable}

\clearpage
\begin{deluxetable}{ll}
\tablewidth{0 pt}
\tablecaption{Complete Subsamples}
\tablehead{
\colhead{Name}               & \colhead{Catalog}}
\startdata
Markarian 341   & Mrk  \\
NGC 266         &  ...\\
Markarian 352   & Mrk\\
Markarian 1     & Mrk\\
NGC 513         & ...\\
Markarian 993   & Mrk, Cfa\\
Markarian 359   & Mrk\\
Markarian 1157  & Mrk\\
Markarian 1158  &Mrk\\
Markarian 573   & Mrk, Cfa\\
UGC 1395        & Cfa\\
NGC 841         &...\\
Markarian 1040  &Mrk\\
NGC 1167        &...\\
UGC 3157        &...\\
MS 04595$+$0327 &...\\
2E 0507+1626    &...\\
UGC 3995        &...\\
Markarian 1419  &Mrk\\
Markarian 461   &Mrk, Cfa\\
IRAS 14082+1347 &...\\
Markarian 1510  &Mrk\\
NGC 5548        &Mrk, Cfa\\
ARK 539         &...\\
NGC 7469        &Mrk, Cfa\\
NGC 7591        &...\\
NGC 7679        &Mrk\\
NGC 7682        &Cfa \\
\enddata
\tablecomments{Those galaxies in our sample included in the CfA Seyfert catalog \citep{huc83} and/or Markarian catalog \citep{mar89}, labelled in column 2 as CfA and Mrk respectively.}

\end{deluxetable}

\begin{deluxetable}{lcccccl} 
\tablewidth{0pt} 
\tablecaption{Observing Parameters}
\tablehead{ 
\multicolumn{1}{c}{} &
\multicolumn{2}{c}{Phase Center} &
\multicolumn{1}{c}{Flux} &
\multicolumn{1}{c}{Secondary} &
\multicolumn{1}{c}{Bandpass } &
\multicolumn{1}{c}{Observing} \\
\multicolumn{1}{c}{Name} &
\multicolumn{1}{c}{R.A.} &
\multicolumn{1}{c}{Decl.} &
\multicolumn{1}{c}{calibrator} &
\multicolumn{1}{c}{calibrator} &
\multicolumn{1}{c}{calibrator}  &
\multicolumn{1}{c}{date}
}
\startdata
Markarian 341   & 00 37 16.6 & 24 02 16 & 0137+331 & 0029+349 & 0029+349 & 2000/7/28; 8/20\\
Markarian 352  & 00 59 53.3 & 31 49 37 & 0137+331 & 0119+321 & 0119+321 & 2004/7/2\\
Markarian 1      & 01 16 04.7 & 33 09 53 & 0137+331 & 0137+331 & 0137+331 & 2000/8/21\\
NGC 513         &01 24 26.8 & 33 47 58& 0137+331 & 0137+331 & 0137+331 & 2001/10/30\\
Markarian 993   & 01 25 34.5 & 32 04 24 & 0137+331 & 0137+331 & 0137+331   & 2000/8/20\\
Markarian 359  &01 27 19.5 & 19 09 35& 0137+331 & 0129+236 & 0129+236     & 2004/7/2\\
Markarian 1157  & 01 33 20.6 & 35 41 42 & 0137+331 & 0137+331 & 0137+331   & 2001/10/30\\
Markarian 1158  & 01 34 59.5 & 34 58 46 & 0137+331 & 0137+331 & 0137+331& 2001/11/2\\
Markarian 573   & 01 43 57.8 & 02 20 60 & 0137+331 & 0149+059  & 0137+331   & 2004/7/2\\
UGC1395         & 01 55 24.5 & 06 43 56 & 0137+331 & 0141+138 & 0141+138 & 2000/8/20\\
NGC 841         & 02 11 17.3 & 37 29 49 & 0137+331 & 0336+323 & 0137+331  & 2000/8/20\\
Markarian 1040  & 02 28 35.0 & 31 18 60 & 0137+331 & 0137+331 & 0137+331 & 2000/7/28; 8/20\\
NGC 1167        & 03 01 42.4 & 35 12 21 & 0137+331 & 0336+323 & 0336+323   & 2000/8/20\\
UGC3157         & 04 46 48.9 & 18 24 53 & 0137+331 & 0521+166 & 0521+166   & 2000/8/21\\
MS 04595+0327   & 05 02 09.0 & 03 31 51& 0137+331 & 0503+020 & 0503+020   & 2001/10/30\\
2E 0507+1626    & 05 10 27.9 & 16 28 22 & 0137+331 &  0521+166   & 0521+166 &  2001/11/2 \\
UGC3995         & 07 44 28.4 & 29 17 06 & 0137+331 & 0741+312 & 0137+331   & 2000/8/20; 8/21\\
Markarian 1419  & 09 40 36.4 & 03 34 37 & 1331+305 & 1008+075 & 1331+305 & 2003/2/24\\
Markarian 461   & 13 47 17.7 & 34 08 55 &1331+305  & 1331+305  & 1331+305   & 2003/2/24\\
IRAS 14082+1347   & 14 10 41.4 & 13 33 28 & 1331+305 & 1347+122 & 1331+305   & 2003/2/21\\
Markarian 1510  & 14 17 58.6 & 26 24 47 & 1331+305 & 1331+305  & 1331+305  & 2004/7/8\\
NGC 5548        & 14 17 59.5 & 25 08 12 & 1331+305 & 1331+305 & 1331+305 & 2003/2/19\\
ARK 539         & 18 28 52.0 & 50 28 48 & 0137+331 & 1829+487 & 1829+487   & 2000/8/20; 8/21\\
NGC 7469        &  23 03 27.7 & 08 54 14 &  0137+331    &  2312+093     & 2312+093   &  2001/10/30 \\
NGC 7591        & 23 17 58.6 & 06 33 60  & 0137+331   & 2255+132 & 2255+132 & 2000/7/28; 8/21\\
NGC 7679        & 23 28 43.9 & 03 27 02 & 0137+331 & 2255+132 & 2255+132 & 2000/7/28; 8/21\\
NGC 7682        & 23 28 43.9 & 03 27 02 & 0137+331 & 2255+132 & 2255+132   & 2000/7/28; 8/21\\

\enddata
\tablecomments{Units of right ascension are hours, minutes, and seconds, and units of declination are degrees, arcminutes, and arcseconds.  The secondary calibrators, located close to the individual target sources, are used for amplitude and phase calibration.}
\end{deluxetable}

\clearpage
\begin{deluxetable}{lccccccl}
\tablewidth{0pt} 
\tablecaption{HI Gas Properties of Active Galaxies in Ensemble Sample}
\tablehead{
\multicolumn{1}{c}{} &
\multicolumn{1}{c}{S(HI)}&
\multicolumn{1}{c}{$\rm v_{sys}(HI)$}&
\multicolumn{1}{c}{M(HI)} &
\multicolumn{1}{c}{HI} &
\multicolumn{1}{c}{Group}&
\multicolumn{1}{c}{Optically}&
\multicolumn{1}{c}{Complete}\\
\multicolumn{1}{c}{Name} &
\multicolumn{1}{c}{(mJy~kms$^{-1}$)} &
\multicolumn{1}{c}{(kms$^{-1}$)} &
\multicolumn{1}{c}{($10^9 {\rm \ M_{\odot}}$)} &
\multicolumn{1}{c}{disturbed?}&
\multicolumn{1}{c}{}&
\multicolumn{1}{c}{disturbed?}&
\multicolumn{1}{c}{Subsample}
}
\startdata
Markarian 341   &... & ... & ... & yes & I &   yes&Mrk\\
Markarian 1     &1096$\pm$188 & 4771& 1.34$\pm$0.23  &yes &I&  no&Mrk\\
NGC 513         &1403$\pm$31 &5803& 2.57$\pm$0.06  & yes &II&   no&\\
Markarian 993   &5890$\pm$423 &4623& 8.18$\pm$0.58 & weak&III&no&Mrk,Cfa\\
Markarian 1157  &5518$\pm$34 &4503&5.94$\pm$0.07 & yes &I&no&Mrk\\
Markarian 1158  &1485$\pm$36 &4561&1.62$\pm$0.04 & yes&II&no&Mrk\\
Markarian 573  &713$\pm$130 &5083& 0.87$\pm$0.17 & yes &I&no&Mrk,Cfa\\
UGC 1395         &2249$\pm$363 &5134& 3.13$\pm$0.49 & yes&II& yes&\\
NGC 841         &5629$\pm$338 &4521& 6.08$\pm$0.36 & yes&I&no& \\
Markarian 1040  &33891$\pm$870&4947& 41.74$\pm$1.06& yes&I&yes&Mrk\\
NGC 1167        &10043$\pm$887 &4925& 12.29$\pm$1.03 & no &III&no&\\
UGC 3157         &5207$\pm$309 &4601& 5.93$\pm$0.36 & weak&III&yes&\\
MS 04595+0327   &817 $\pm$27 &4817&0.85$\pm$0.03 & yes &II&no&\\
UGC 3995        &6170$\pm$367 &4705& 6.67$\pm$0.39 & no &III&no&\\
Markarian 1419  &1681$\pm$271 &4916& 1.81 $\pm$0.27& weak&III&yes&Mrk\\
Markarian 461   &$1878\pm$134 &4850& 2.31$\pm$0.16 & yes&I&no&Mrk,Cfa\\
IRAS 14082+1347 &  25$\pm$ 68 &4854& 0.035$\pm$0.049& probably &III& no&\\ 
NGC 5548        & 793$\pm$224 & 5122& 1.1$\pm$0.29& yes &I&no&Cfa \\
ARK 539         & 445$\pm$115 &5123& 0.548$\pm$0.13& yes &I&no&\\
NGC 7469        &8700$\pm$38 & 4923 &10$\pm$1.2  & yes &I&no&Cfa\\
NGC 7591        &17240$\pm$403&4919& 23.98$\pm$0.55 & yes &I&no& \\
NGC 7679        &6366$\pm$276 &5129& 8.85$\pm$0.34& yes &I&yes&\\
NGC 7682        &7230$\pm$328 &5066& 10.05$\pm$0.46 & yes  &I&no&Cfa\\
\enddata
\tablecomments{Assignment of groups based on observed spatial-kinematic distribution of HI gas (see text): Group~I are those that show HI tidal features tracing interactions with neighboring galaxies (which also are detected in HI); Group~II are those that show both spatial and kinematic disturbances in HI, but for which the observed HI disturbances cannot be directly (but can often be indirectly) traced to tidal interactions with neighboring galaxies; and Group~III are those that show either weak or no HI disturbances.  Optical disturbances assessed based on DSS2 images, and for some confirmed from HI maps. In the last column,  those galaxies remarked with "Mrk" or "Cfa" comprise a complete subsample of Seyfert galaxies in either Markarian \citep{mar89} or CFA catalogue \citep{huc83} within out selected redshift interval. Note that Mrk 352, Mrk 359, and Mrk 1510
are not included in this table because these data were discarded due to either nondetection or RFI. The exclusion of these
three galaxies makes the Markarian subsample less "complete". }
\end{deluxetable}

\clearpage
\begin{deluxetable}{llllcccc}
\tablecolumns{8} 
\tablewidth{0pt} 
\tablecaption{Other Detected Galaxies in Active Galaxy Field} 
\tablehead{ 
\multicolumn{1}{c}{Active} & 
\multicolumn{1}{c}{Neighbouring} & 
\multicolumn{1}{c}{R.A.} &
\multicolumn{1}{c}{Decl.} &
\multicolumn{1}{c}{} &
\multicolumn{1}{c}{} &
\multicolumn{1}{c}{Separation} &
\multicolumn{1}{c}{Hubble} \\
\multicolumn{1}{c}{Galaxy} & \multicolumn{1}{c}{Galaxy} &
\multicolumn{1}{c}{(J2000)} & \multicolumn{1}{c}{(J2000)} &
\multicolumn{1}{c}{z} & 
\multicolumn{1}{c}{$m_B$} &
\multicolumn{1}{c}{(kpc)} &
\multicolumn{1}{c}{type}
}
\startdata
Markarian 341 & NGC 169     & 00 36 51.6 & 23 59 27 & 0.015434 & 12.99 & 7 & Sab \\
Markarian 1  &  NGC451    & 01 16 12.3& 33 03 51 & 0.016278 &14.30 & 39 & Sc \\
NGC 513    & PGC 212747  & 01 24 36.0& 33 45 34 & ...& ... & 79 & S?\\
Markarian 993 & HI 0126+321 & 01 26 12.7& 32 08 14 & ... & ... &175 & ...\\
Markarian 1157 & PGC 5848  & 01 34 16.7 & 35 35 16 & 0.015000 & ...& 202 & Sc \\
Markarian 573 &PGC 1226059  & 01 44 01.7 & 02 24 37 & ...& 17.73& 83 &..\\
UGC 1395 &  HI 0155+066  & 01 55 13.5 & 06 36 25 & ...& ... & 48   & ..\\
                   & IC 1749           & 01 56 11.1& 06 44 41 & 0.017038& 14.18    &326 & S0\\
NGC 841 & NGC 834       & 02 11 01.3& 37 39 59 & 0.015324& 13.06 &209 & Sbc  \\
                &NGC845         & 02 12 19.8& 37 28 38 & 0.014774 & 13.20 &244 & Sb \\
NGC 1167 & HI 0302+352    & 03 02 09.0 & 35 16 34 & ...& ... &  147  & ... \\
UGC 3157  & HI 0446+185    & 04 46 18.0& 18 31 32.5 & ...&... & 95 &  ... \\
                    &PGC 3097128   & 04 45 25.6& 18 25 05 & 0.015474& ...& 303& S?\\
MS 04595+0327 & MS Clump  & 05 02 07.7& 03 34 30 & ... & ... & 54 & ...\\
UGC 3995 & UGC 3995A   & 07 44 07.1 & 29 14 57 & 0.015812 & 14.08\tablenotemark{a} & 11 & ...\\
Markarian 461 & PGC 048870   & 13 47 10.7& 34 05 20 & ...& 14.76& 82  & S? \\
                         & HI 1346+341  & 13 46 19.2& 34 11 06 & ...&  ... &  259 &...\\ 
NGC 5548 & PGC 1725892  & 14 17 33.9& 25 06 52 & 0.017200 & 15.59 &132 & S? \\
                   & SDSS~J141824\tablenotemark{b}    & 14 18 24.7& 25 06 51 & ... & 17.73 &130&  ...   \\
ARK 539     & PGC~2369294     & 18 29 01.0 & 50 18 48 & ... & 15.90 & 91 & S?  \\
            & HI 1828+503     & 18 28 53.8 & 50 19 55 & ... & ...   & 58 & ... \\
            & PGC~2370360     & 18 29 07.4 & 50 20 57 & ... & 17.23 & 76 & ... \\
NGC 7469 & IC 5283     & 23 03 18.0& 08 53 37 & 0.016024 & 13.93 & 31 & Sc \\
NGC 7591 & PGC 214933   & 23 18 13.6& 06 33 26 & ...& ... & 39 &  S?\\
                  & HI 2318+065   & 23 18 30.8& 06 35 32 & ...& ... & 77 & ...  \\
                  & HI 2318+064   & 23 18 56.1& 06 28 58 & ...& ... & 249 & ...\\ 
NGC 7679 & UGC12628   & 23 29 22.0& 03 23 23 &0.015607& 14.76 &255 & Sc\\
\enddata
\tablenotetext{a}{From Table~II in \citet{marz99}, where UGC~3995A is referred to as UGC~3995B.}
\tablenotetext{b}{Full name SDSS~J141824.74+250650.7.}
\tablecomments{Units of right ascension are hours, minutes, and seconds, and units of declination are degrees, arcminutes, and arcseconds.  Coordinates are from the Hyperleda database except for UGC~3995A (from NED) and those not cataloged in the optical (from our HI maps).  Optical redshifts are from NED where available.  Apparent corrected $B$-band magnitudes and Hubble types from the Hyperleda database except otherwise noted.  S? denotes unclassified spiral galaxy.}
\end{deluxetable}

\begin{deluxetable}{llccccc}
\tablecolumns{7} 
\tablewidth{0pt} 
\tablecaption{HI Gas Properties of other Galaxies in Active Galaxy Field} 
\tablehead{ 
\multicolumn{1}{c}{Active} &
\multicolumn{1}{c}{Neighbouring} &
\multicolumn{1}{c}{S(HI)}&
\multicolumn{1}{c}{$\rm v_{sys}(HI)$}&
\multicolumn{1}{c}{M(HI)} &
\multicolumn{1}{c}{HI} &
\multicolumn{1}{c}{} \\
\multicolumn{1}{c}{Galaxy} &
\multicolumn{1}{c}{Galaxy} &
\multicolumn{1}{c}{(mJy kms$^{-1}$)} &
\multicolumn{1}{c}{(km s$^{-1}$)} &
\multicolumn{1}{c}{$10^9 \rm \ M_{\odot}$} &
\multicolumn{1}{c}{disturbed?} &
\multicolumn{1}{c}{Group}
}
\startdata
Markarian 341 & NGC 169  & 7200$\pm$508 & 4578 & 10.01$\pm$0.7& yes & I\\
Markarian 1 & NGC 451     &1633$\pm$196& 4833  & 1.99$\pm$0.24 & yes & I\\
NGC 513 & PGC212747     &  80$\pm$13 &  6071  &0.14$\pm$0.02  & yes & II \\
Markarian 993 & HI 0126+321  &1268$\pm$415 & 4163& 1.75$\pm$0.57& no & III \\
Markarian 1157 & PGC 5848     & 2200$\pm$200& 4694 & 2.3$\pm$0.2 & yes & I \\
Markarian 573 & PGC 1226059   & 77.7$\pm$124 & 5011 & 0.10$\pm$0.17& no & I\\
UGC 1395 & HI 0155+066  &127$\pm$107& 5046 & 0.17$\pm$0.14& no & II\\
                   &IC1749            &2650$\pm$331& 5049 & 3.68$\pm$0.40 & yes & ...  \\
NGC 841 & NGC 834   &3848$\pm$430 & 4563 & 4.16$\pm$0.45& yes & I\\
                 & NGC 845   &1115$\pm$430 & 4398 & 1.20$\pm$0.45 & no &...\\
NGC 1167 & HI 0302+352  & 188$\pm$200 & 4862 & 0.26$\pm$0.27 & no & ... \\
UGC 3157 & HI 0446+185  & 915$\pm$268 & 4747 & 1.04$\pm$0.30 & no  & III \\
                  & PGC 3097128 &7181$\pm$938& 4622 & 8.17$\pm$1.06 &  no &...\\
MS 04595+0327 & MS Clump  & 260$\pm$27 & 4775 & 0.30$\pm$0.03 & no  & II\\
UGC 3995  & UGC 3995A & ...& ... &...& no & ...\\
Markarian 461 & PGC 048870    & 202$\pm$ 90 & 4891 & 0.25$\pm$0.10  & yes & I\\
                         & HI 1346+341    & 258$\pm$122 & 4850 & 0.32$\pm$0.15& no &... \\
NGC 5548 & PGC 1725892 & 136$\pm$146 & 5122 & 0.18$\pm$0.20 & yes & I \\
                  &  SDSS~J141824\tablenotemark{b}   & 117$\pm$187 & 4916 & 0.16$\pm$0.26  & yes & I\\
ARK 539     & PGC~2369294     & 260$\pm$103 & 4830 & 0.067$\pm$0.07 & yes & ...   \\
                     & HI 1828+503     & 55$\pm$ 65 & 4851 & 0.067$\pm$0.07 & no & ...   \\
                     & PGC~2370360     & 1070$\pm$163 & 5102 & 1.31$\pm$0.19 & yes & I \\
NGC 7469    & IC 5283      & 470$\pm$32 &  4870 & 0.60$\pm$0.04 & yes & I \\
NGC 7591  & PGC 214933  &1637$\pm$163 & 4772 & 2.27$\pm$0.22 &yes & I\\
                    &  HI 2318+065   & 273$\pm$139 & 5024 & 0.37$\pm$0.18 &yes  &I\\
                     & HI 2318+064       & 429$\pm$379 & 4835 & 0.59$\pm$0.49 & no&...\\
NGC 7679/7682   & UGC 12628  &2531$\pm$183 & 4647 & 3.52$\pm$0.25 & no & ...\\
\enddata
\tablenotetext{a}{Full name SDSS~J141824.74+250650.7.}
\tablecomments{Neighboring galaxies in Group~I are those identified from HI maps to be tidally interacting with their corresponding active galaxies, in Group~II are those identified as probably tidally interacting with their corresponding active galaxies, and in Group~III as possible candidates for producing the observed HI disturbances in their corresponding active galaxies (see text).}
\end{deluxetable}


\begin{figure}
\begin{center}
\includegraphics[angle=0, scale=0.485]{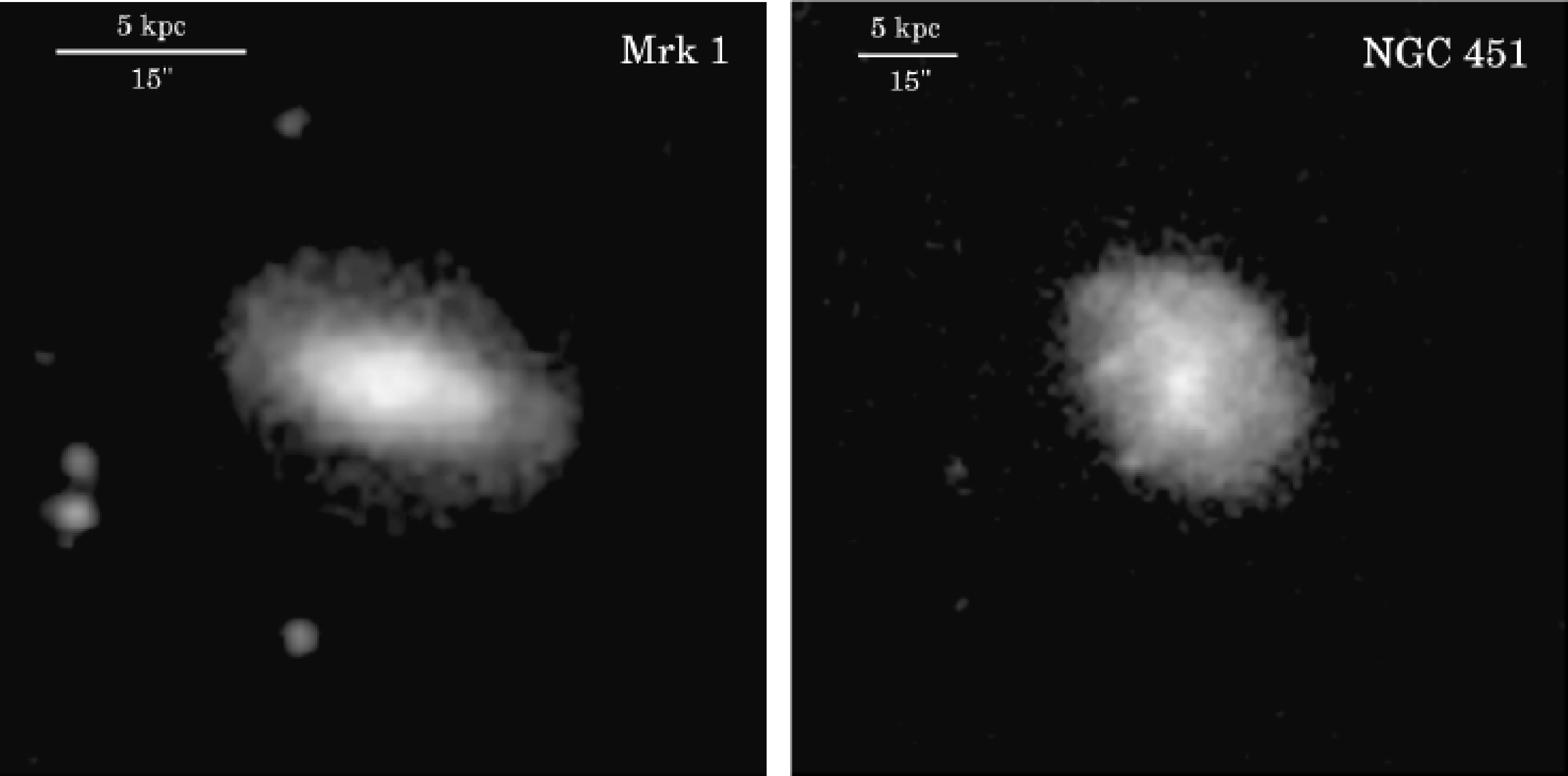}
\hspace*{-1.4 cm}
\includegraphics[angle=0, scale=0.482]{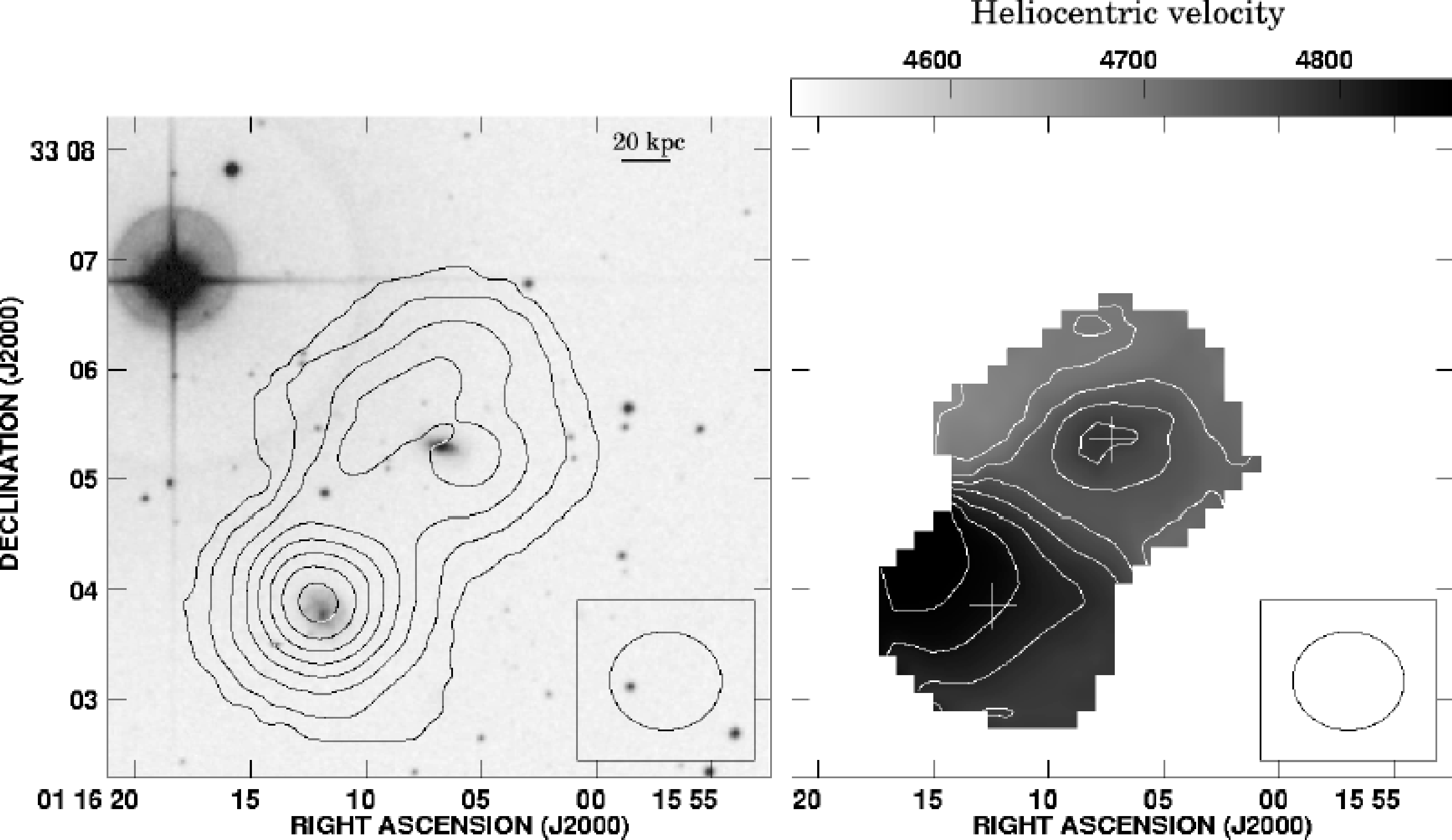} \\
\caption{Upper panel: Optical image of Markarian~1 (Mrk~1) (active galaxy) and NGC~451 from the 2nd Digitized Sky Survey (DSS2).  Lower panels: (\textsl{Left}) Contours of integrated HI intensity (zeroth moment) overlaid on the DSS2 image, and (\textsl{Right}) map of intensity-weighted HI mean velocity (first moment).  Mrk~1 lies at the center of the map, whereas NGC~451 is located south-east of center.  In the zeroth moment map, contours are plotted at 5, 10, 15, 20, 25, 30, 35 $\times 25 {\rm \ mJy \ beam^{-1} \ km \ s^{-1}}$ ($1.25 \times 10^{19} {\rm \ cm^{-2}}$). In the first moment map, velocities are indicated by the scale wedge, and contours plotted at intervals of $25 {\rm \ km \ s^{-1}}$. The ellipse at the lower right corner of the lower panels indicates the half-power width of the synthesized beam, and has a size of 61$\arcsec \times 54\arcsec$.
}
\end{center}
\end{figure}

\begin{figure}
\begin{center}
\vspace*{-1.2 cm}
\hspace*{-1.8 cm}
\includegraphics[angle=0, scale=0.7]{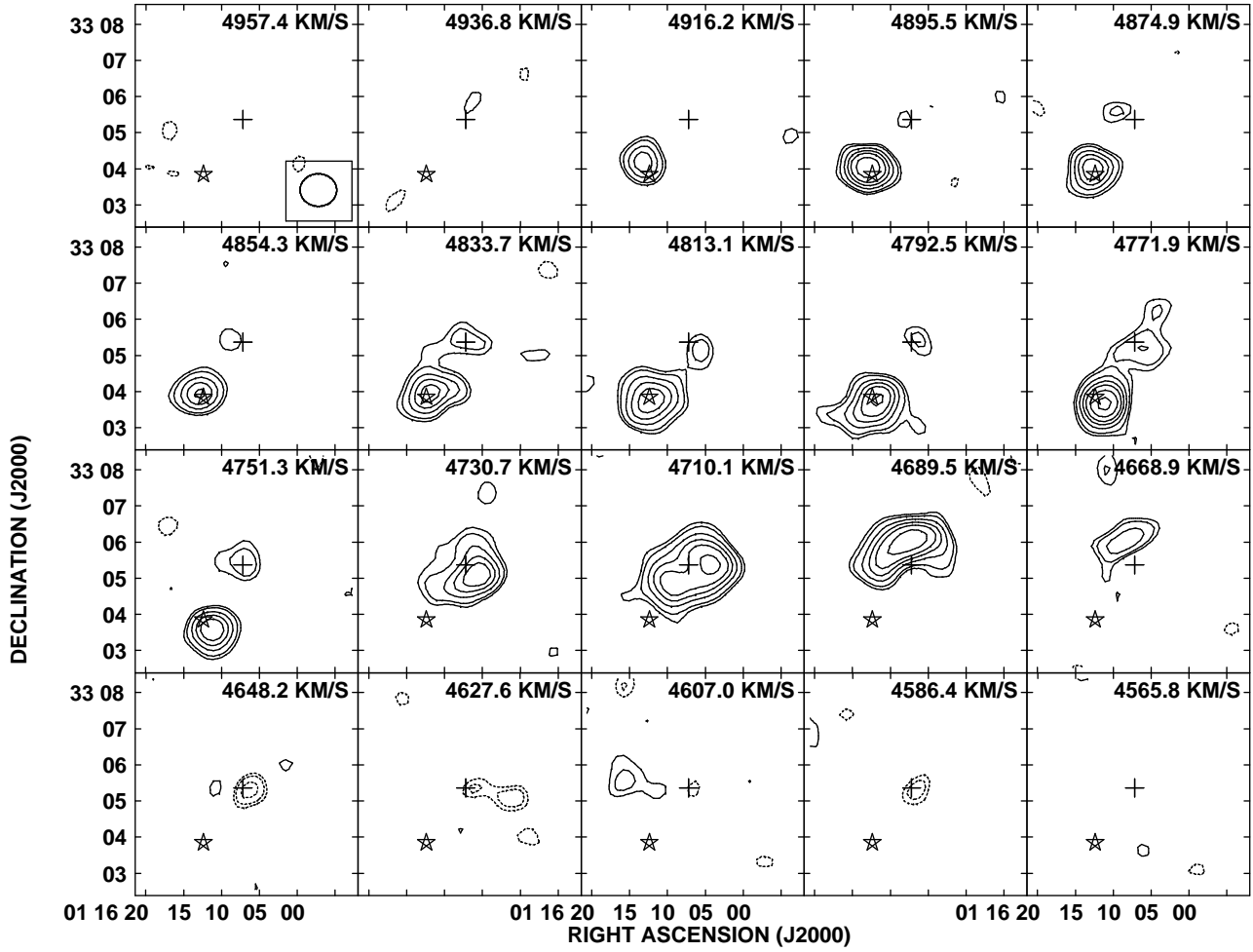}
\vspace*{-1.0 cm}
\caption{HI channel maps of Mrk~1 and NGC~451. Contour levels are plotted at $-6$, $-4$, $-3$, 3, 4, 6, 8, 10, 13, $17 \times 0.43 {\rm \ mJy \ beam^{-1}}$ ($1 \sigma$), which corresponds to a HI column density of 4.43 $\times 10^{18} {\rm \ cm^{-2}}$. The central heliocentric velocity is shown for each channel. The cross marks the position of Mrk~1, and star the position of NGC 451. The synthesized beam is shown by the ellipse at the lower right corner of the top left panel.}
\end{center}
\end{figure}


\begin{figure}
\begin{center}
\includegraphics[angle=0, scale=0.485]{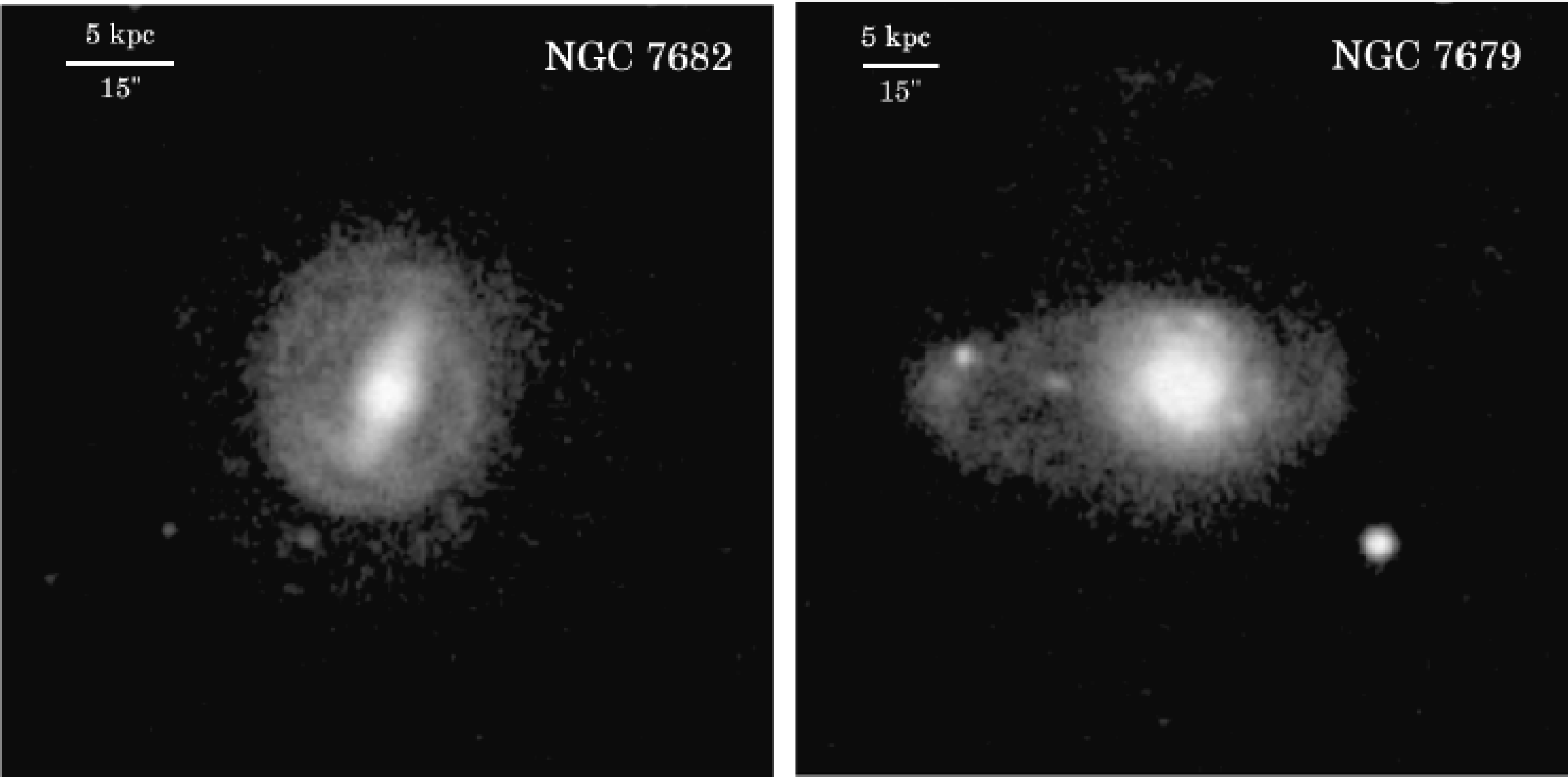}
\hspace*{-1.4 cm}
\includegraphics[angle=0, scale=0.482]{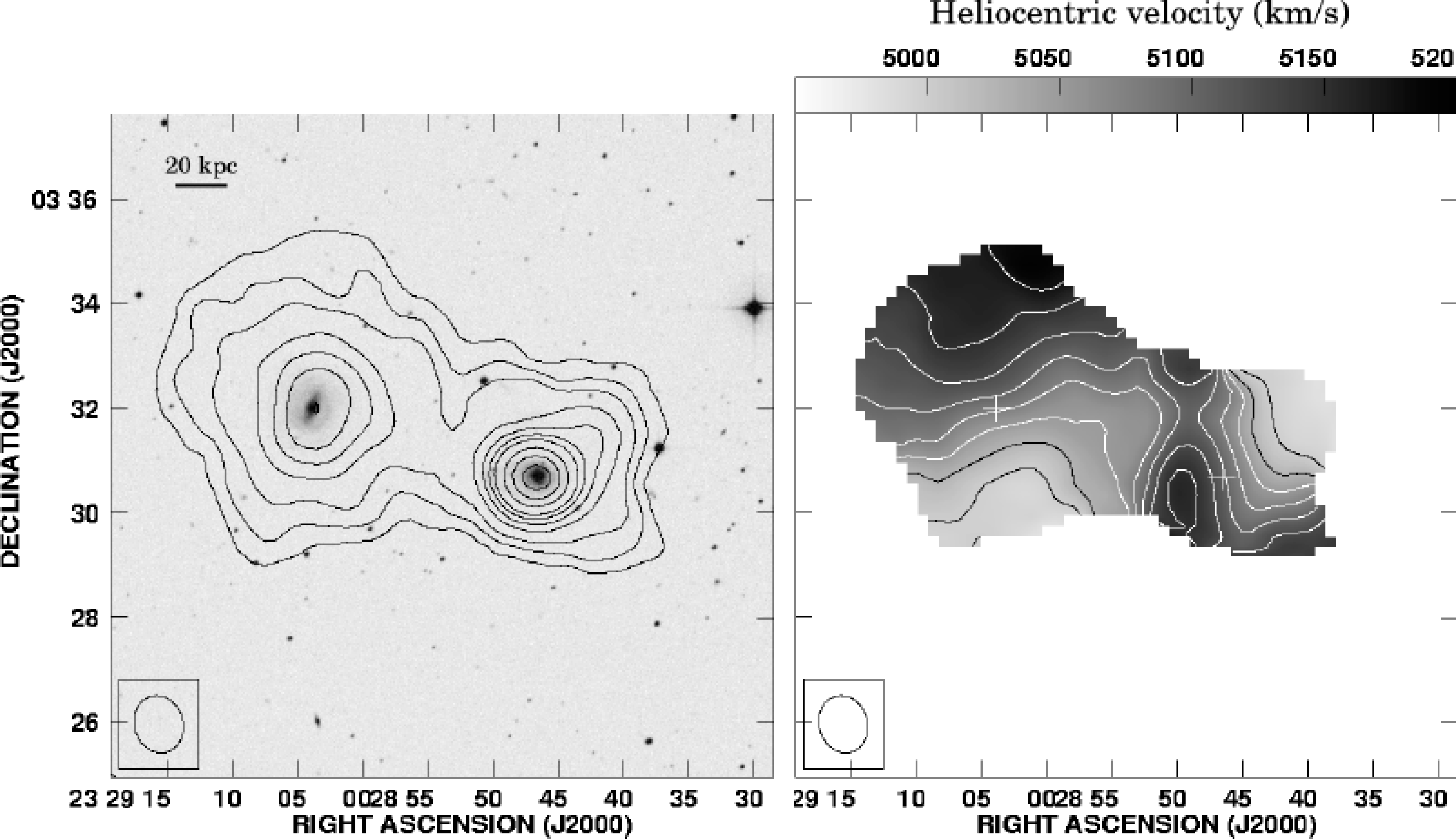} \\
\caption{Upper panel: Optical image of NGC~7679 and NGC~7682 (both active galaxies) from the DSS2.  Lower panels: (\textsl{Left}) Contours of integrated HI intensity (zeroth moment) overlaid on the DSS2 image, and (\textsl{Right}) map of intensity-weighted HI mean velocity (first moment).  NGC~7679 lies to the
west and NGC~7682 to the east of center.  In the zeroth moment map, contours are plotted at 1, 5, 10, 20, 30, 40, 50, 70, 90 $\times 20 {\rm \ mJy \ beam^{-1} \ km \ s^{-1}}$ ($9.12 \times 10^{18} {\rm \ cm^{-2}}$).  In the first moment map, velocities are indicated by the scale wedge, and contours plotted at intervals of $25 {\rm \ km \ s^{-1}}$.  The ellipse at the lower left corner of the lower panels indicates the half-power width of the synthesized beam, and has a size of 68$\arcsec \times 53\arcsec$.
}
\end{center}
\end{figure}

\begin{figure}
\begin{center}
\vspace*{-0.3 cm}
\hspace*{-2.0 cm}
\includegraphics[angle=0, scale=0.72]{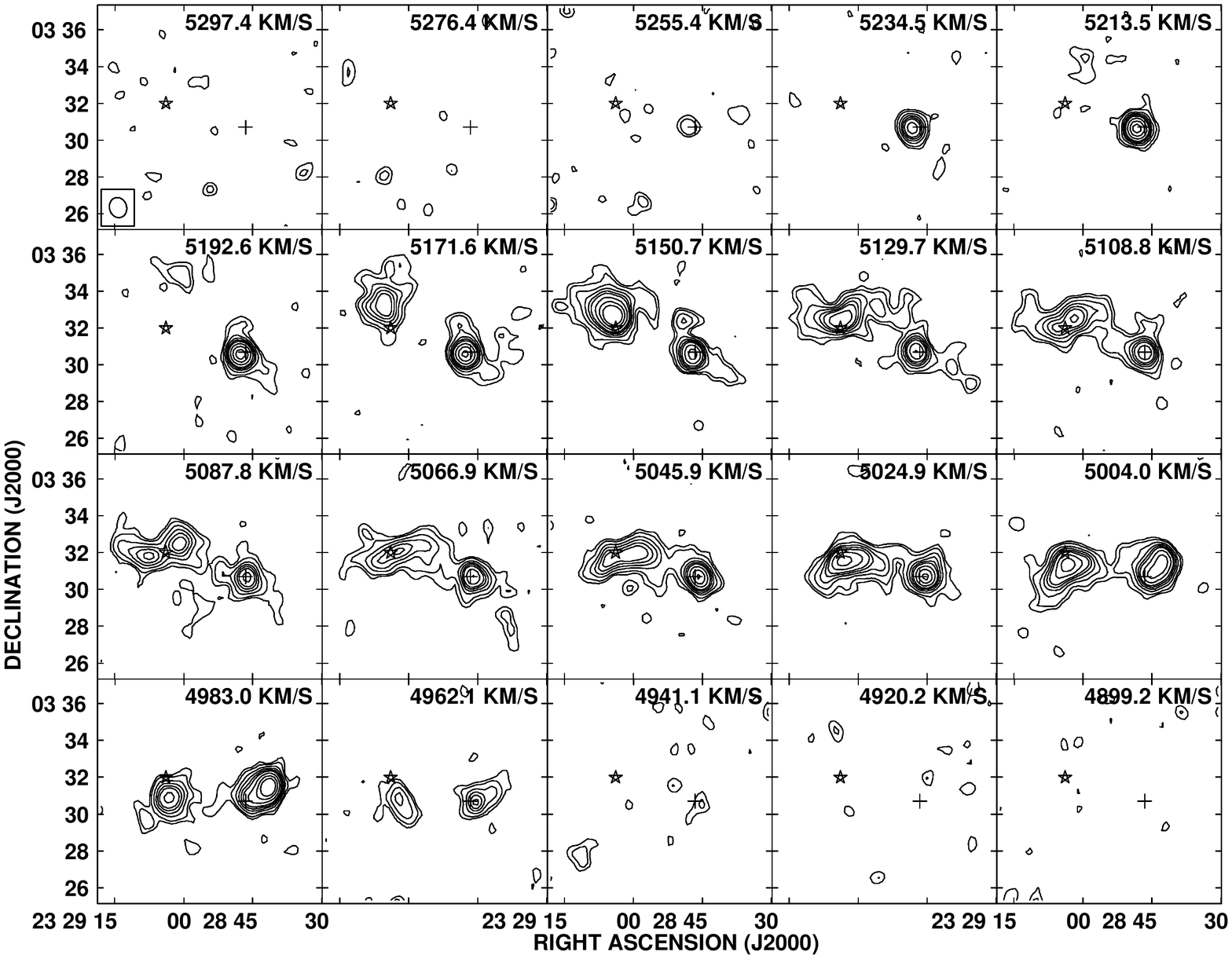}
\vspace*{0.0 cm}
\caption{
HI channel maps of NGC~7679 and NGC~7682. Contour levels are plotted at  2, 3, 5, 7, 9, 11, 15, 20, $25 \times 0.45 {\rm \ mJy \ beam^{-1}}$ ($1 \sigma$), which corresponds to a HI column density of 4.23 $\times 10^{18} {\rm \ cm^{-2}}$. The central heliocentric velocity is shown for each channel. The cross marks the position of NGC~7679, and star the position of NGC~7682. The synthesized beam is shown by the ellipse ar the lower left corner of the top left panel.}
\end{center}
\end{figure}

\begin{figure}
\begin{center}
\includegraphics[angle=0, scale=0.53]{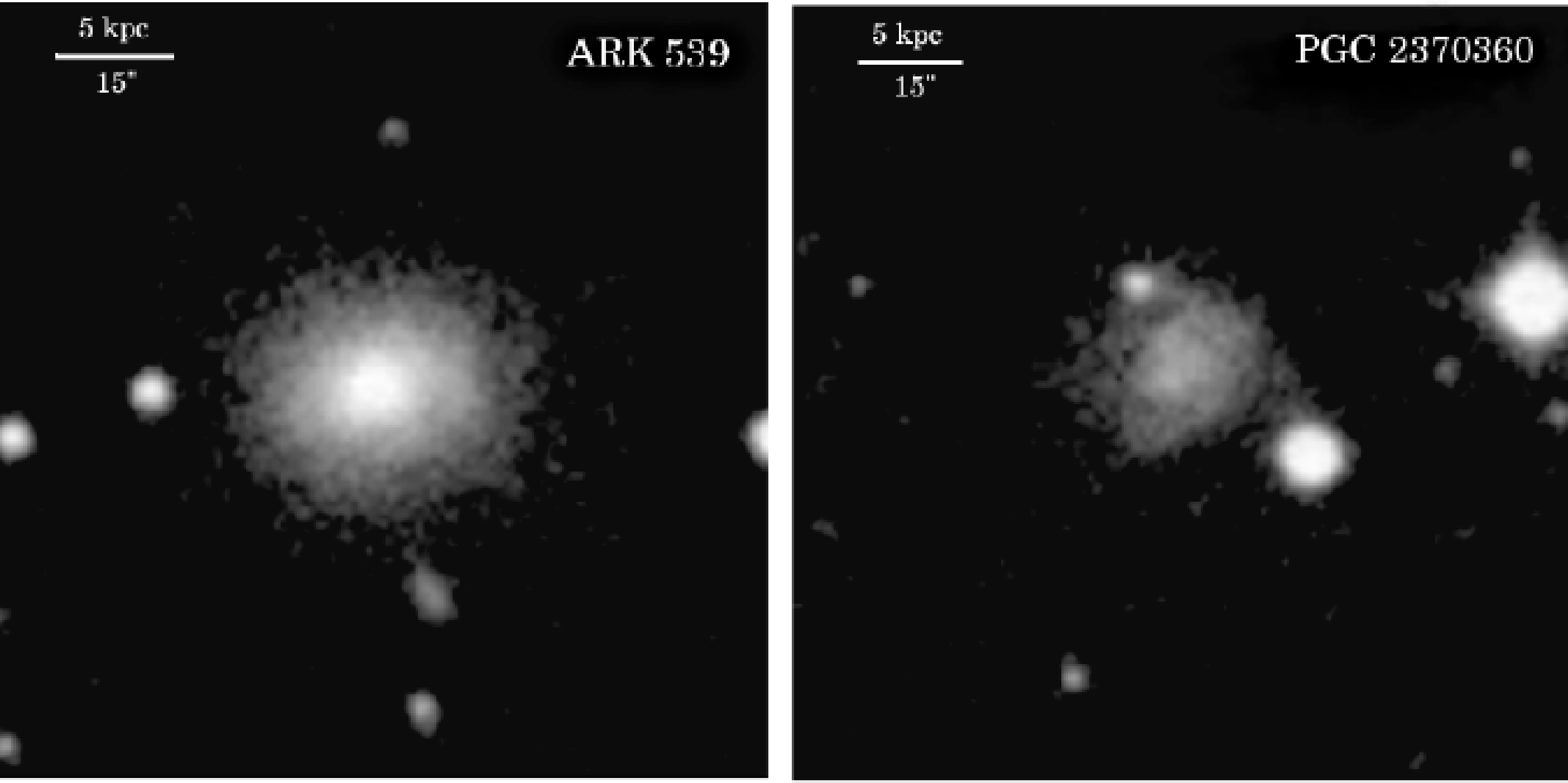}
\hspace*{-1.4 cm}
\includegraphics[angle=0, scale=0.482]{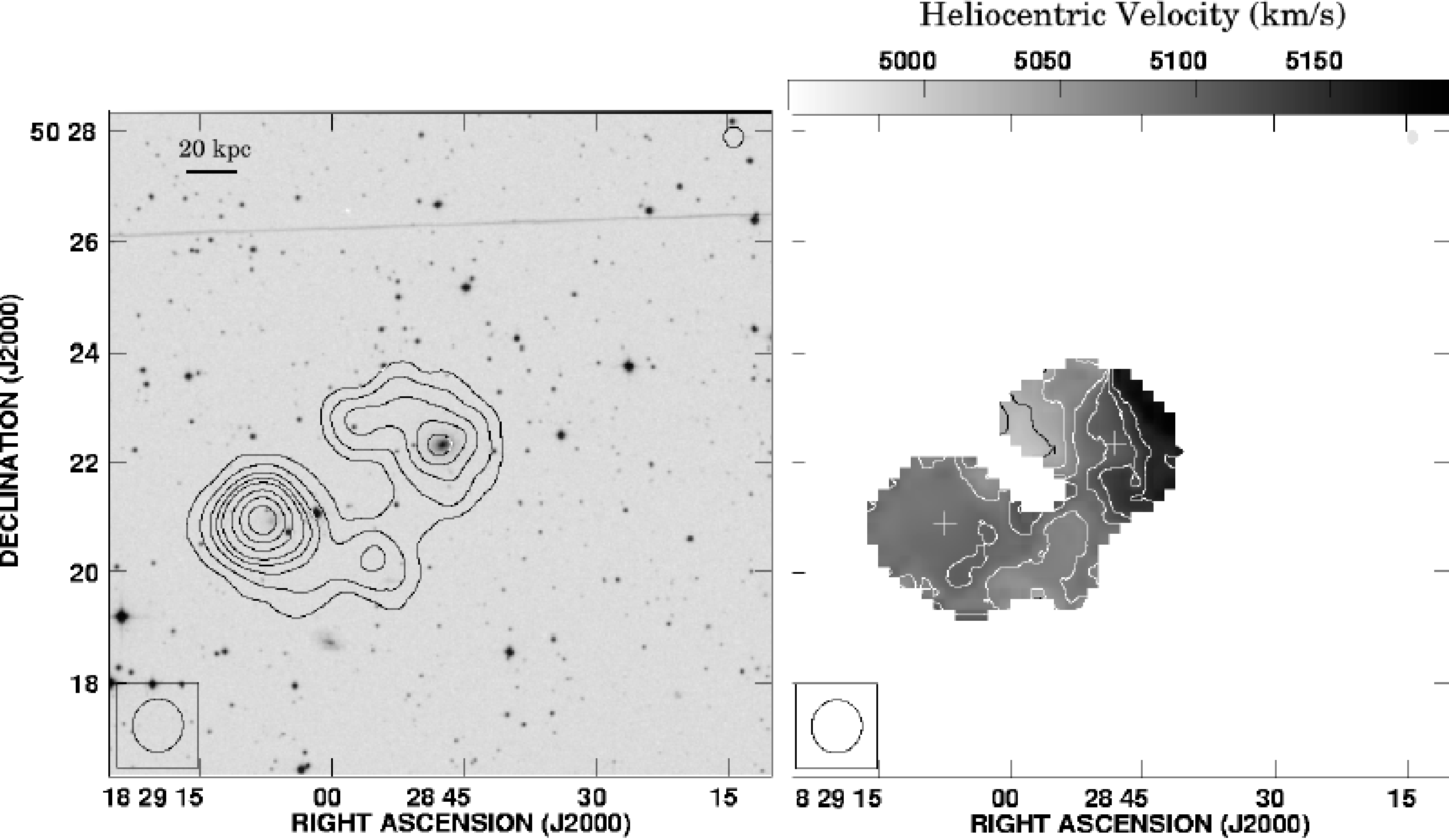} \\
\caption{
Upper panel: Optical image of ARK~539 (active galaxy) and PGC~2370360 from the DSS2.  Lower panels: (\textsl{Left}) Contours of integrated HI intensity (zeroth
moment) overlaid on the DSS2 image, and (\textsl{Right}) map of intensity-weighted HI mean velocity (first moment). ARK~539 lies at the center of the map, whereas PGC~2370360 is located south-east of center.  In the zeroth moment map, contours are plotted at 1, 5, 10, 15, 20, 30, 40, $50 \times 15 {\rm \ mJy \ beam^{-1} \ km \ s^{-1}}$ ($7.86 \times 10^{18} {\rm \ cm^{-2}}$). In the first moment map, velocities are indicated by the scale wedge, and contours plotted at intervals of $25 {\rm \ km \ s^{-1}}$.  The ellipse at the lower left corner of the lower panels indicates the half-power width of the synthesized beam, and has a size of 58$\arcsec \times 54 \arcsec$.}
\end{center}
\end{figure}

\begin{figure}
\begin{center}
\vspace*{-0.3 cm}
\hspace*{-1.7 cm}
\includegraphics[angle=0, scale=0.7]{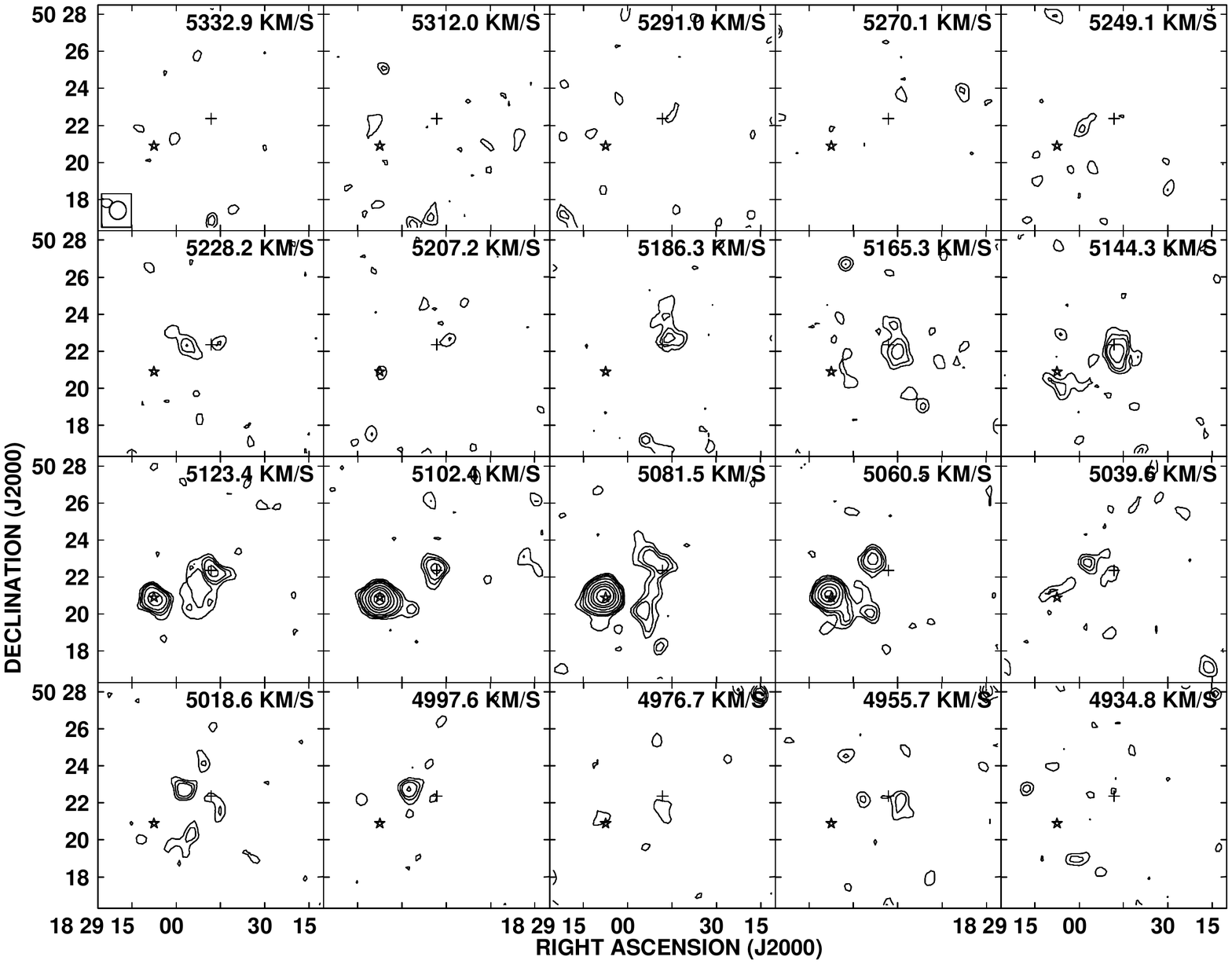}
\vspace*{0.0 cm}
\caption{
HI channel maps of ARK 539 and PGC~2370360. Contour levels are plotted at  2, 3, 4, 6, 10, 15, 20, $28 \times 0.38 {\rm \ mJy \ beam^{-1}}$ ($1 \sigma$), which corresponds to a HI column density of 4.10 $\times 10^{18} {\rm \ cm^{-2}}$. The central heliocentric velocity is shown for each channel. The cross marks the position of ARK~539, and star the position of PGC~2370360. The synthesized beam is shown by the ellipse at the lower left corner of the top left panel.}
\end{center}
\end{figure}

\begin{figure}
\begin{center}
\includegraphics[angle=0, scale=0.534]{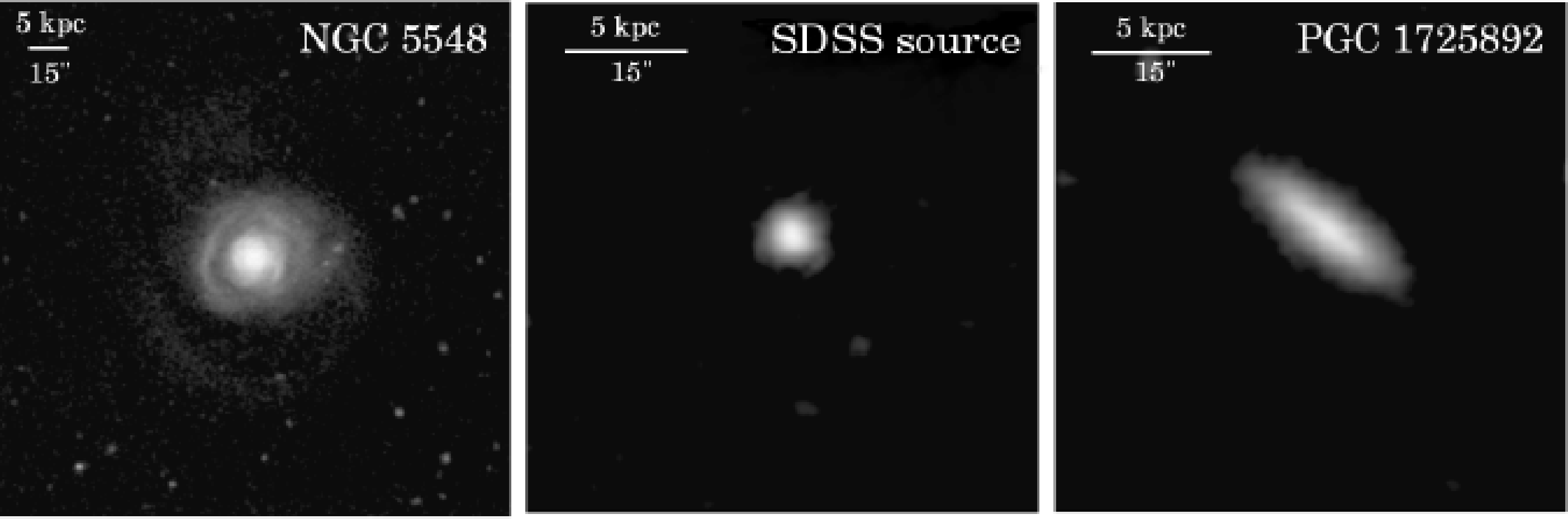}
\hspace*{-1.4 cm}
\includegraphics[angle=0, scale=0.482]{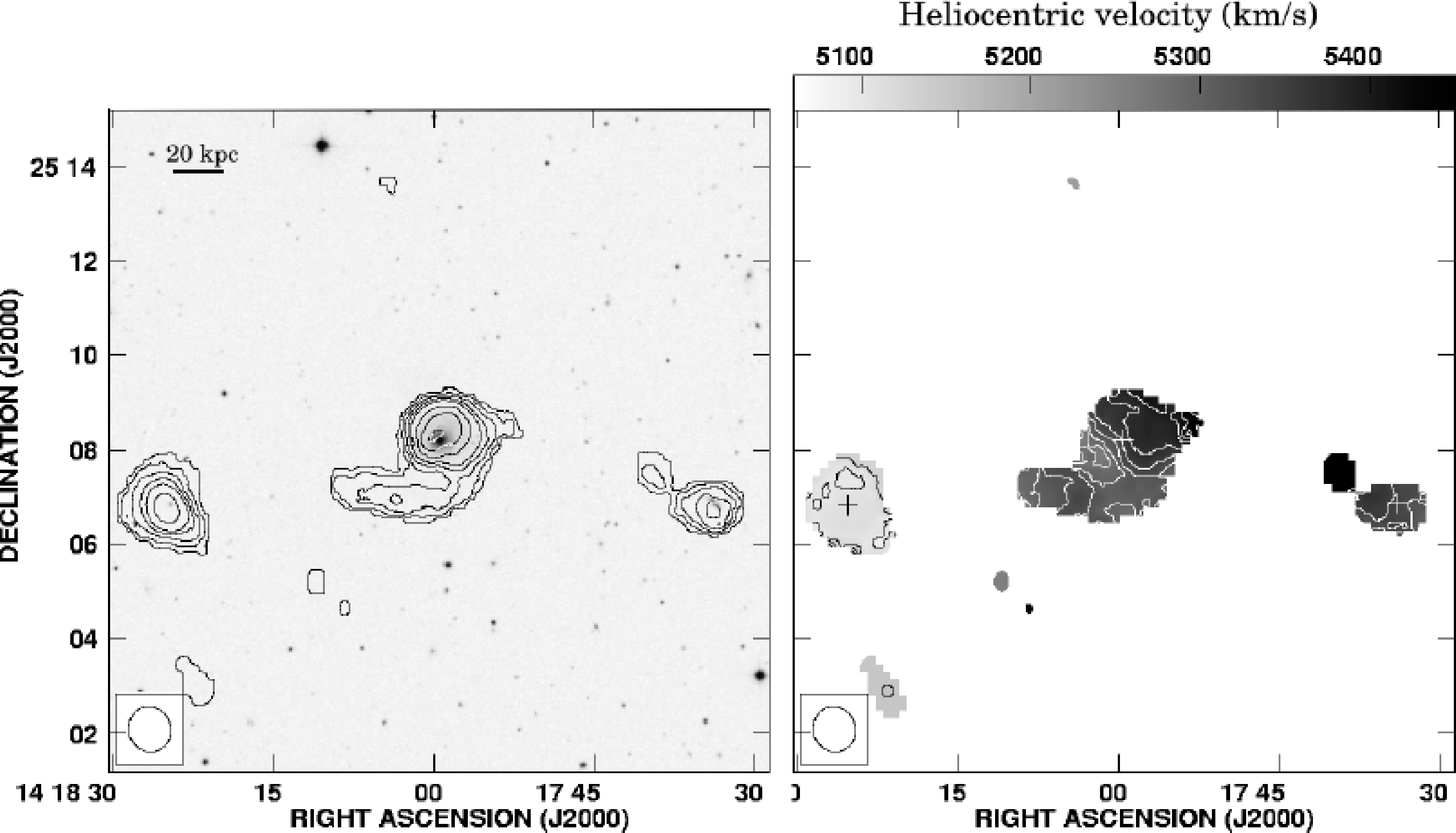} \\
\caption{
Upper panel: Optical image of NGC~5548 (active galaxy), SDSS~J141824.74$+$250650.7, and PGC~1725892 from the DSS2.  Lower panels: (\textsl{Left}) Contours of integrated HI intensity (zeroth moment) overlaid on the DSS2 image, and (\textsl{Right}) map of intensity-weighted HI mean velocity (first moment).  NGC~5548 lies at the center of the map, whereas SDSS~J141824.74$+$250650.7 is located south-east and PGC 1725892 south-west of center.  In the zeroth moment map, contours are plotted at 1, 5, 10, 15, 20, 30, 40, $50 \times 35 {\rm \ mJy \ beam^{-1} \ km \ s^{-1}}$ ($1.94 \times 10^{19} {\rm \ cm^{-2}}$). In the first moment map, velocities are indicated by the scale wedge, and contours plotted at intervals of $25 {\rm \ km \ s^{-1}}$. The ellipse at the lower left corner of the lower panels indicates the half-power width of the synthesized beam, and has a size of 56$\arcsec \times 53\arcsec$.}
\end{center}
\end{figure}

\begin{figure}
\begin{center}
\vspace*{-1.6 cm}
\hspace*{-3.3 cm}
\includegraphics[angle=0, scale=0.80]{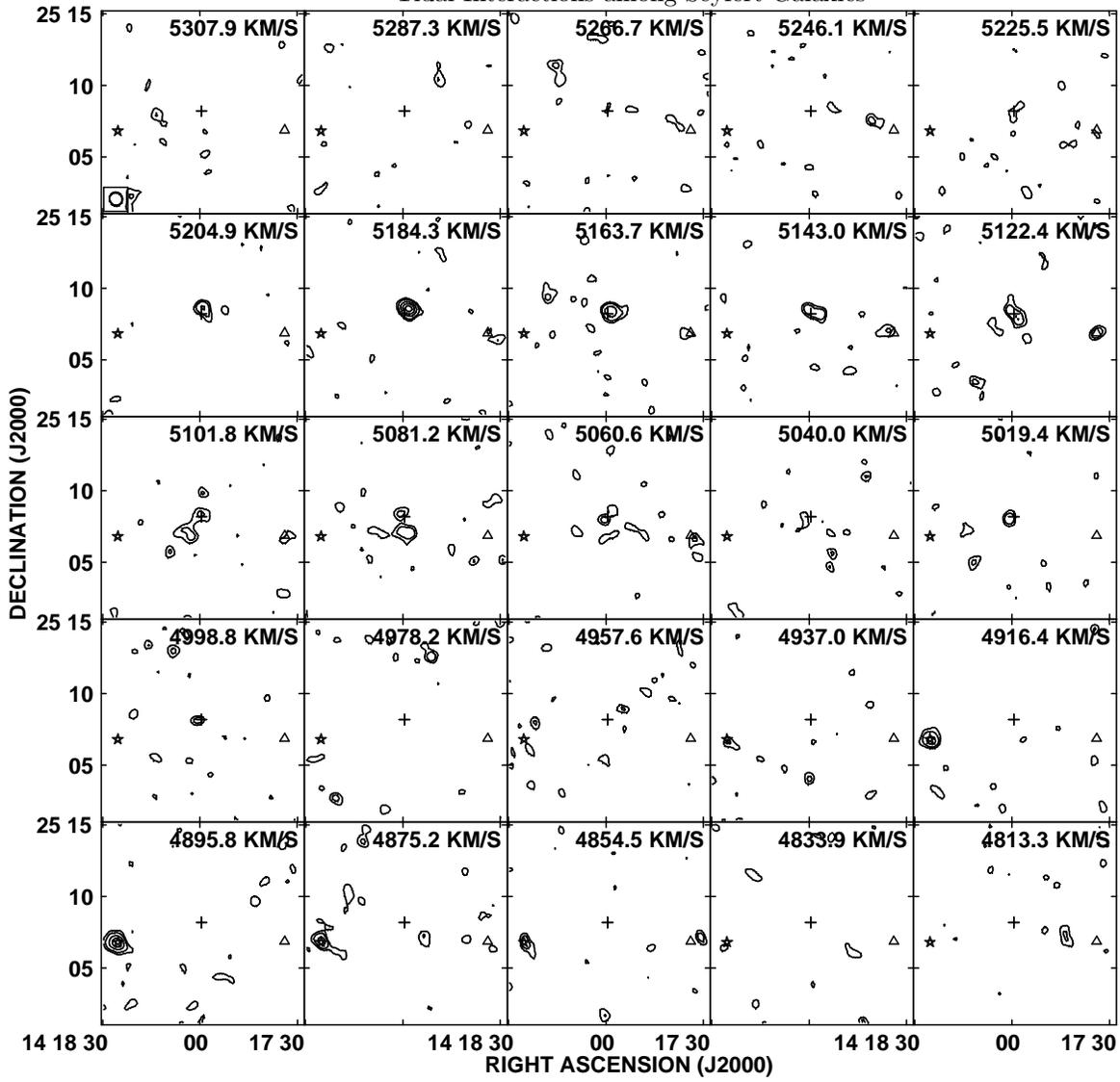}
\vspace*{0.0 cm}
\caption{HI channel maps of NGC~5548 (active galaxy), SDSS~J141824.74$+$250650.7, and PGC~1725892. Contour levels are  plotted at 2, 3, 5, 7, 9, 11 $\times 0.45 {\rm \ mJy \ beam^{-1}}$ ($1 \sigma$), which corresponds to a HI column density of $5.18 \times 10^{18} {\rm \ cm^{-2}}$. The central heliocentric velocity is shown for each channel. The cross marks the position of NGC~5548, star the position of SDSS~J141824.74$+$250650.7, and triangle the position of PGC~1725892. The synthesized beam is shown by the ellipse at the lower left corner of the top left panel.}
\end{center}
\end{figure}

\begin{figure}
\begin{center}
\includegraphics[angle=0, scale=0.485]{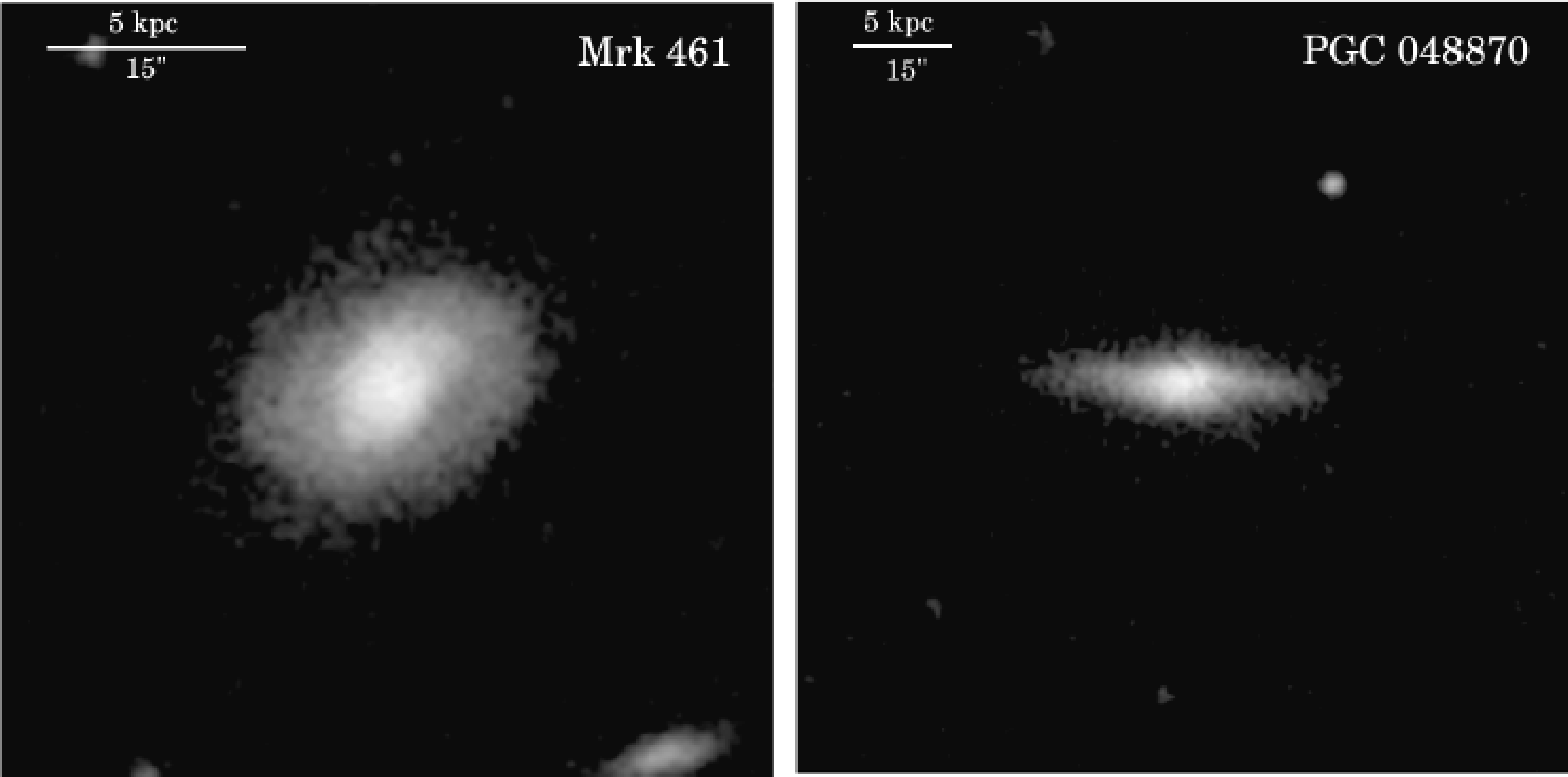}
\hspace*{-1.4 cm}
\includegraphics[angle=0, scale=0.482]{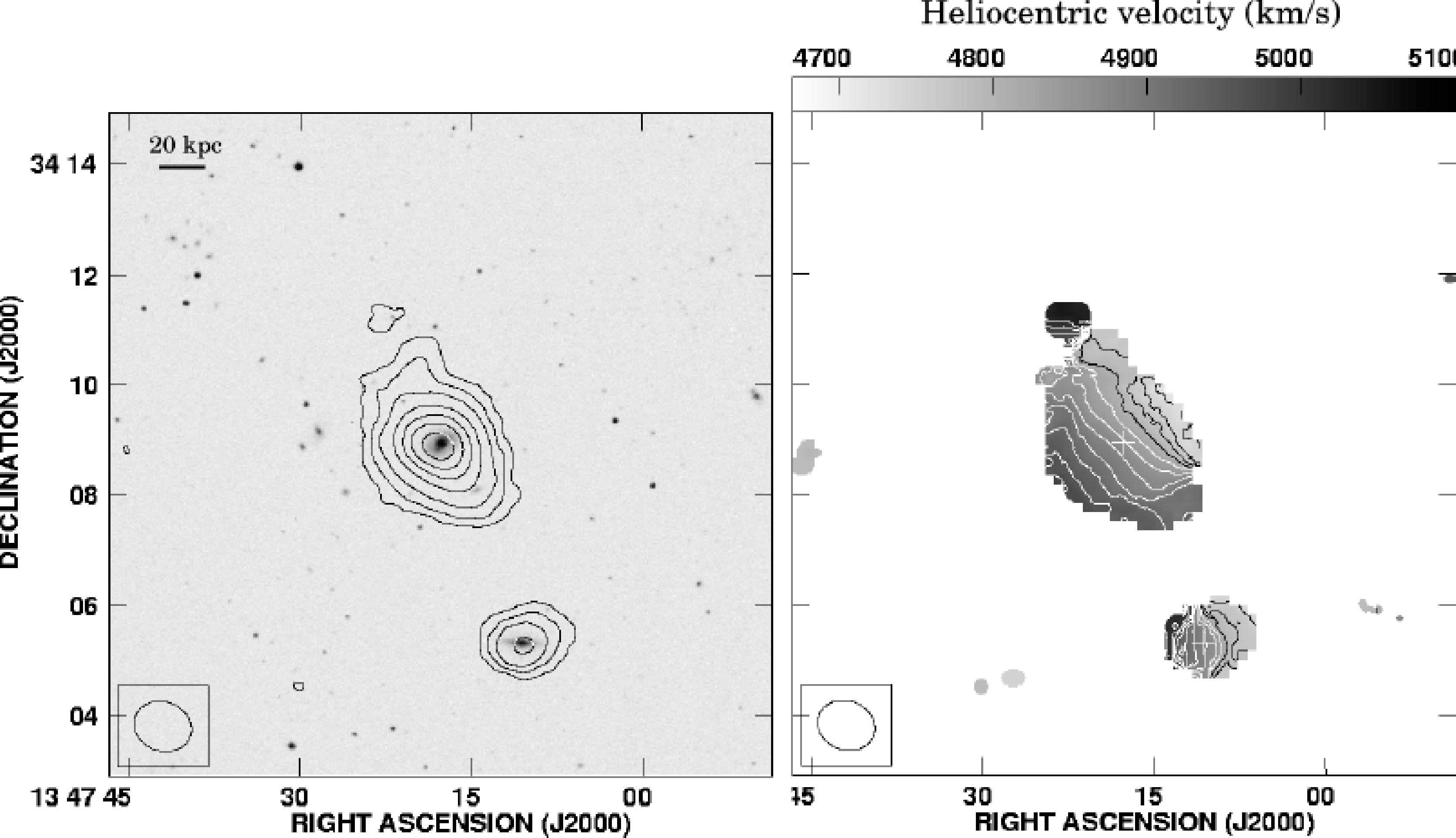} \\
\caption{Upper panel: Optical image of Markarian~461 (Mrk~461) (active galaxy) and PGC~48870 from the DSS2.  Lower panels: (\textsl{Left}) Contours of integrated HI intensity (zeroth moment) overlaid on the DSS2 image, and (\textsl{Right}) map of intensity-weighted HI mean velocity (first moment).  Mrk~461 lies at the center of the map, whereas PGC~48870 is located south-west of center.  In the zeroth moment map, contours are plotted at 1, 5, 10, 20, 30, 40, $50 \times 20 {\rm \ mJy \ beam^{-1} \ km \ s^{-1}}$ ($4.49 \times 10^{18} {\rm \ cm^{-2}}$). In the first moment map, velocities are indicated by the scale wedge, and contours plotted at intervals of $25 {\rm \ km \ s^{-1}}$. The ellipse at the lower left corner of the lower panels indicates the half-power width of the synthesized beam, and has a size of 64$\arcsec \times 54\arcsec$.}
\end{center}
\end{figure}

\begin{figure}
\begin{center}
\vspace*{-0.3 cm}
\hspace*{-2.0 cm}
\includegraphics[angle=0, scale=0.7]{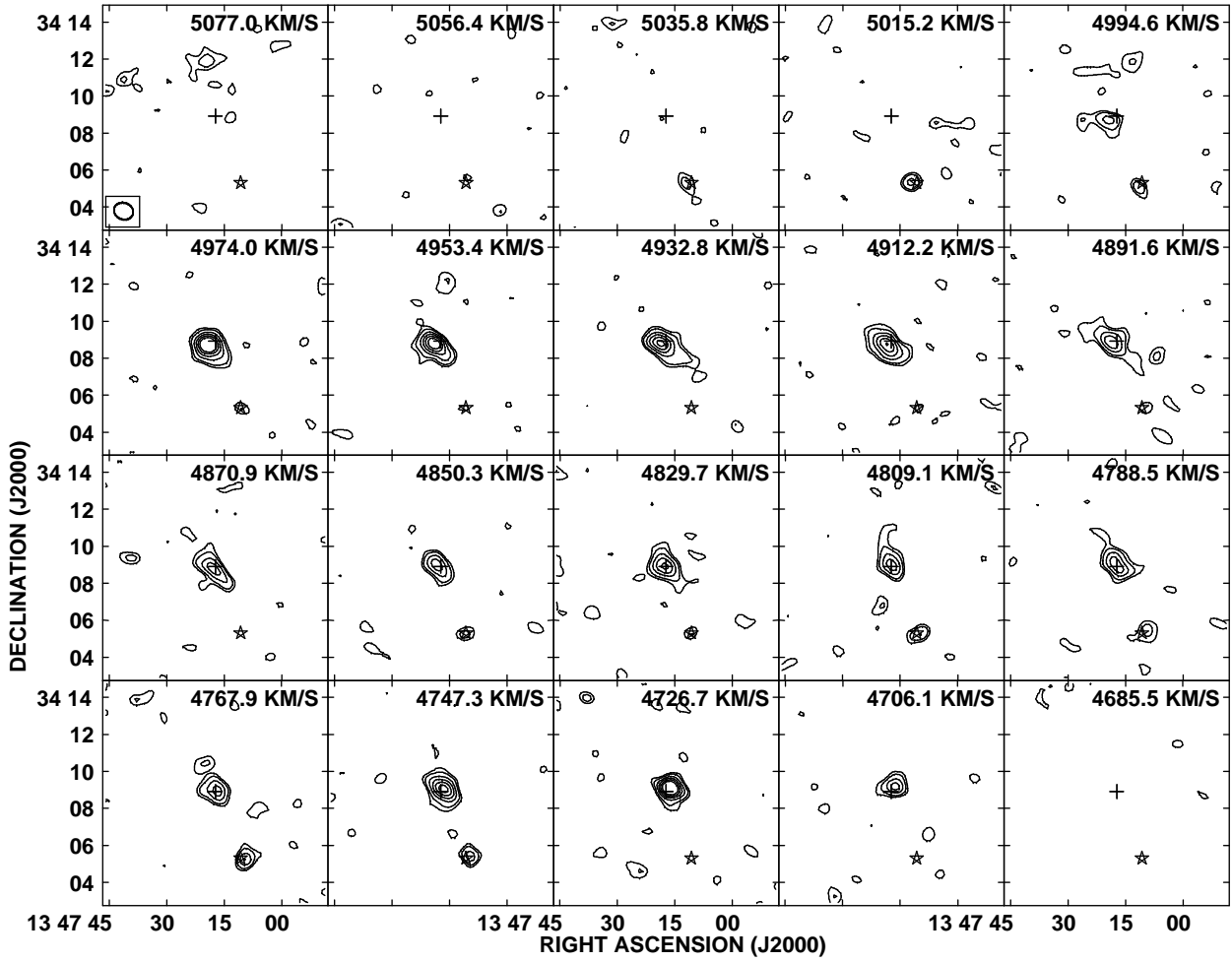}
\vspace*{0.0 cm}
\caption{HI channel maps of Mrk~461 and PGC~48870. Contour levels are 
plotted at 2, 3, 5, 7, 9, $15 \times 0.43 {\rm \ mJy \ beam^{-1}}$ ($1 \sigma$), which corresponds to a HI column density of $4.40 \times 10^{18} {\rm \ cm^{-2}}$. The central heliocentric velocity is shown for each channel. The cross marks the position of Mrk~461, and star the position of PGC~48870. The synthesized beam is shown by the ellipse at the lower left corner of the top left panel.}
\end{center}
\end{figure}

\begin{figure}
\begin{center}
\includegraphics[angle=0, scale=0.485]{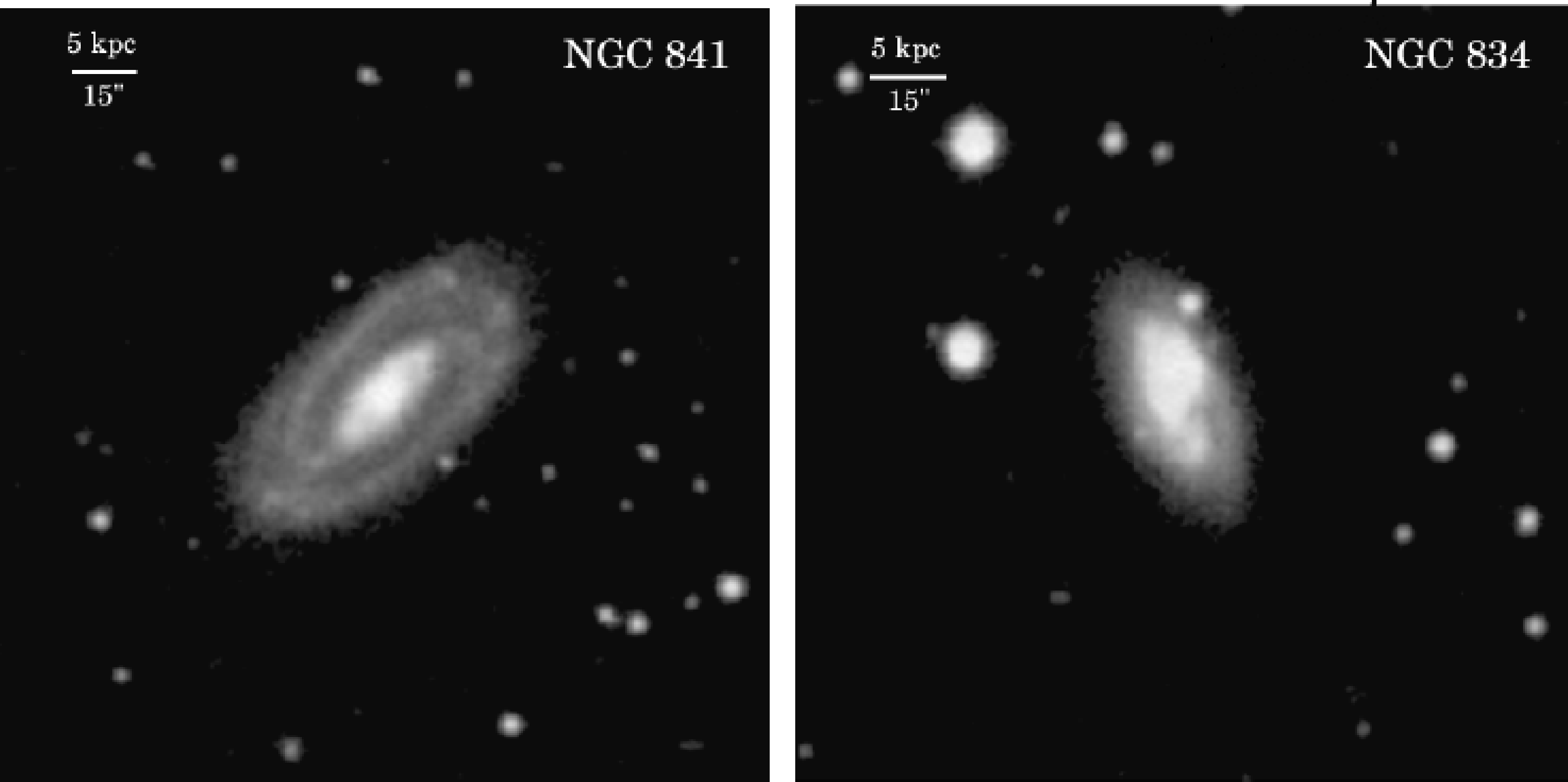}
\hspace*{-1.4 cm}
\includegraphics[angle=0, scale=0.482]{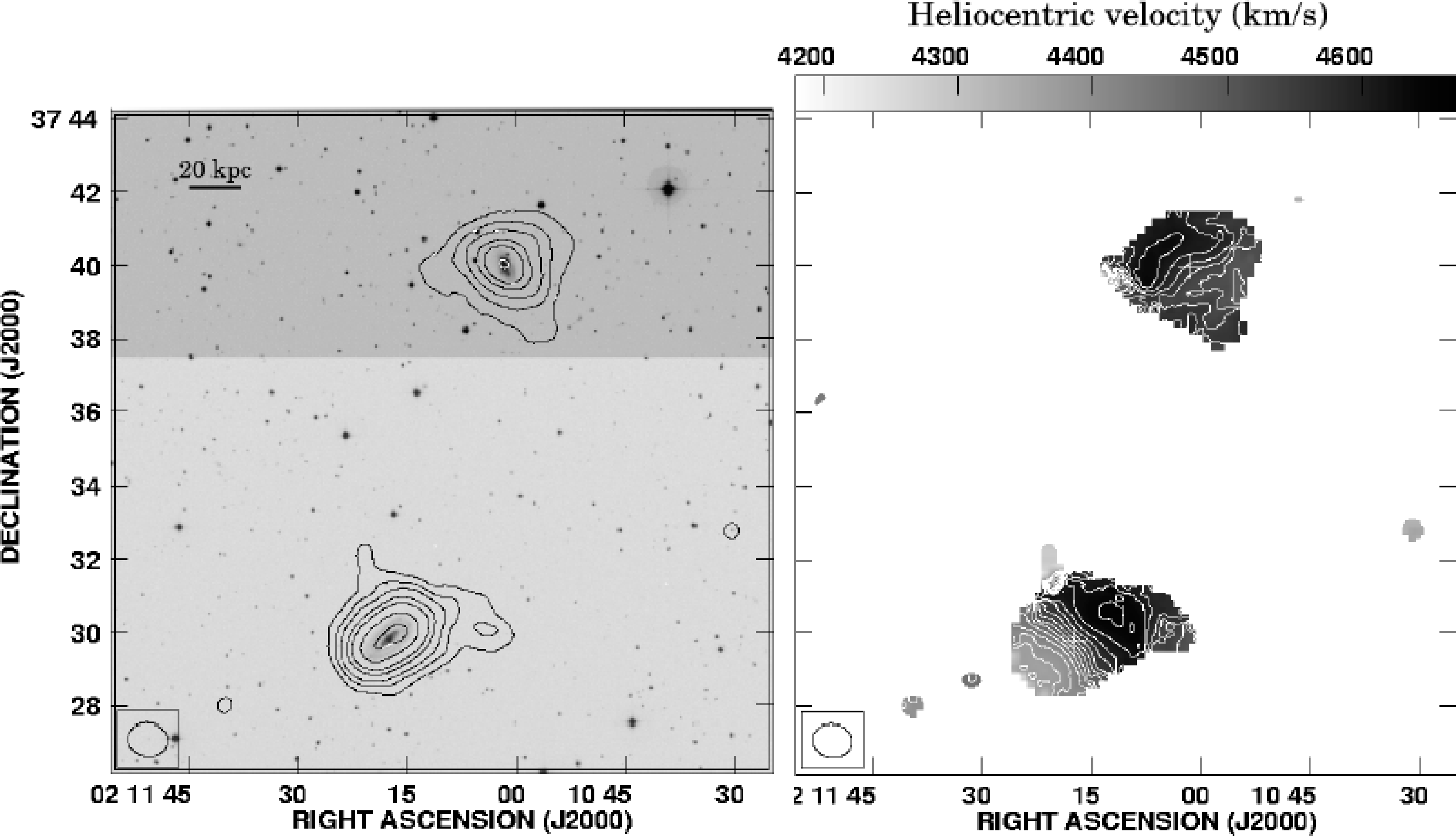} \\
\caption{Upper panel: Optical image of NGC~841 (active galaxy) and NGC~834 from the DSS2.  Lower panels: (\textsl{Left}) Contours of integrated HI intensity (zeroth moment) overlaid on the DSS2 image, and (\textsl{Right}) map of intensity-weighted HI mean velocity (first moment).  NGC~841 lies south-east of center, whereas NGC~834 is located north-west of center. In the zeroth moment map, contours are plotted at 1, 5, 10, 20, 30, 50, $70 \times 30 {\rm \ mJy \ beam^{-1} \ km \ s^{-1}}$ ($1.39 \times 10^{19} {\rm \ cm^{-2}}$).  In the first moment map, velocities are indicated by the scale wedge, and contours plotted at intervals of $25 {\rm \ km \ s^{-1}}$.  The ellipse at the lower left corner of the lower panels indicates the half-power width of the synthesized beam, and has a size of 65$\arcsec \times 54\arcsec$.}
\end{center}
\end{figure}

\begin{figure}
\begin{center}
\vspace*{-0.3 cm}
\hspace*{-2.0 cm}
\includegraphics[angle=0, scale=0.7]{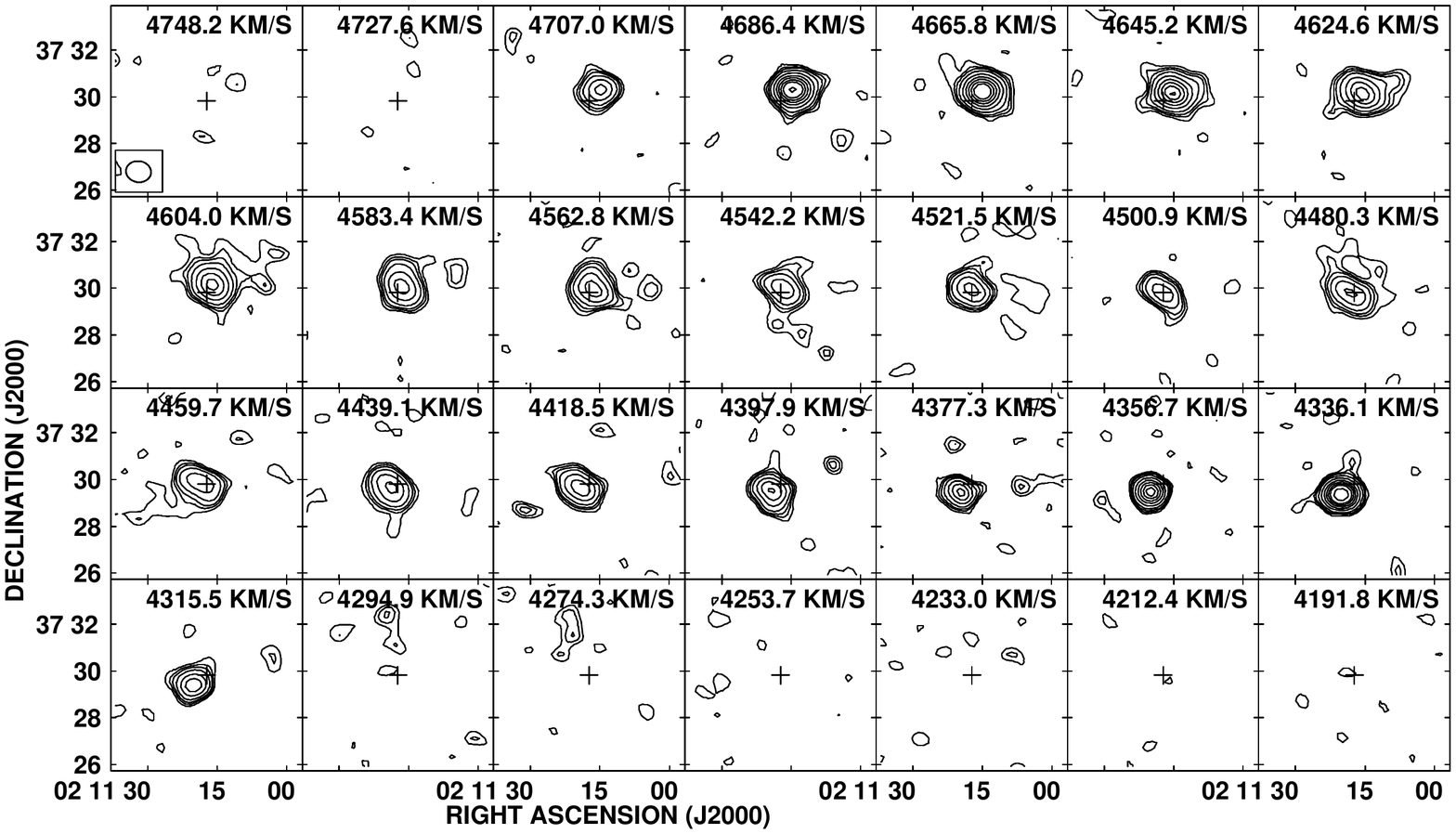}
\vspace*{0.0 cm}
\caption{
HI channel maps of NGC~841. Contour levels are plotted at 2, 3, 4, 6, 10, 15, 20, 25, 30, $40 \times 0.42 {\rm \ mJy \ beam^{-1}}$ ($1 \sigma$), which corresponds to a HI column density of $4.01 \times 10^{18} {\rm \ cm^{-2}}$.  The central heliocentric velocity is shown for each channel.  The cross marks the position of NGC~841.  The synthesized beam is shown by the ellipse at the lower left corner in the top left panel.}
\end{center}
\end{figure}

\begin{figure}
\begin{center}
\vspace*{-0.3 cm}
\hspace*{-2 cm}
\includegraphics[angle=0, scale=0.7]{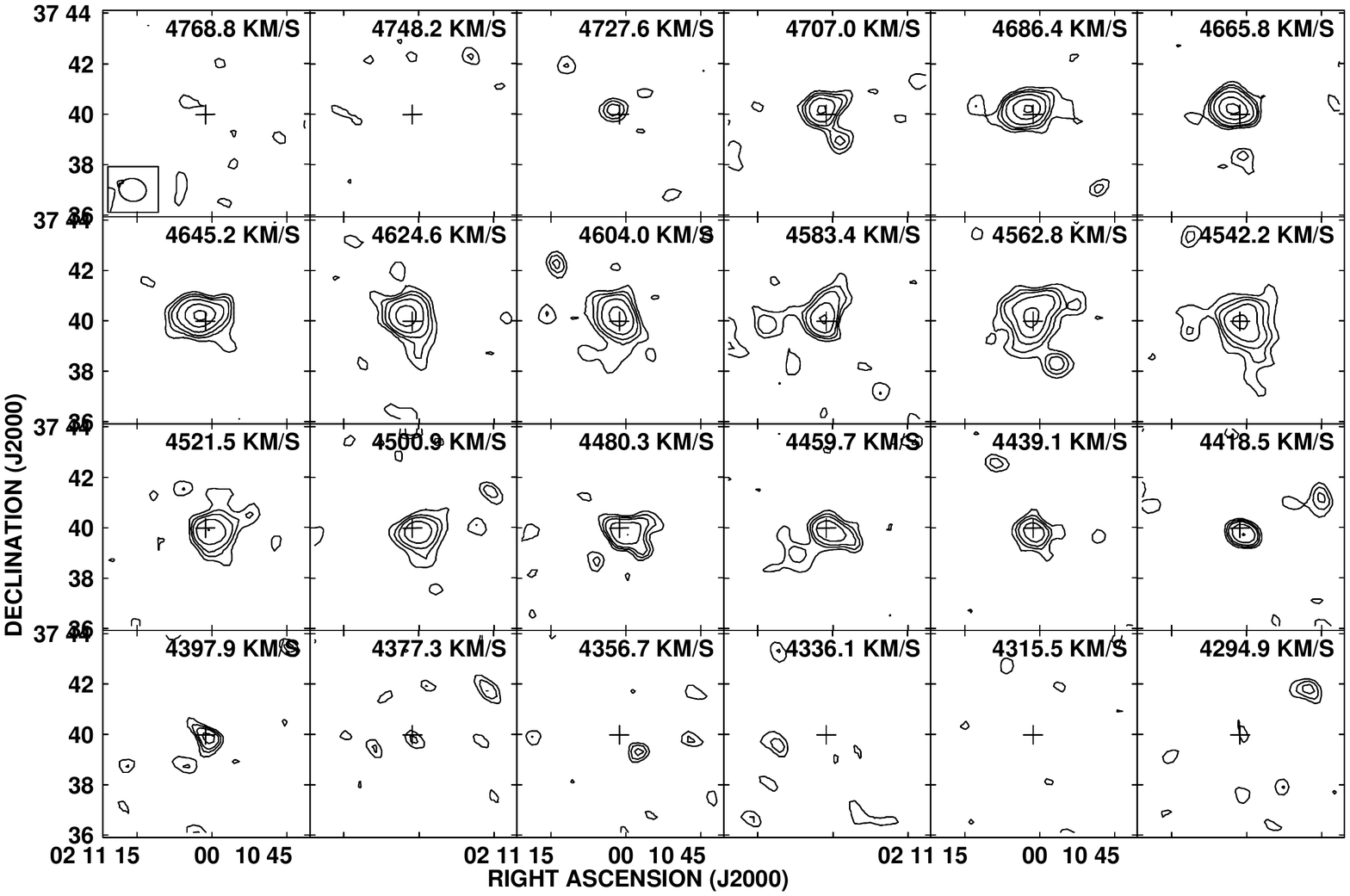}
\vspace*{0.0 cm}
\caption{
HI channel maps of NGC~834. Contour levels are plotted at 2, 3, 4, 6, $10 \times 0.42 {\rm \ mJy \ beam^{-1}}$ ($1 \sigma$), which corresponds to a HI column density of $4.01 \times 10^{18} {\rm \ cm^{-2}}$.  The central heliocentric velocity is shown for each channel.  The cross marks the position of NGC~834. The synthesized beam is shown by the ellipse at the lower left corner of the top left panel.}
\end{center}
\end{figure}

\begin{figure}
\begin{center}
\includegraphics[angle=0, scale=0.53]{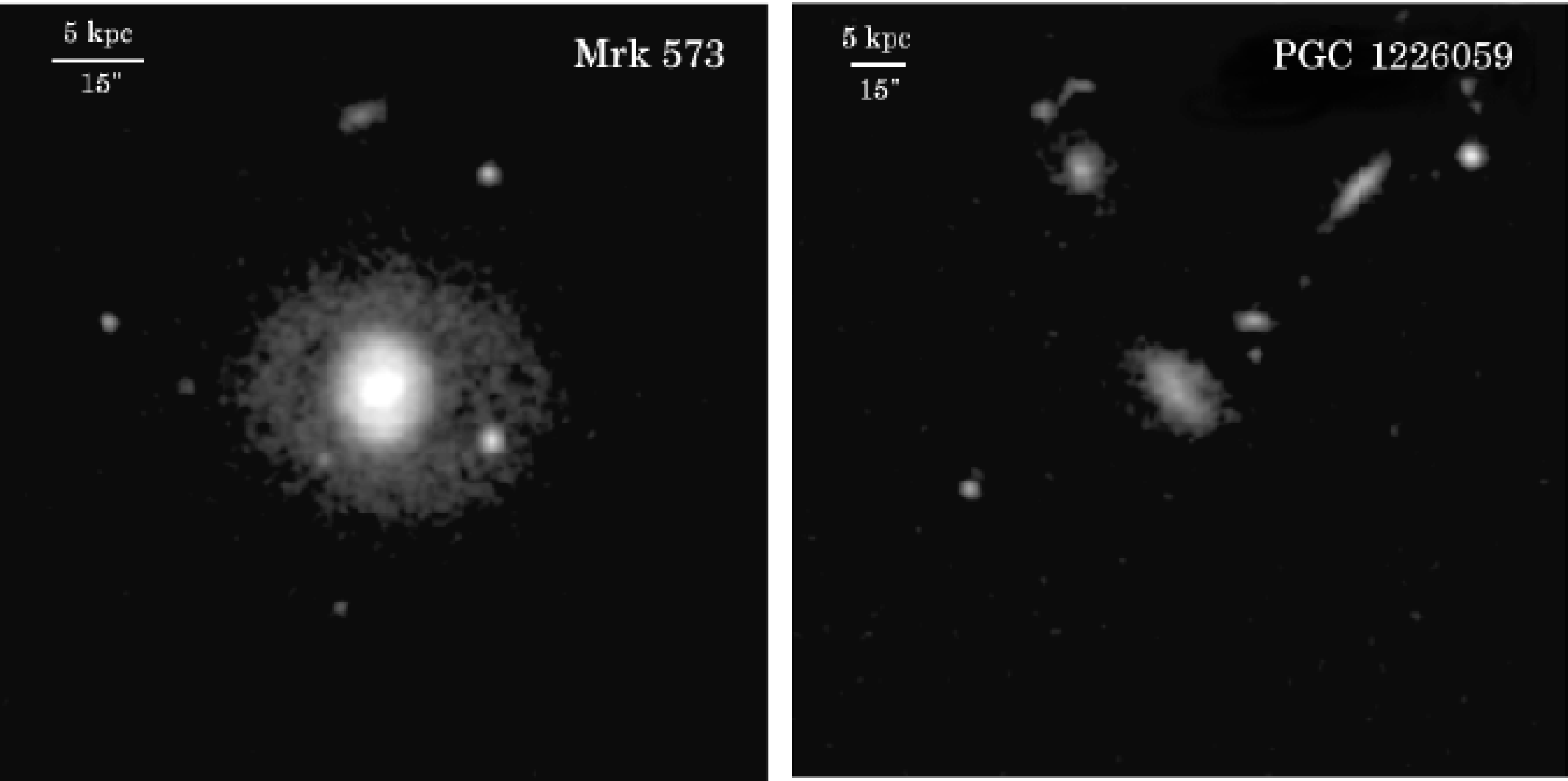}
\hspace*{-1.4 cm}
\includegraphics[angle=0, scale=0.482]{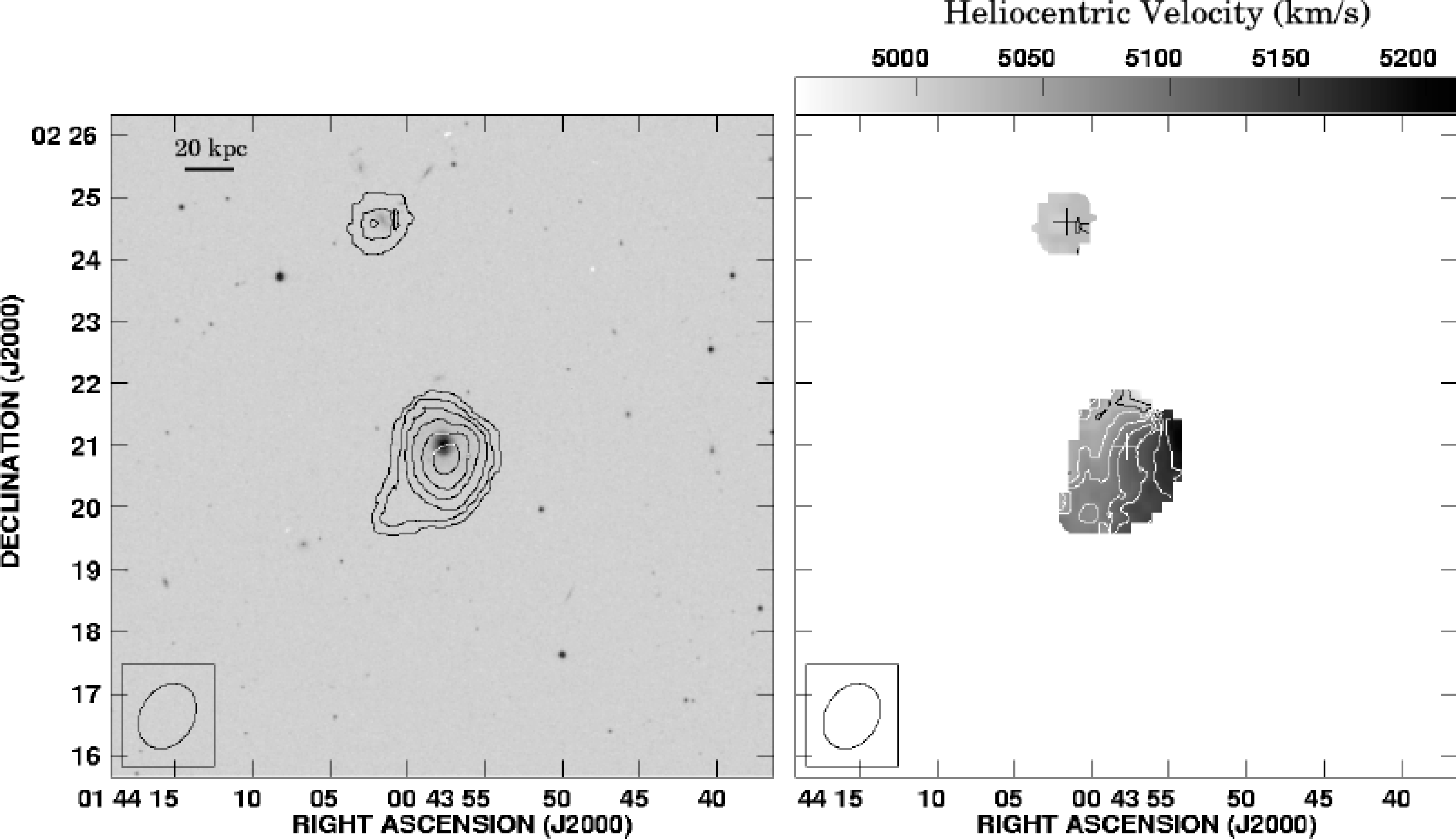} \\
\caption{Upper panel: Optical image of Markarian~573 (Mrk~573) (active galaxy) and PGC~1226059 from the DSS2.  Lower panels: (\textsl{Left}) Contours of integrated HI intensity (zeroth moment) overlaid on the DSS2 image, and (\textsl{Right}) map of intensity-weighted HI mean velocity (first moment). Mrk~573 lies at the center of the map, whereas PGC~1226059 is located north-east of center.  In the zeroth moment map, contours are plotted at 1, 5, 10, 15, 20, $25 \times 20 {\rm \ mJy \ beam^{-1} \ km \ s^{-1}}$ ($9.62 \times 10^{18} {\rm \ cm^{-2}}$). In the first moment map, velocities are indicated by the scale wedge, and contours plotted at intervals of $25 {\rm \ km \ s^{-1}}$. The ellipse at the lower left corner of the lower panels is the half-power width of the synthesized beam, and has a size of $67\arcsec \times 51\arcsec$.}
\end{center}
\end{figure}

\begin{figure}
\begin{center}
\vspace*{-0.3 cm}
\hspace*{-3.1 cm}
\includegraphics[angle=0, scale=0.80]{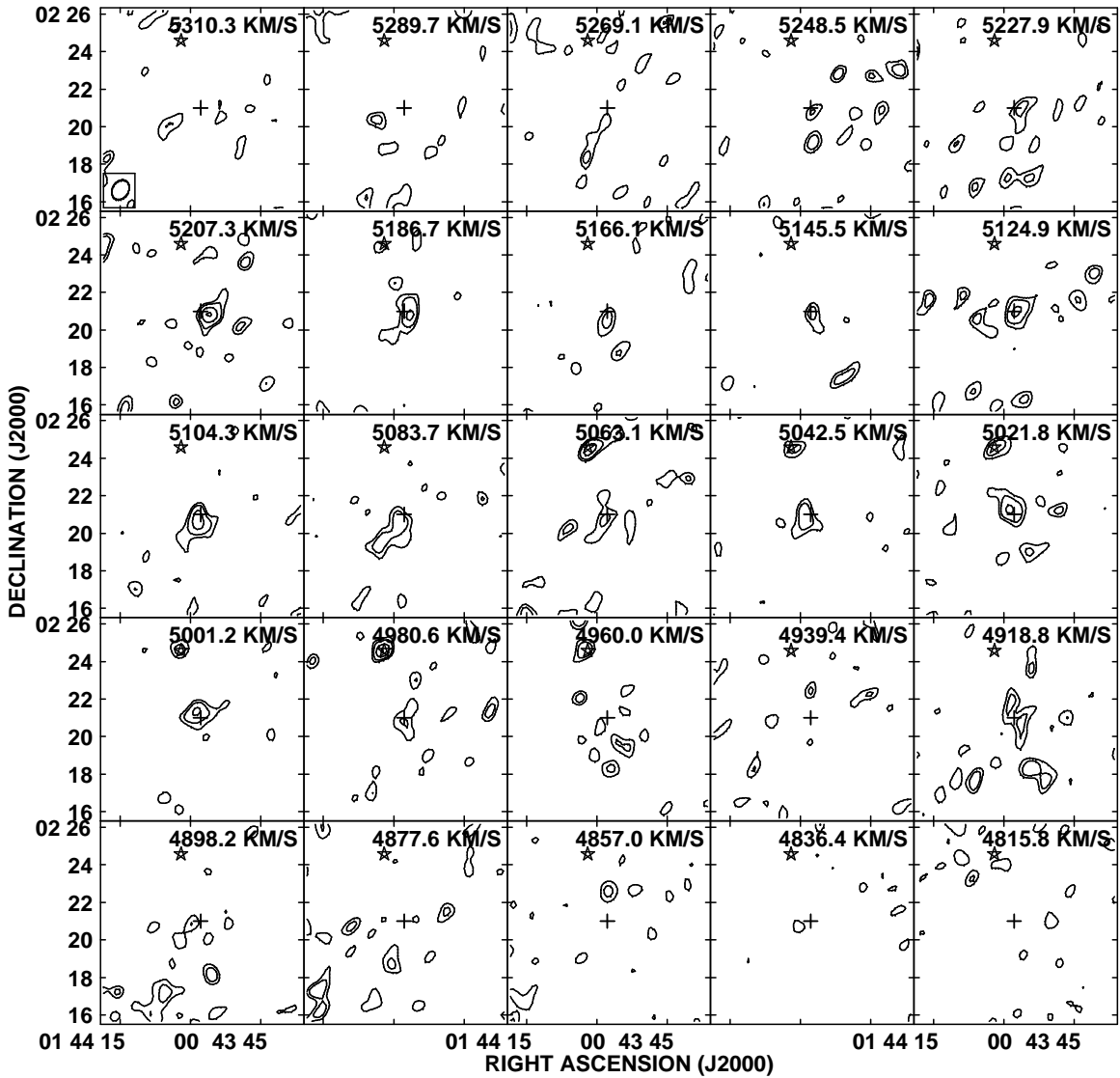}
\vspace*{0.0 cm}
\caption{
HI channel maps of Mrk~573 and PGC~1226059. Contour levels are plotted at 2, 3, 5, $7 \times 0.5 {\rm \ mJy \ beam^{-1}}$ ($1 \sigma$), which corresponds to a HI column density of $4.95 \times 10^{18} {\rm \ cm^{-2}}$. The central heliocentric velocity is shown for each channel. The cross marks the position of Mrk~573, and star the position of PGC~1226059. The synthesized beam is shown by the ellipse at the lower left corner of the top left panel.}
\end{center}
\end{figure}

\begin{figure}
\begin{center}
\includegraphics[angle=0, scale=0.53]{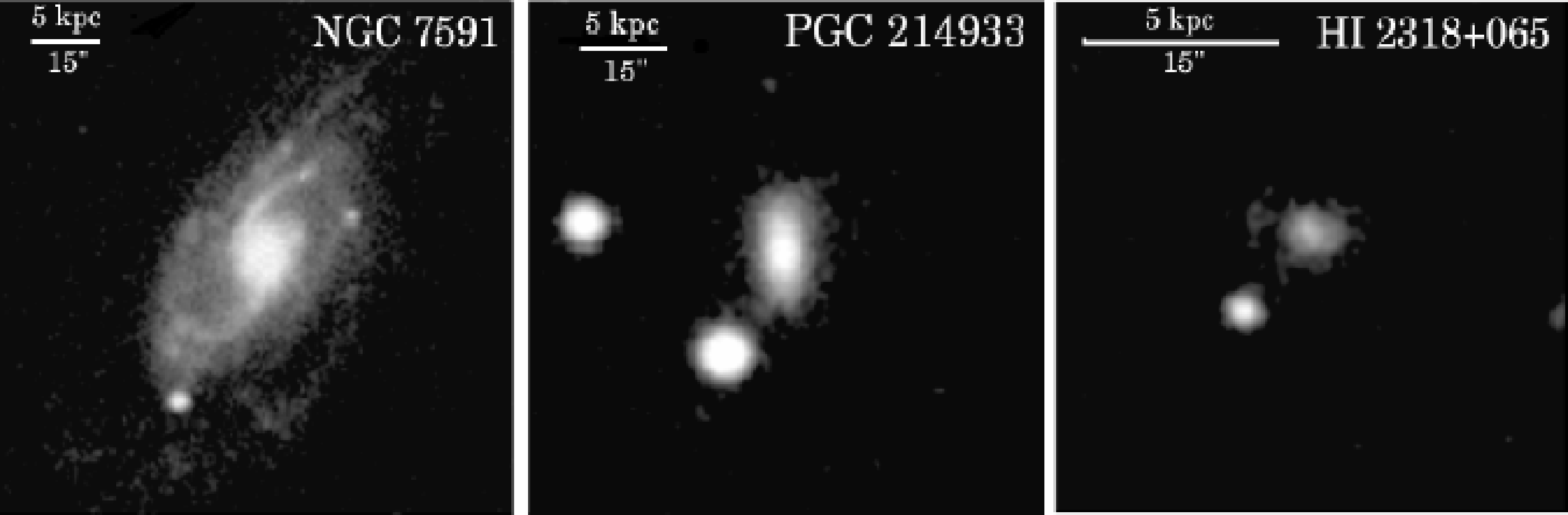}
\hspace*{-1.4 cm}
\includegraphics[angle=0, scale=0.482]{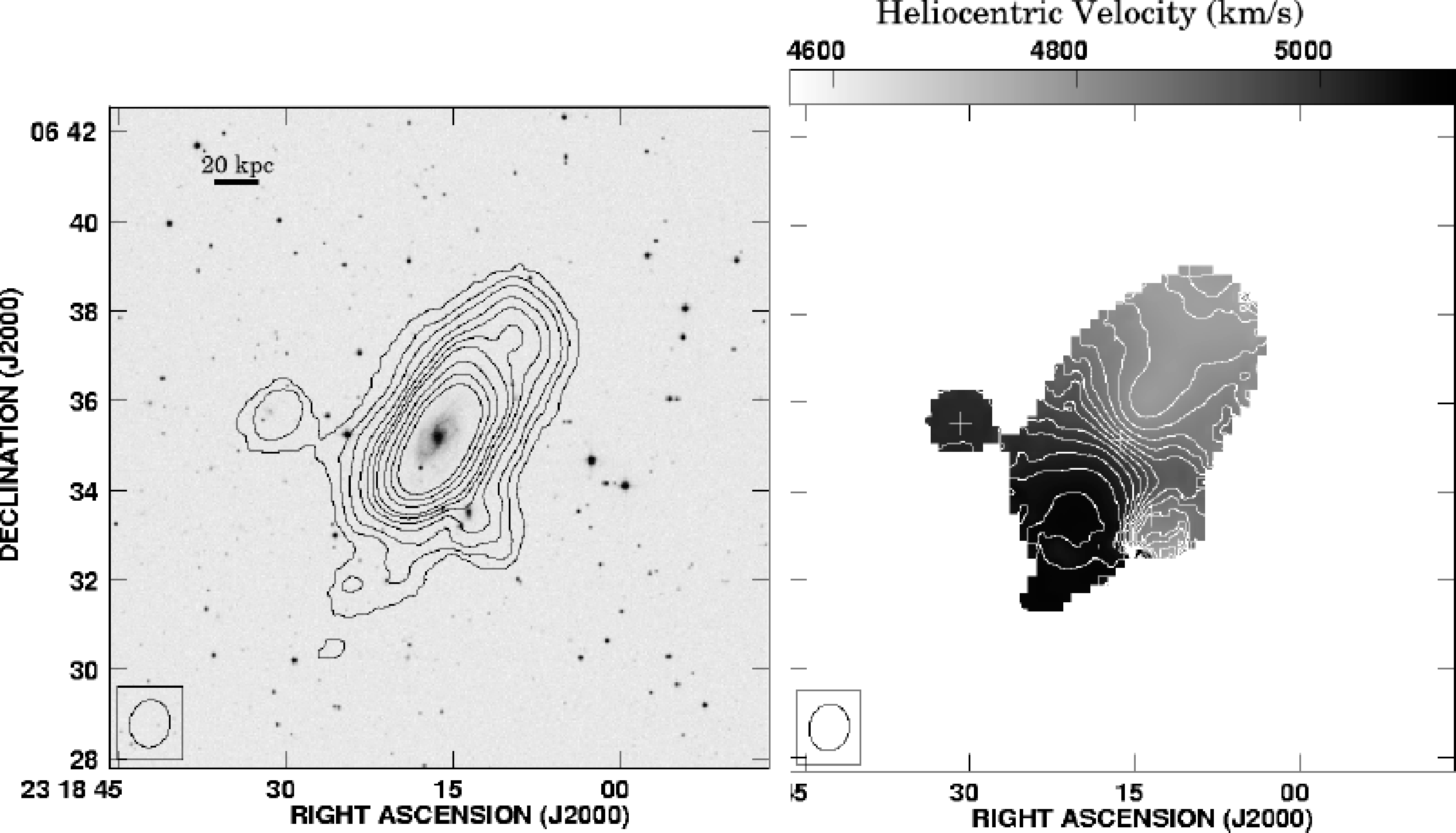} \\
\caption{Upper panel: Optical image of NGC~7591 (active galaxy), PGC~214933, and
HI~2318$+$065 from the DSS2.  Lower panels: (\textsl{Left}) Contours of integrated HI intensity (zeroth moment) overlaid on the DSS2 image, and (\textsl{Right}) map of intensity-weighted HI mean velocity (first moment).  NGC~7591 lies at the center of the map, whereas PGC~214933 is located south-west and HI~2318$+$065 east of center. In the zeroth moment map, contours are plotted at 1, 5, 10, 20
, 30, 40, 50, 70, 90, 110 $\times 20 {\rm \ mJy \ beam^{-1} \ km \ s^{-1}}$ ($7.39 \times 10^{18} {\rm \ cm^{-2}}$). In the first moment map, velocities are indicated by the scale wedge, and contours plotted at intervals of $25 {\rm \ km \ s^{-1}}$. The ellipse at the lower left corner of the lower panels indicates the half-power width of the synthesized beam, and has a size of 63$\arcsec \times 53\arcsec$.
}
\end{center}
\end{figure}

\begin{figure}
\begin{center}
\vspace*{-0.3 cm}
\hspace*{-3.0 cm}
\includegraphics[angle=0, scale=0.78]{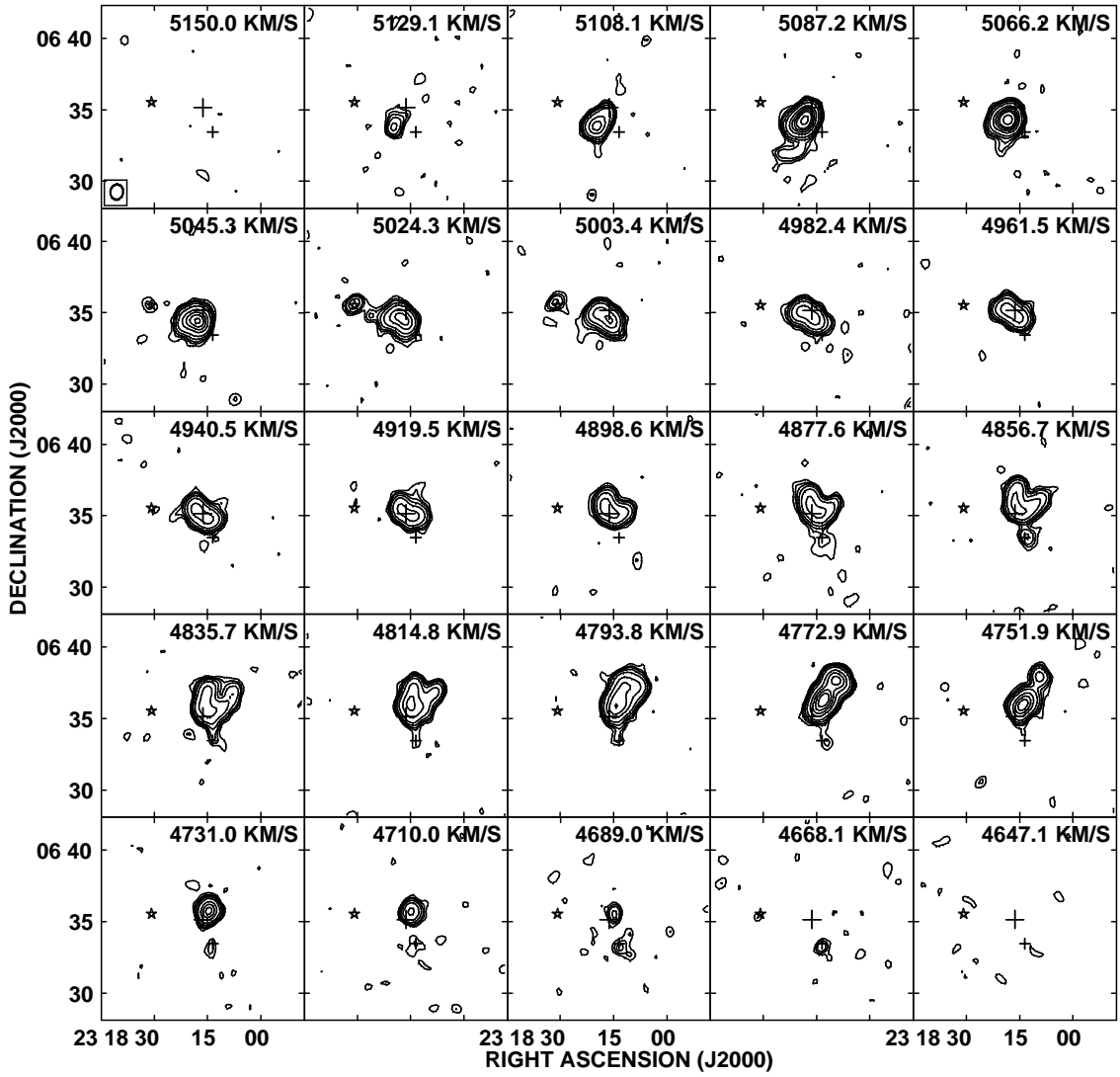}
\vspace*{0.0 cm}
\caption{
HI channel maps of NGC~7591, PGC~214933, and HI~2318$+$065.  Contour levels are plotted at 2, 3, 4, 6, 10, 20, 30, 40, 50, $70 \times 0.45 {\rm \ mJy \ beam^{-1}}$ ($1 \sigma$), which corresponds to a HI column density of $3.43 \times 10^{18} {\rm \ cm^{-2}}$. The central heliocentric velocity is shown for each channel. The bigger cross marks the position of NGC~7591, star the position of HI~2318$+$065, and the smaller cross the position of PGC~214993. The synthesized beam is shown by the ellipse at the lower left corner of the top left panel.}
\end{center}
\end{figure}

\begin{figure}
\begin{center}
\includegraphics[angle=0, scale=0.485]{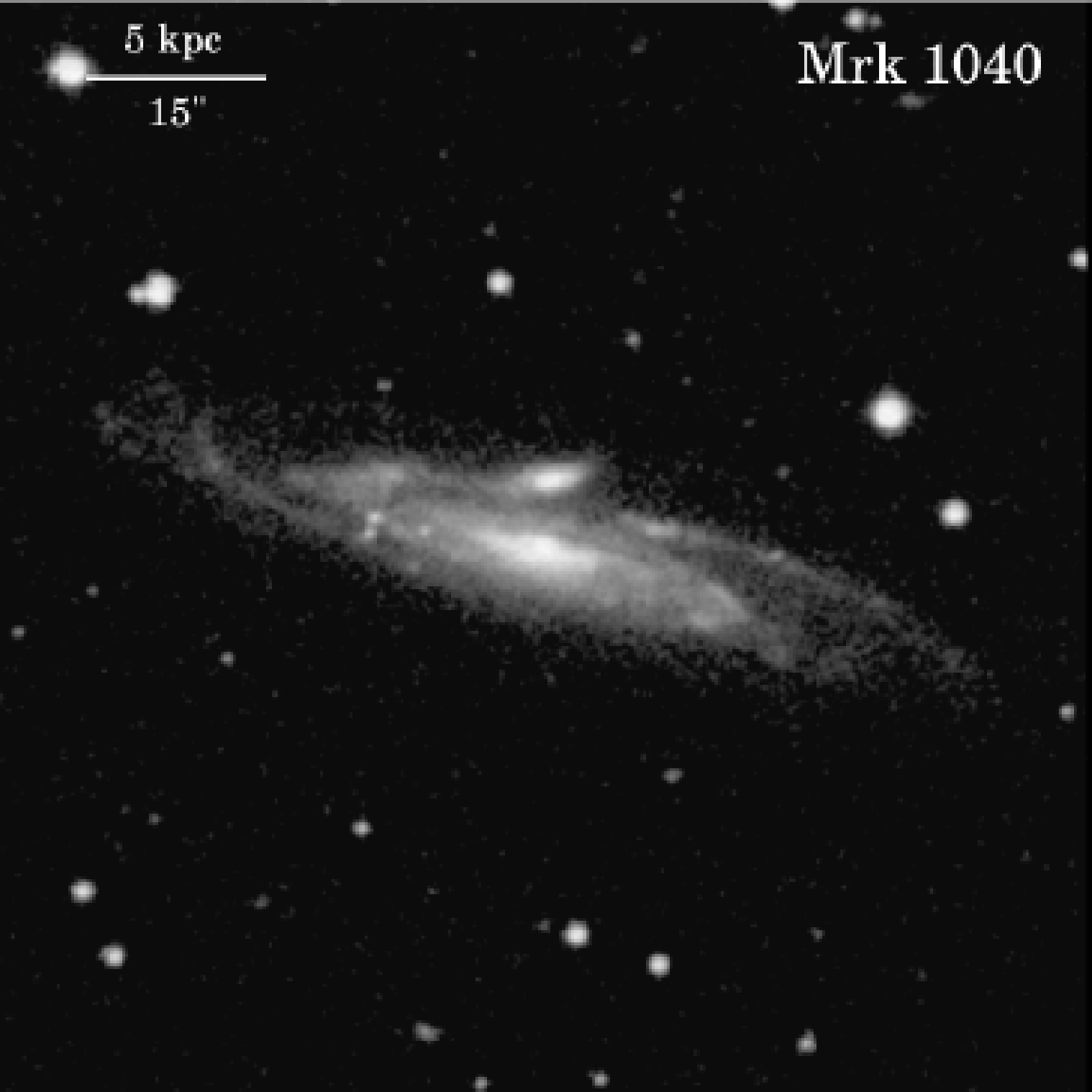}
\hspace*{-1.4 cm}
\includegraphics[angle=0, scale=0.482]{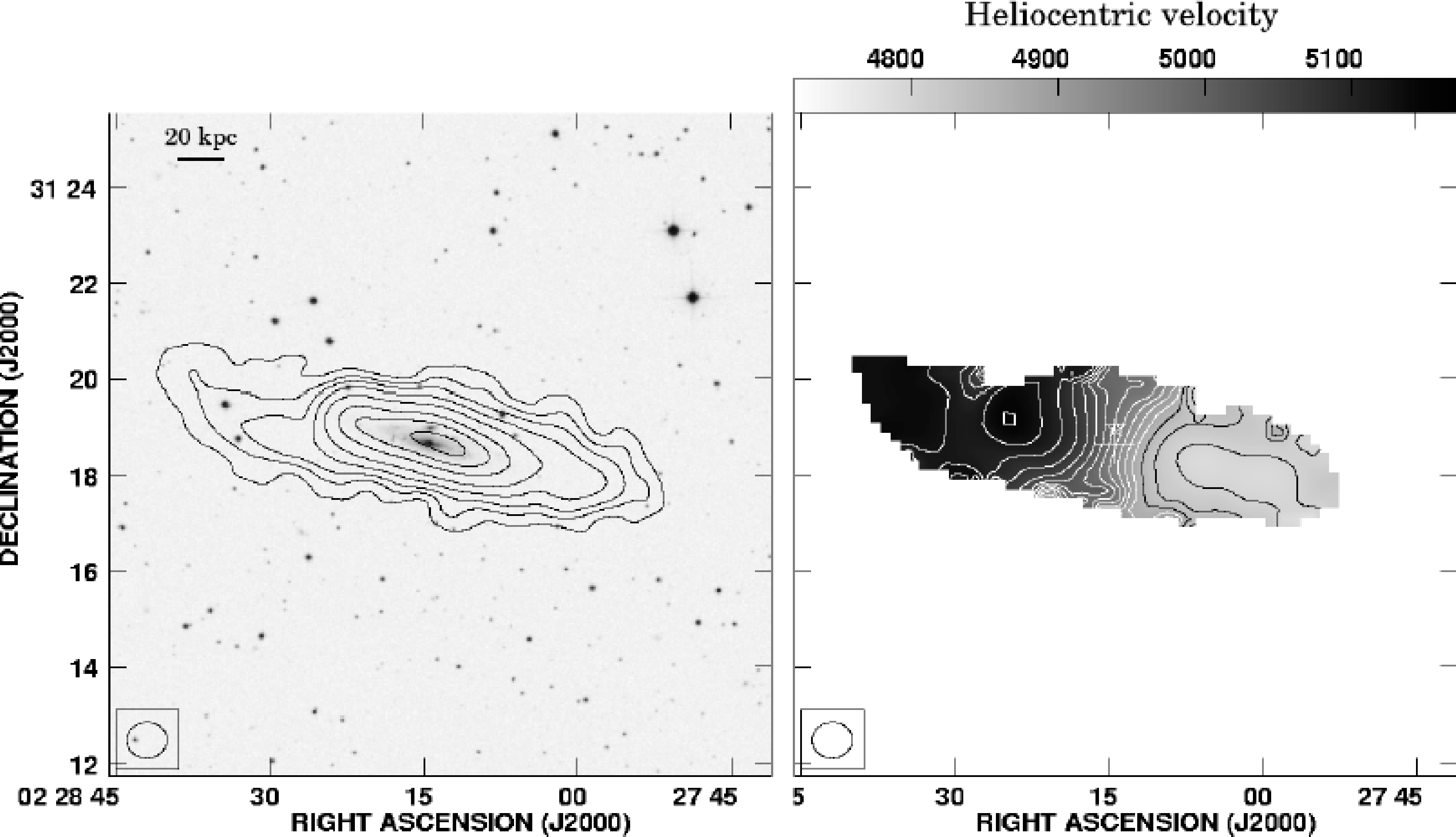} \\
\caption{
Upper panel: Optical image of Markarian~1040 (Mrk 1040) (active galaxy) and its companion galaxy PGC~212995 from the DSS2. Lower panels: (\textsl{Left}) Contours of integrated HI intensity (zeroth moment) overlaid on the DSS2 image, and (\textsl{Right}) map of intensity-weighted HI mean velocity (first moment). Mrk~1040 is the larger galaxy at the center of the map, whereas PGC~212995 is the smaller galaxy located just to the north.  In the zeroth moment map, contours are plotted at 1, 5, 10,20, 40, 80, 120, $160 \times 30 {\rm \ mJy \ beam^{-1} \ km \ s^{-1}}$ ($2.19 \times 10^{19} {\rm \ cm^{-2}}$). In the first moment map, velocities are indicated by the scale wedge, and contours plotted at intervals of $25 {\rm \ km \ s^{-1}}$. The ellipse at the lower left corner of the lower panels indicates the half-power width of the synthesized beam, and has a size of $50\arcsec \times 54\arcsec$.}
\end{center}
\end{figure}

\begin{figure}
\begin{center}
\vspace*{-0.3 cm}
\hspace*{-2 cm}
\includegraphics[angle=0, scale=0.7]{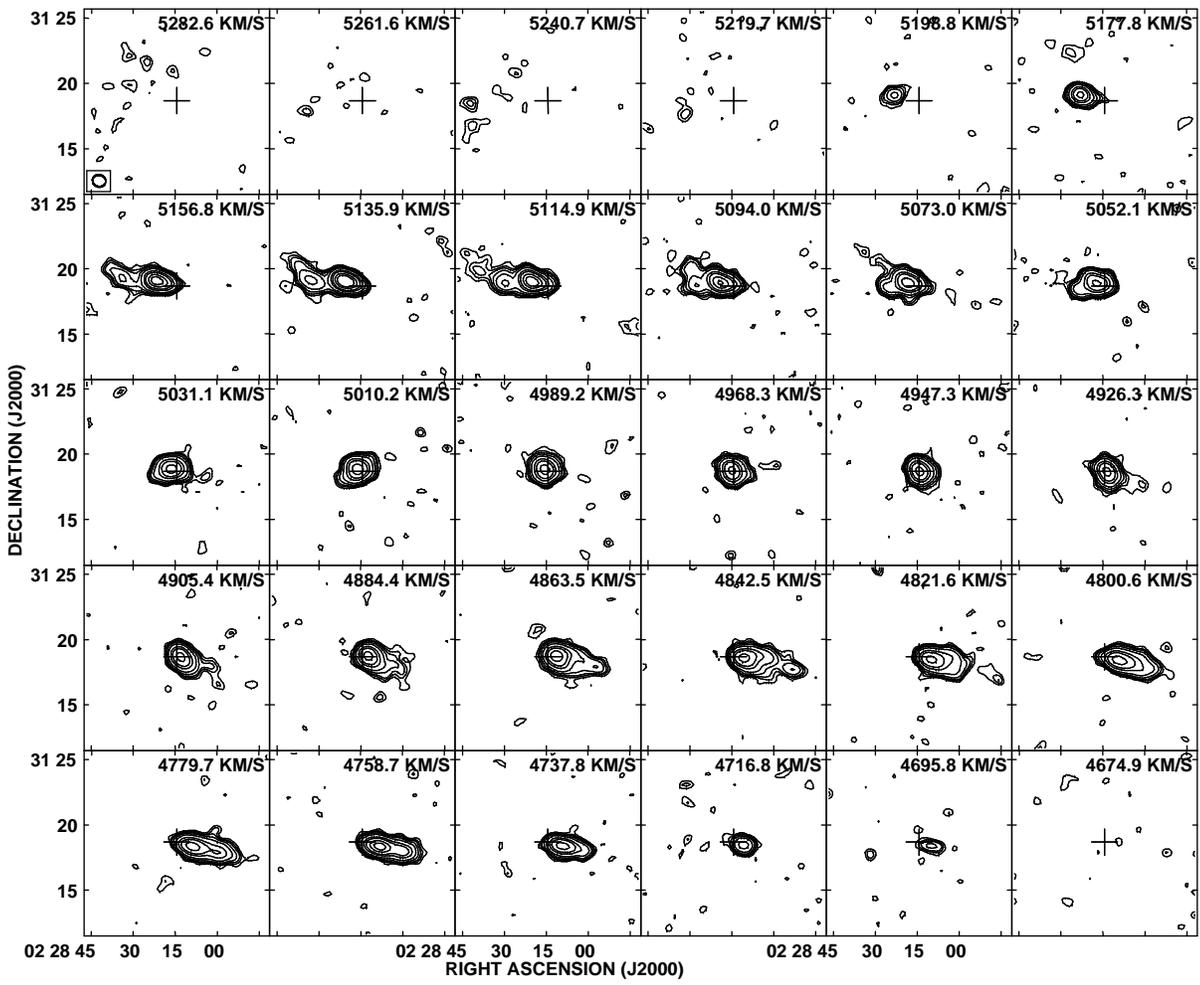}
\vspace*{0.0 cm}
\caption{HI channel maps of Mrk 1040 and PGC~212995. Contour levels are plotted at 2, 3, 4, 6, 10, 20, 30, 50, $70 \times 0.45 {\rm \ mJy 
\ beam^{-1}}$ ($1 \sigma$), which corresponds to a HI column density of $6.77 \times 10^{18} {\rm \ cm^{-2}}$. The central heliocentric velocity is shown for each channel. The cross marks the position of Mrk~1040.  The synthesized beam is shown by the ellipse at the lower left corner of the top left panel.}
\end{center}
\end{figure}

\begin{figure}
\begin{center}
\includegraphics[angle=0, scale=0.485]{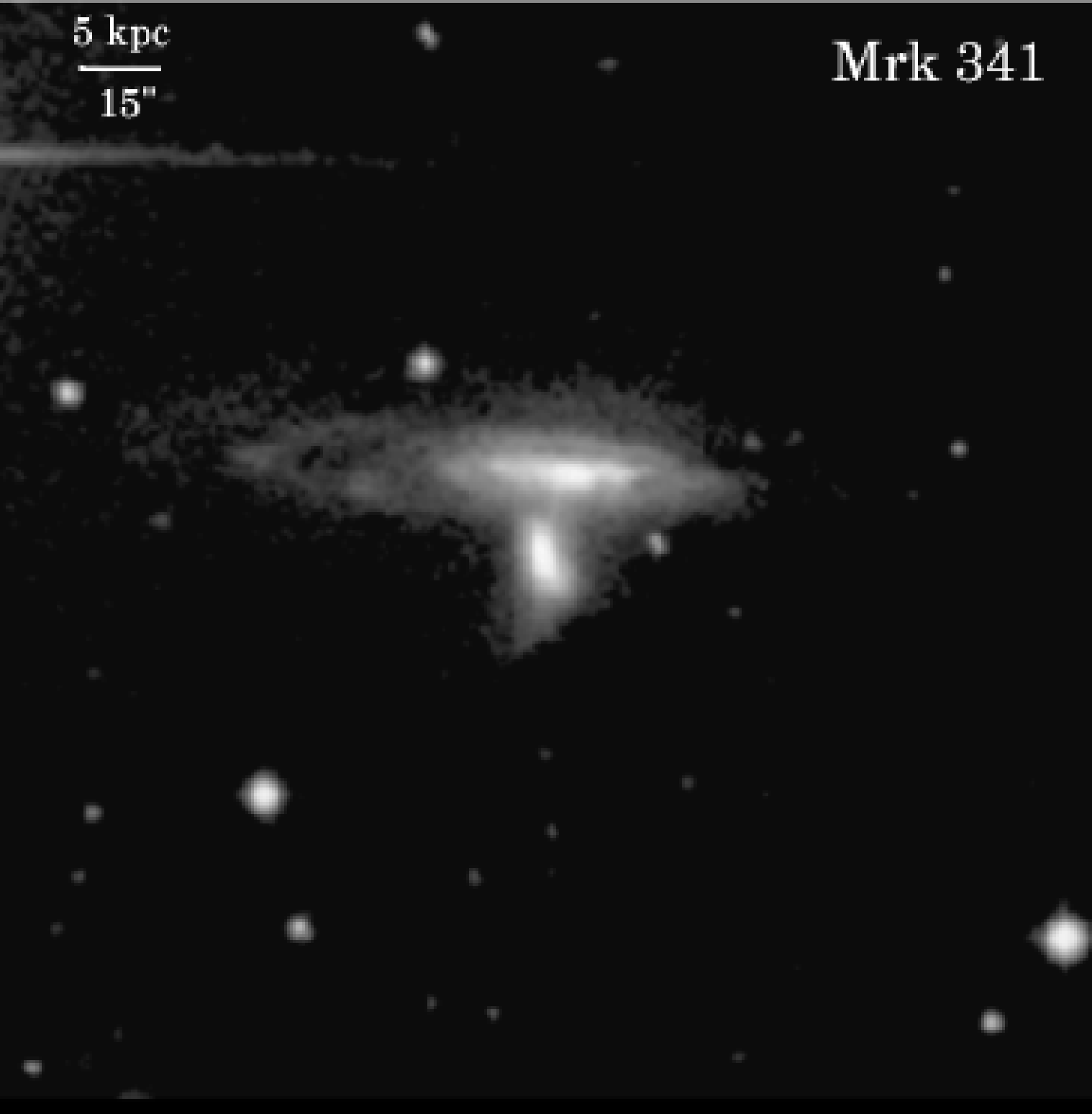}
\hspace*{-1.4 cm}
\includegraphics[angle=0, scale=0.482]{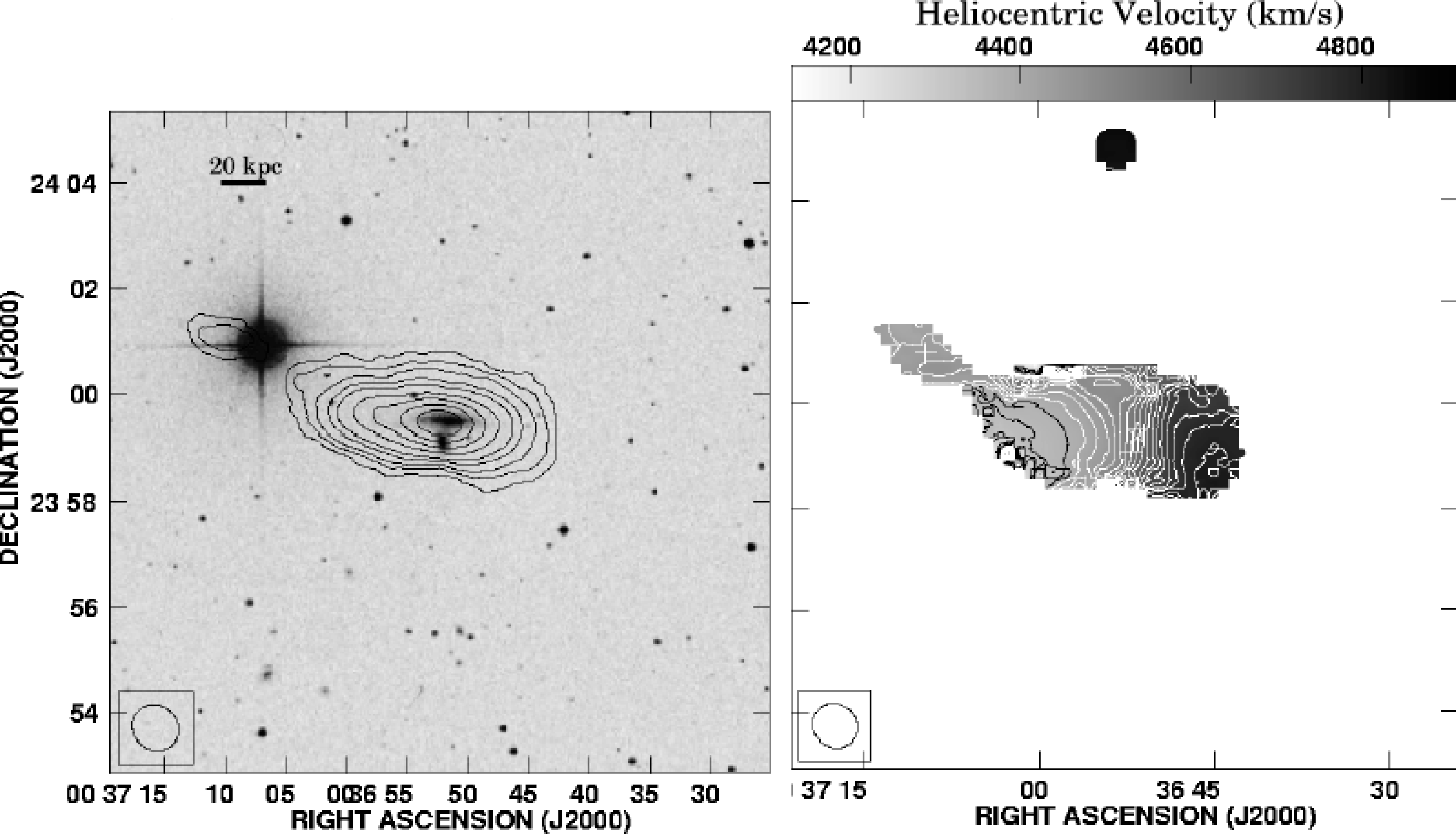} \\
\caption{
Upper panel: Optical image of Markarian~341 (Mrk~341) (active galaxy) and NGC~169 from the DSS2.  Lower panels: (\textsl{Left}) Contours of integrated HI intensity (zeroth moment) overlaid on the DSS2 image, and (\textsl{Right}) map of intensity-weighted HI mean velocity (first moment).  Mrk~341 is the smaller galaxy at the center of the map, and NGC~169 the larger galaxy located just to the north. In the zeroth moment map, contours are plotted at 1, 5, 10, 20, 30, 50, 70, 90, $110 \times 20 {\rm \ mJy \ beam^{-1} \ km \ s^{-1}}$ ($1.24 \times 10^{19} {\rm \ cm^{-2}}$). In the first moment map, velocities are indicated by the scale wedge, and contours plotted at intervals of $25 {\rm \ km \ s^{-1}}$. The ellipse at the lower left corner of the lower panels indicates the half-power width of the synthesized beam, and has a size of $54\arcsec \times 49\arcsec$.}
\end{center}
\end{figure}

\begin{figure}
\begin{center}
\vspace*{-0.3 cm}
\hspace*{-1.8 cm}
\includegraphics[angle=0, scale=0.7]{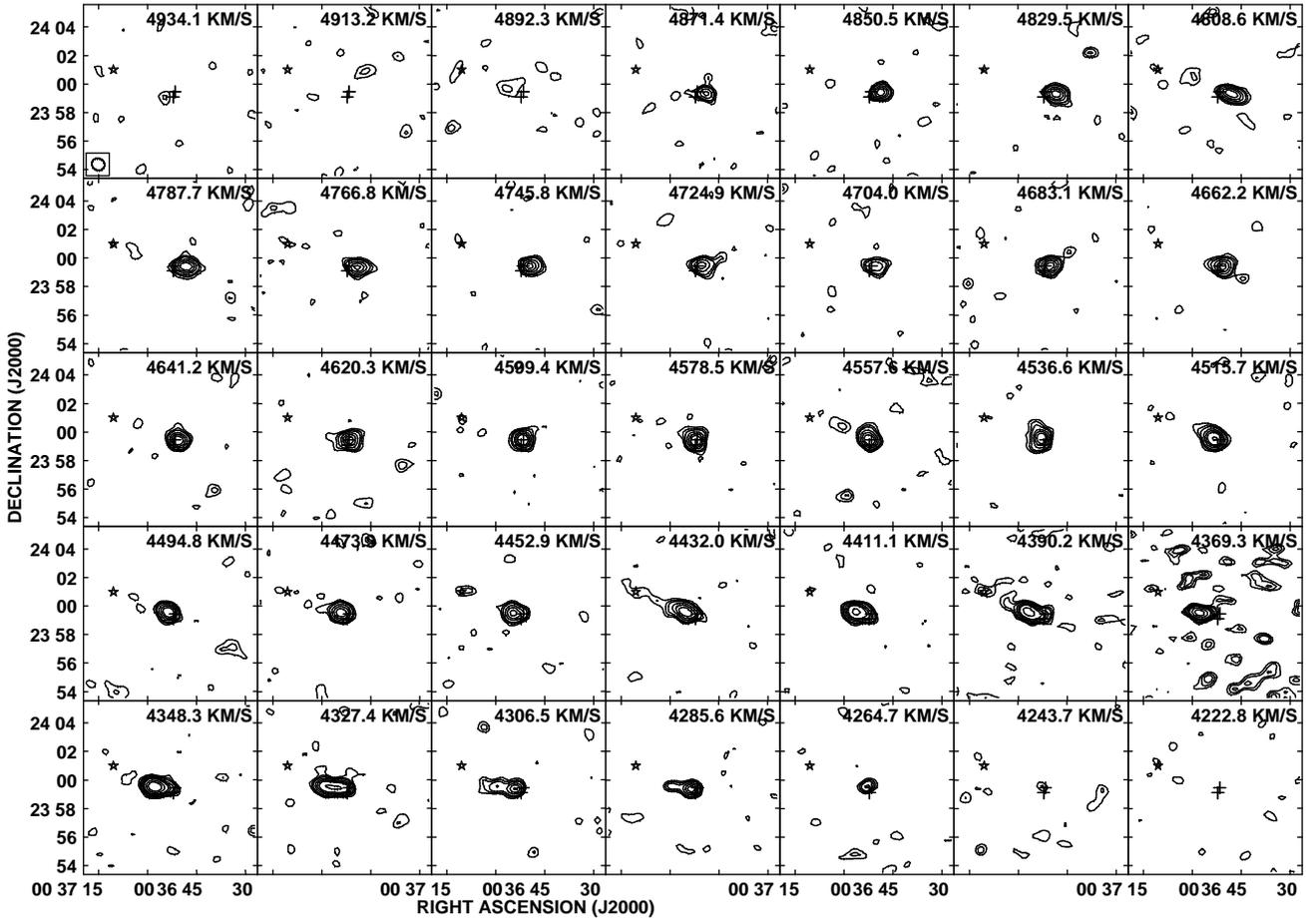}
\vspace*{0.0 cm}
\caption{HI channel maps of Mrk~341 and NGC~169. Contour levels are plotted at 2, 3, 4, 6, 8, 10, 15, $20 \times 0.5 {\rm \ mJy \ beam^{-1}}$ ($1 \sigma$), which corresponds to a HI column density of $6.39 \times 10^{18} {\rm \ cm^{-2}}$. The central heliocentric velocity is shown for each channel. The southern cross marks the position of Mrk~341, and the northern cross the position of NGC~169. The synthesized beam is shown by the ellipse at the lower left corner of the top left panel.}
\end{center}
\end{figure}

\begin{figure}
\begin{center}
\includegraphics[angle=0, scale=0.53]{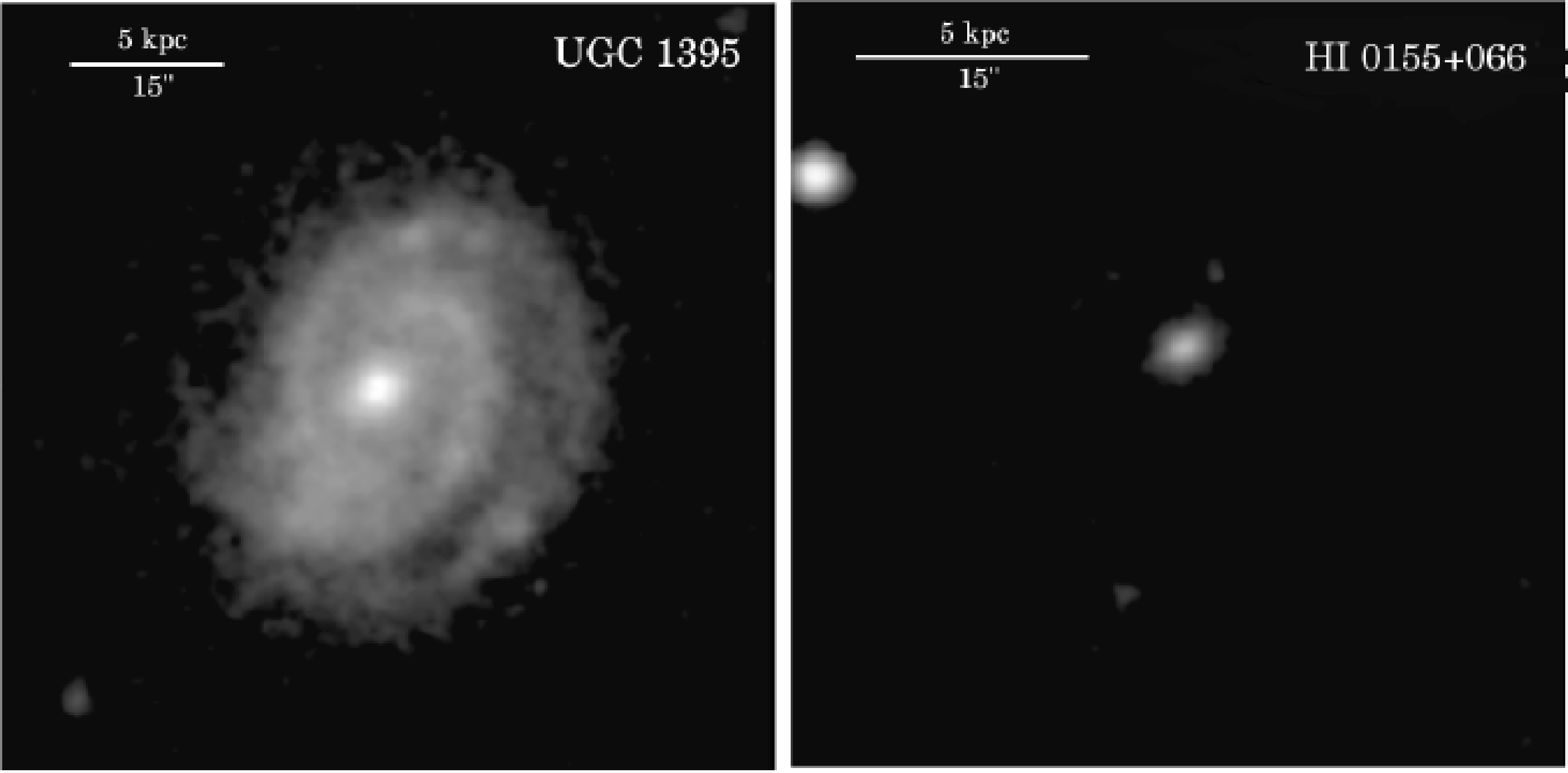}
\hspace*{-1.4 cm}
\includegraphics[angle=0, scale=0.482]{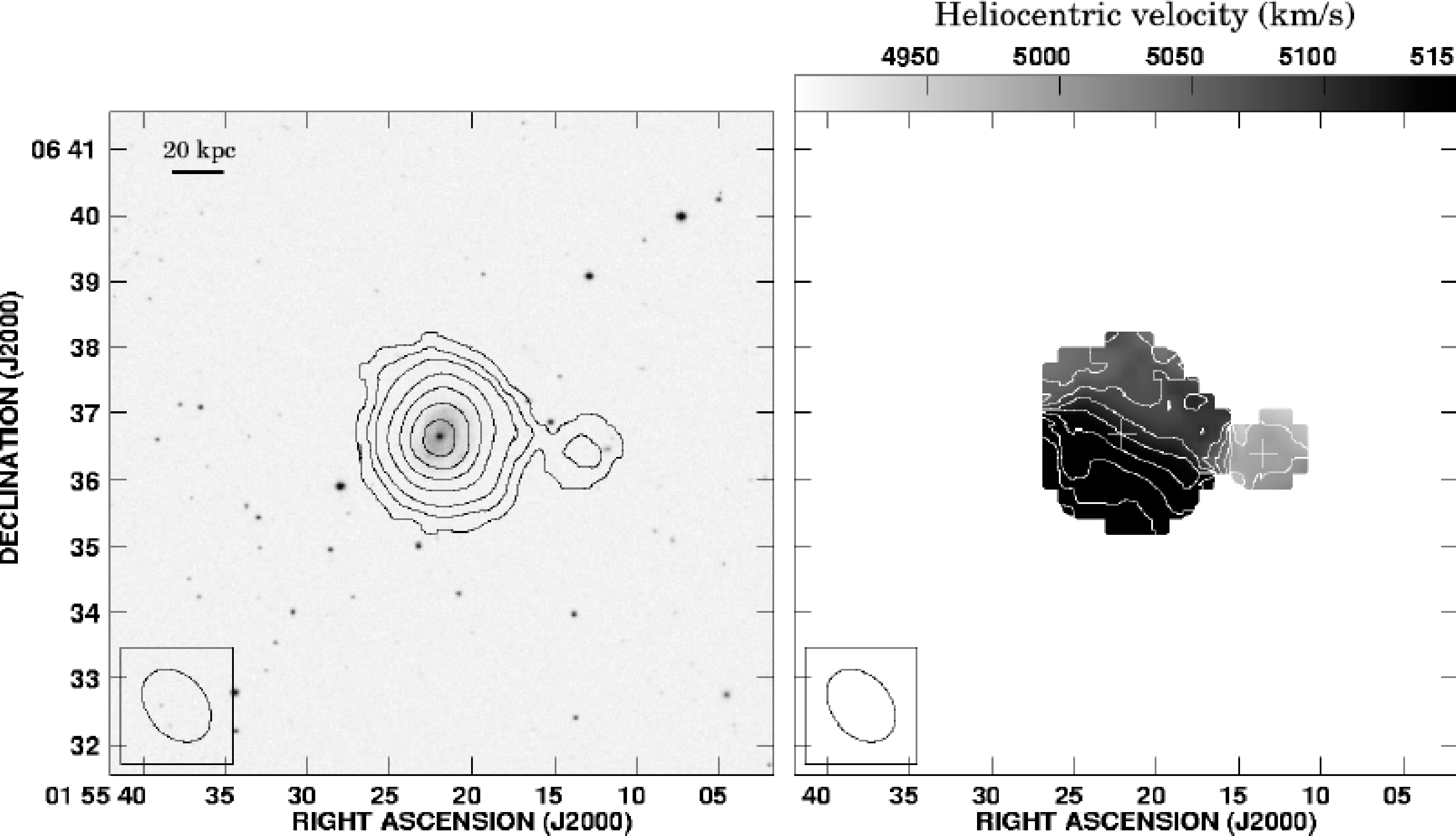} \\
\caption{
Upper panel: Optical image of UGC~1395 (active galaxy) and HI~0155$+$066 from the DSS2.  Lower panels: (\textsl{Left}) Contours of integrated HI intensity (zeroth
moment) overlaid on the DSS2 image, and (\textsl{Right}) map of intensity-weighted HI mean velocity (first moment). UGC~1395 lies at the center of the map, whereas HI~0155$+$066 is located west of center. In the zeroth moment map, contours are plotted at 1, 5, 10, 20, 30, 40, $50 \times 20 {\rm \ mJy \ beam^{-1} \ km \ s^{-1}}$ ($8.5 \times 10^{18} {\rm \ cm^{-2}}$). In the first moment map, velocities are indicated by the scale wedge, and contours plotted at intervals of $25 {\rm \ km \ s^{-1}}$. The ellipse at the lower left corner of the lower panels indicates the half-power width of the synthesized beam, and has a size of $73\arcsec \times 53\arcsec$.}
\end{center}
\end{figure}

\begin{figure}
\begin{center}
\vspace*{-0.3 cm}
\hspace*{-2 cm}
\includegraphics[angle=0, scale=0.7]{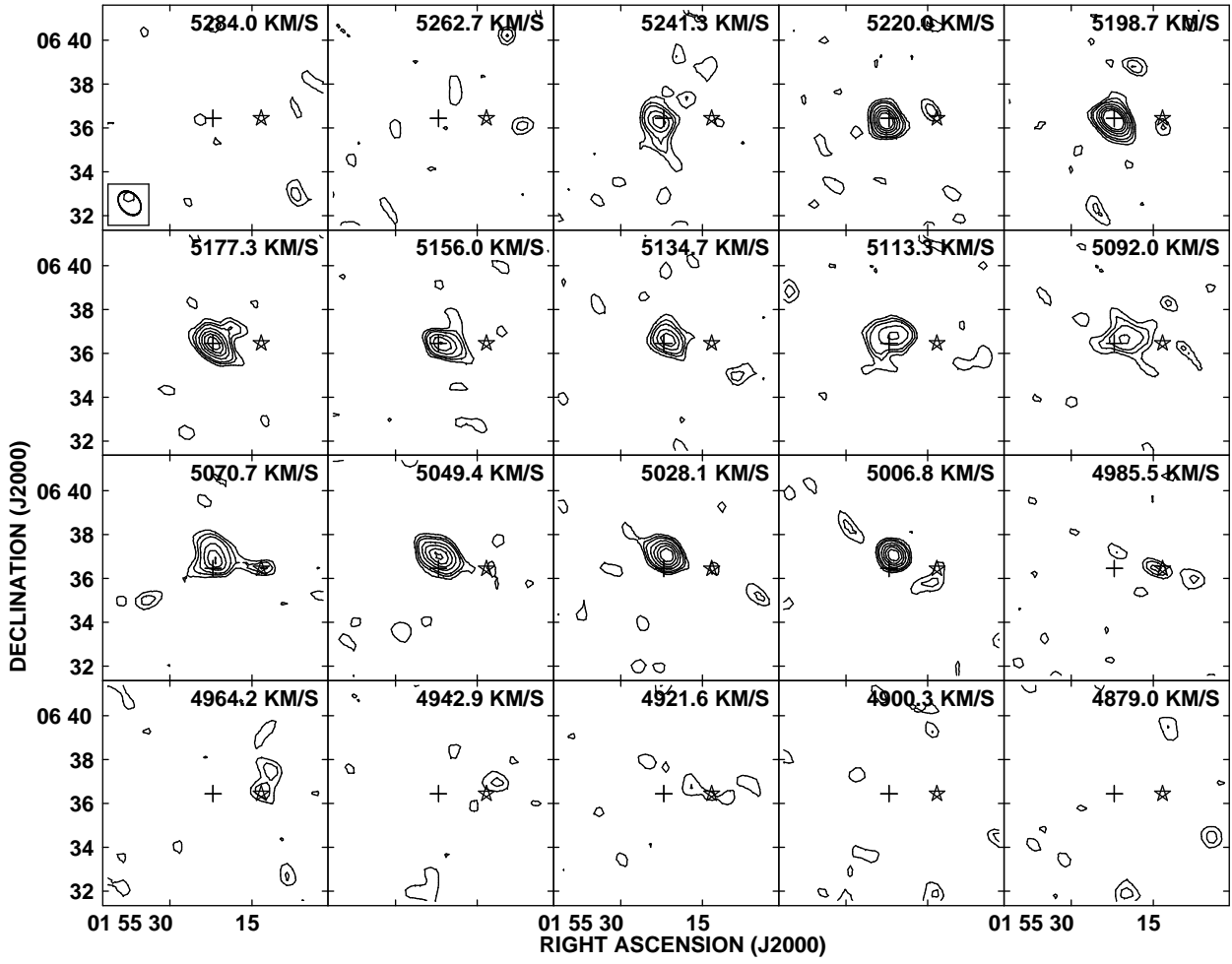}
\vspace*{0.0 cm}
\caption{HI channel maps of UGC~1395 and HI~0155$+$066. Contour levels are plotted at 2, 3, 4, 6, 8, 10, $12 \times 0.46 {\rm \ mJy \ beam^{-1}}$ ($1 \sigma$), which corresponds to a HI column density of $4.03 \times 10^{18} {\rm \ cm^{-2}}$. The central heliocentric velocity is shown for each channel. The cross marks the position of UGC~1395, and star the position of HI~0155$+$066. The synthesized beam is shown by the ellipse at the lower left corner in the top left panel.}
\end{center}
\end{figure}

\begin{figure}
\begin{center}
\includegraphics[angle=0, scale=0.485]{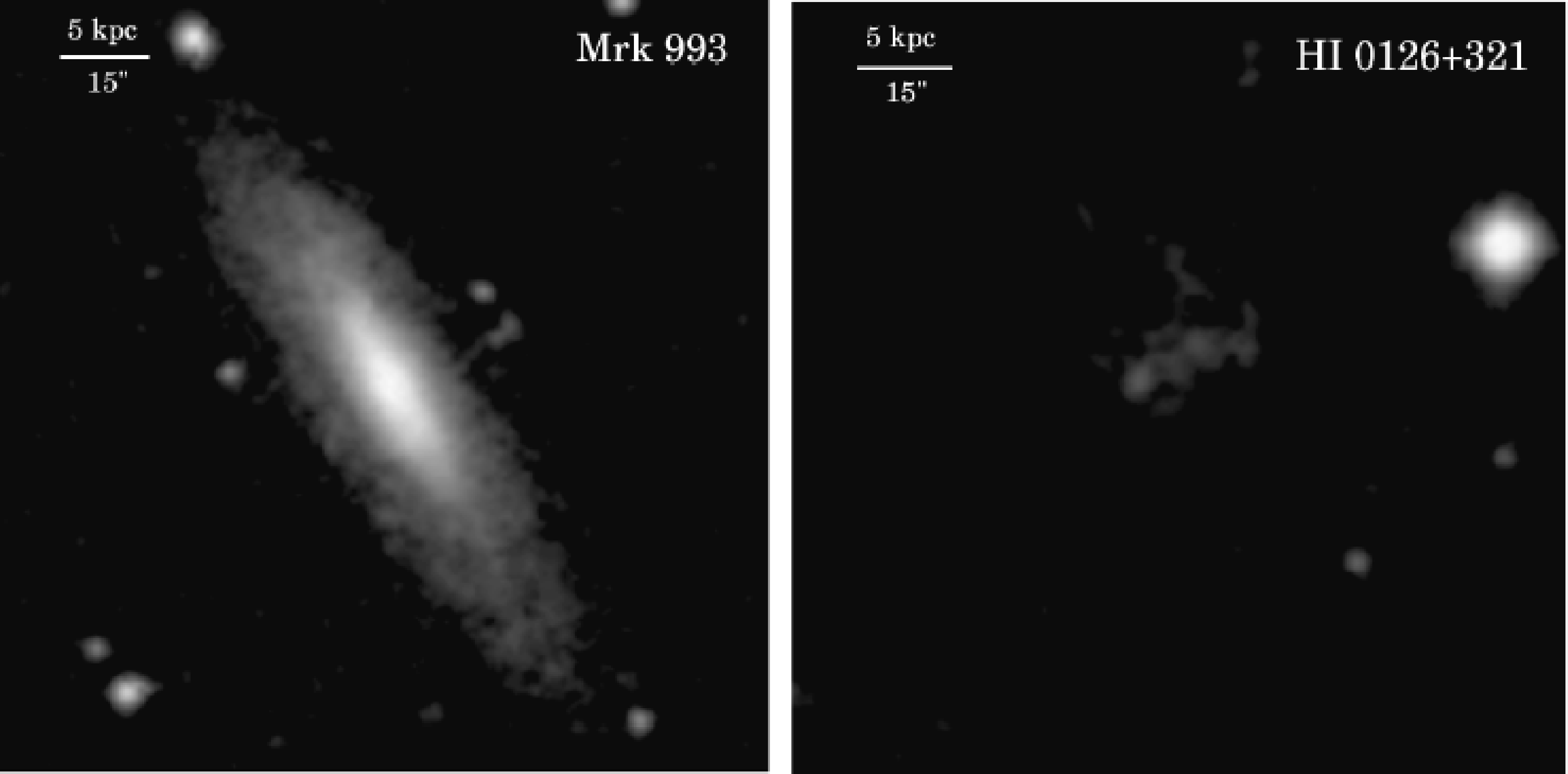}
\hspace*{-1.4 cm}
\includegraphics[angle=0, scale=0.482]{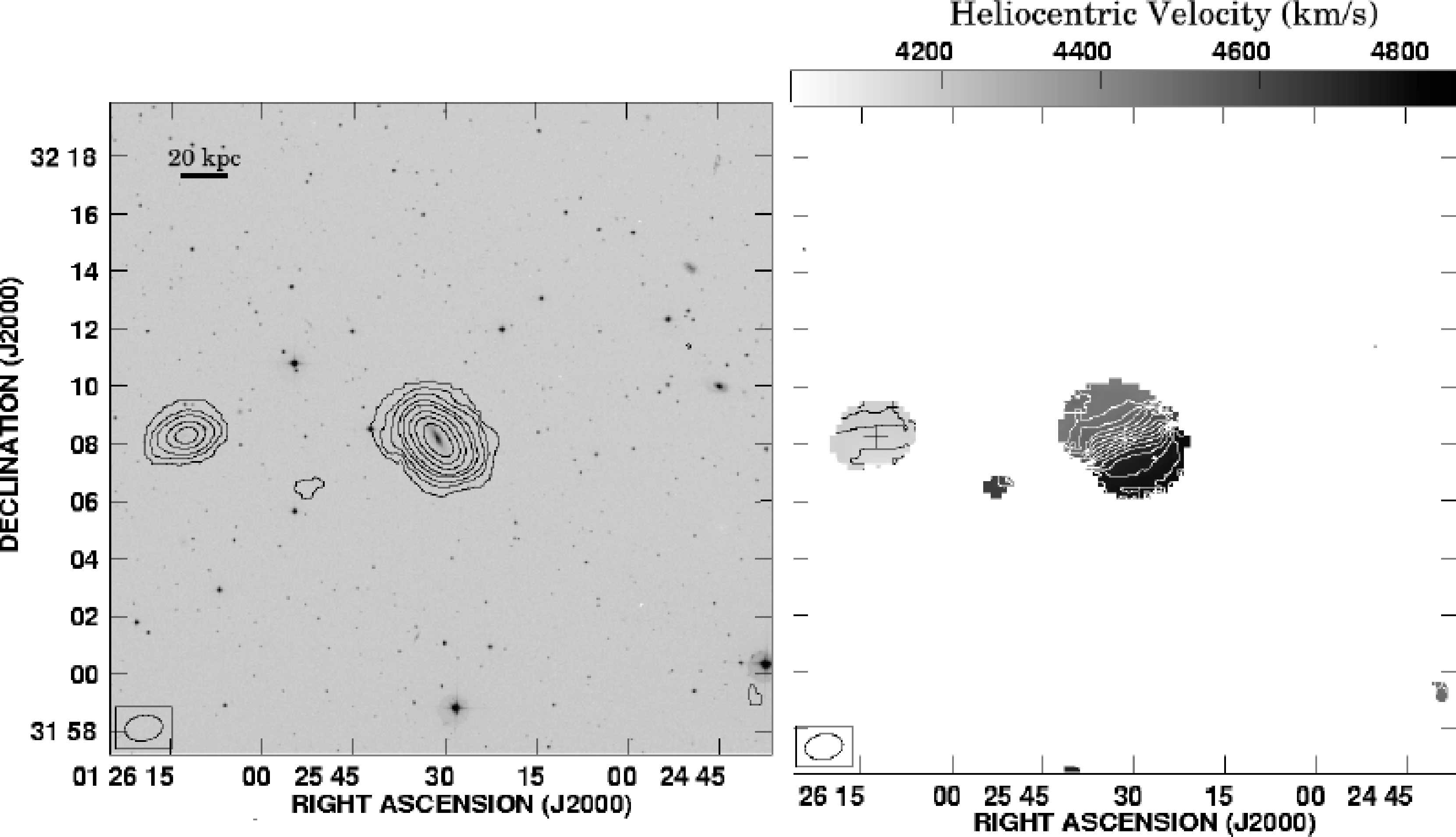} \\
\caption{Upper panel: Optical image of Markarian~993 (Mrk~993) (active galaxy) and HI~0126$+$321 from the DSS2.  Lower panels: (\textsl{Left}) Contours of integrated HI intensity (zeroth moment) overlaid on the DSS2 image, and (\textsl{Right}) map of intensity-weighted HI mean velocity (first moment). Mrk~993 lies at the center of the map, whereas HI~0126$+$321 is located east of center. In the zeroth moment map, contours are plotted at 1, 5, 10, 20, 30, 50, 70, $90 \times 20 {\rm \ mJy \ beam^{-1} \ km \ s^{-1}}$ ($3.92 \times 10^{18} {\rm \ cm^{-2}}$). In the first moment map, velocities are indicated by the scale wedge, and contours plotted at intervals of $25 {\rm \ km \ s^{-1}}$. The ellipse at the lower left corner of the lower panels indicates the half-power width of the synthesized beam, and has a size of $79\arcsec \times 53\arcsec$.}
\end{center}
\end{figure}

\clearpage

\begin{figure}
\begin{center}
\vspace*{-0.3 cm}
\hspace*{-2 cm}
\includegraphics[angle=0, scale=0.7]{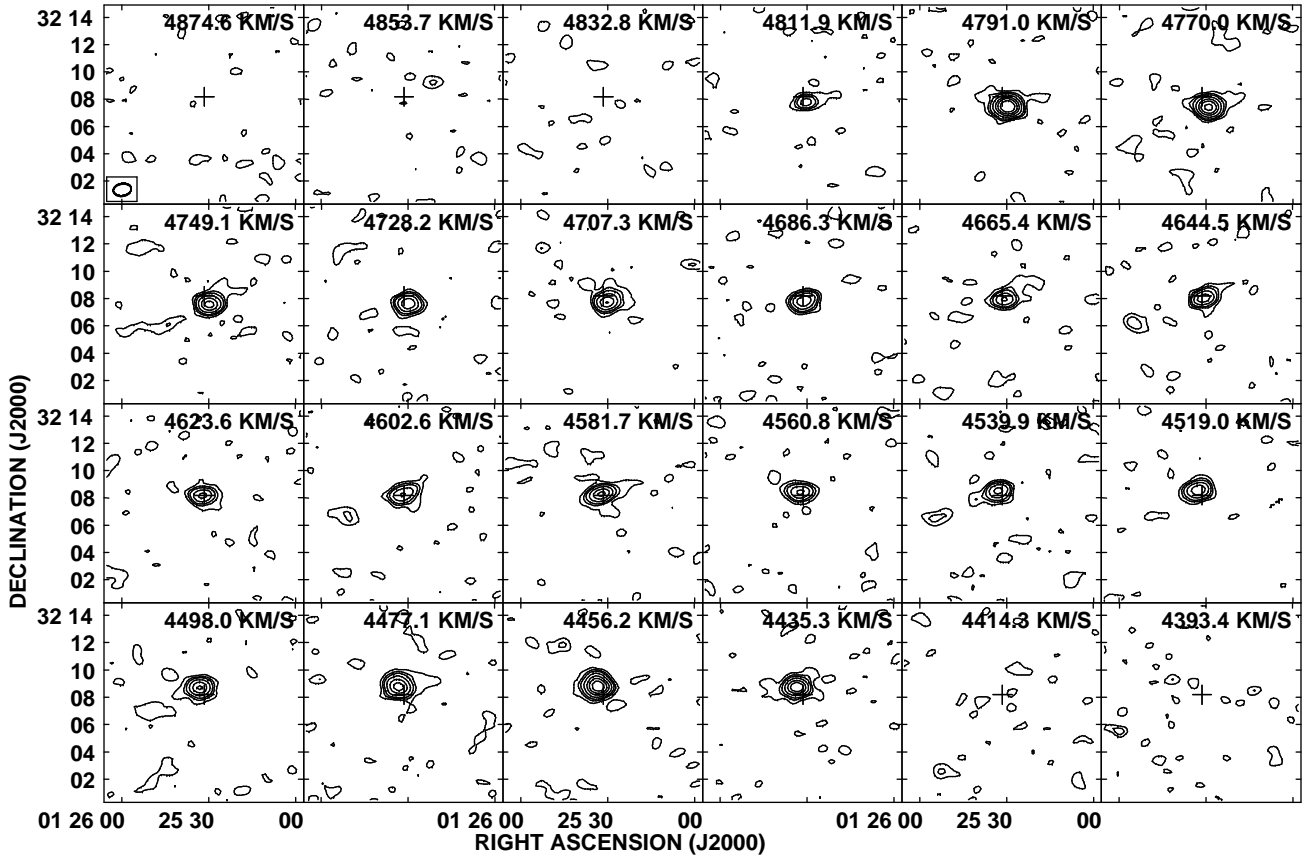}
\vspace*{0.0 cm}
\caption{HI channel maps of Mrk~993.  Contour levels are 
plotted at 2, 3, 4, 6, 10, 11, 20, $25 \times 0.45 {\rm \ mJy 
\ beam^{-1}}$ ($1 \sigma$), which corresponds to a HI column density of $1.82 \times 10^{18} {\rm \ cm^{-2}}$. The central heliocentric velocity is shown for each channel.  The cross marks the position of Mrk~993. The synthesized beam is shown by the ellipse at the lower left corner in the top left panel.}
\end{center}
\end{figure}

\begin{figure}
\begin{center}
\vspace*{-0.3 cm}
\hspace*{-2 cm}
\includegraphics[angle=0, scale=0.7]{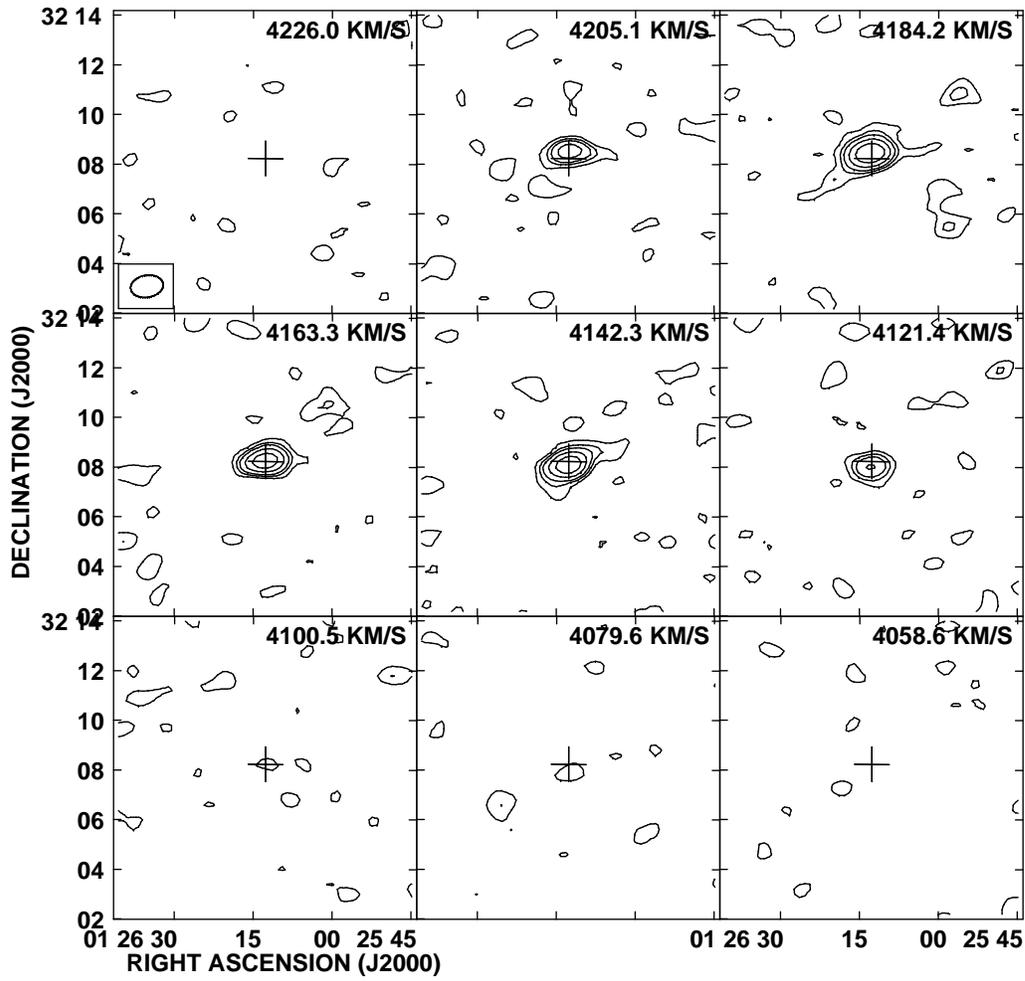}
\vspace*{0.0 cm}
\caption{HI channel maps of HI~0126$+$321. Contour levels are plotted at 2, 4, 6, 10, $16 \times 0.42 {\rm \ mJy \ beam^{-1}}$ ($1 \sigma$), which corresponds to a HI column density of $1.70 \times 10^{18} {\rm \ cm^{-2}}$.  The central heliocentric velocity is shown for each channel.  The cross marks the position of HI~0126$+$321.  The synthesized beam is shown by the ellipse at the lower left corner in the top left panel.}
\end{center}
\end{figure}

\clearpage

\begin{figure}
\begin{center}
\includegraphics[angle=0, scale=0.485]{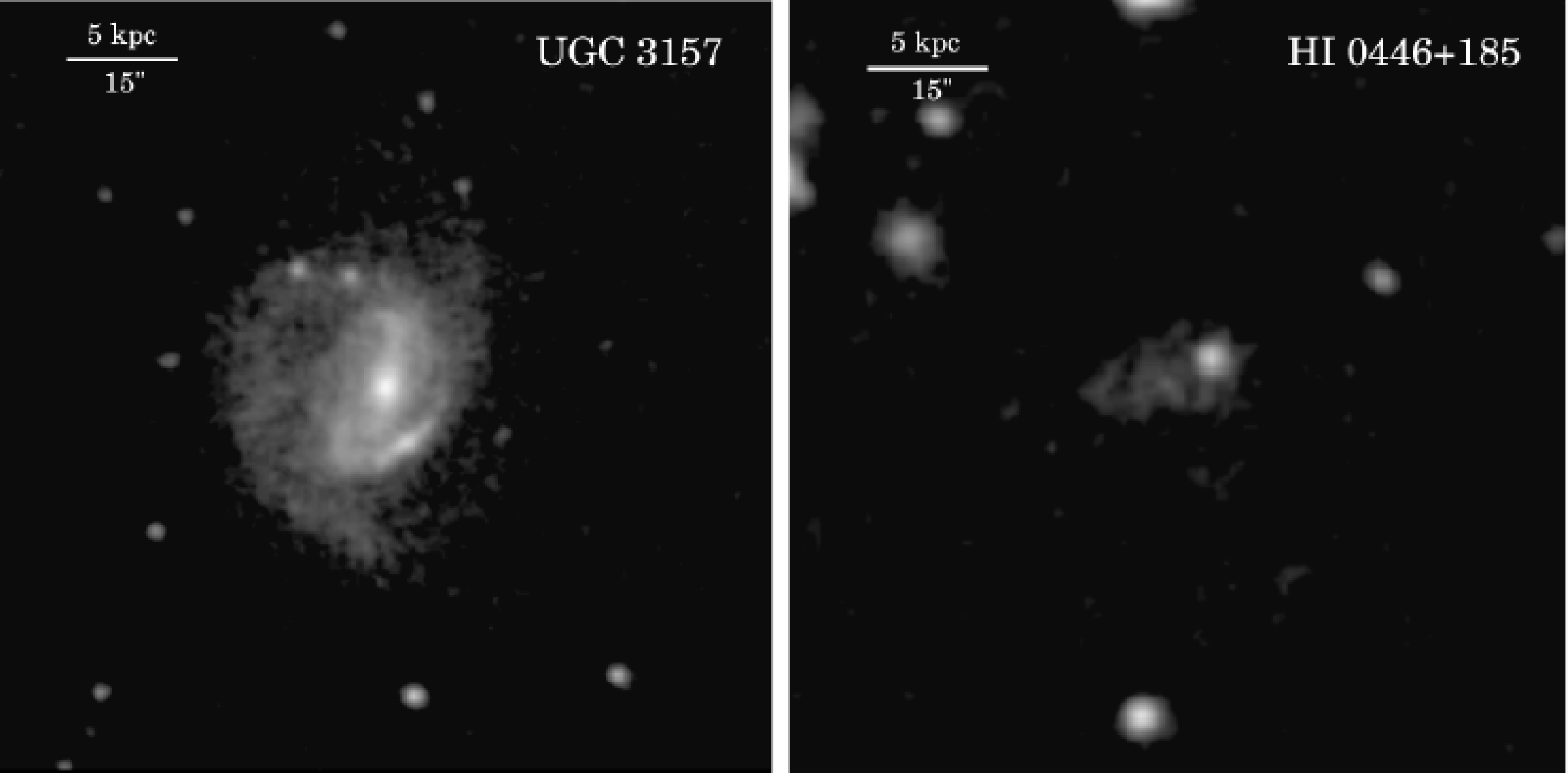}
\hspace*{-1.4 cm}
\includegraphics[angle=0, scale=0.482]{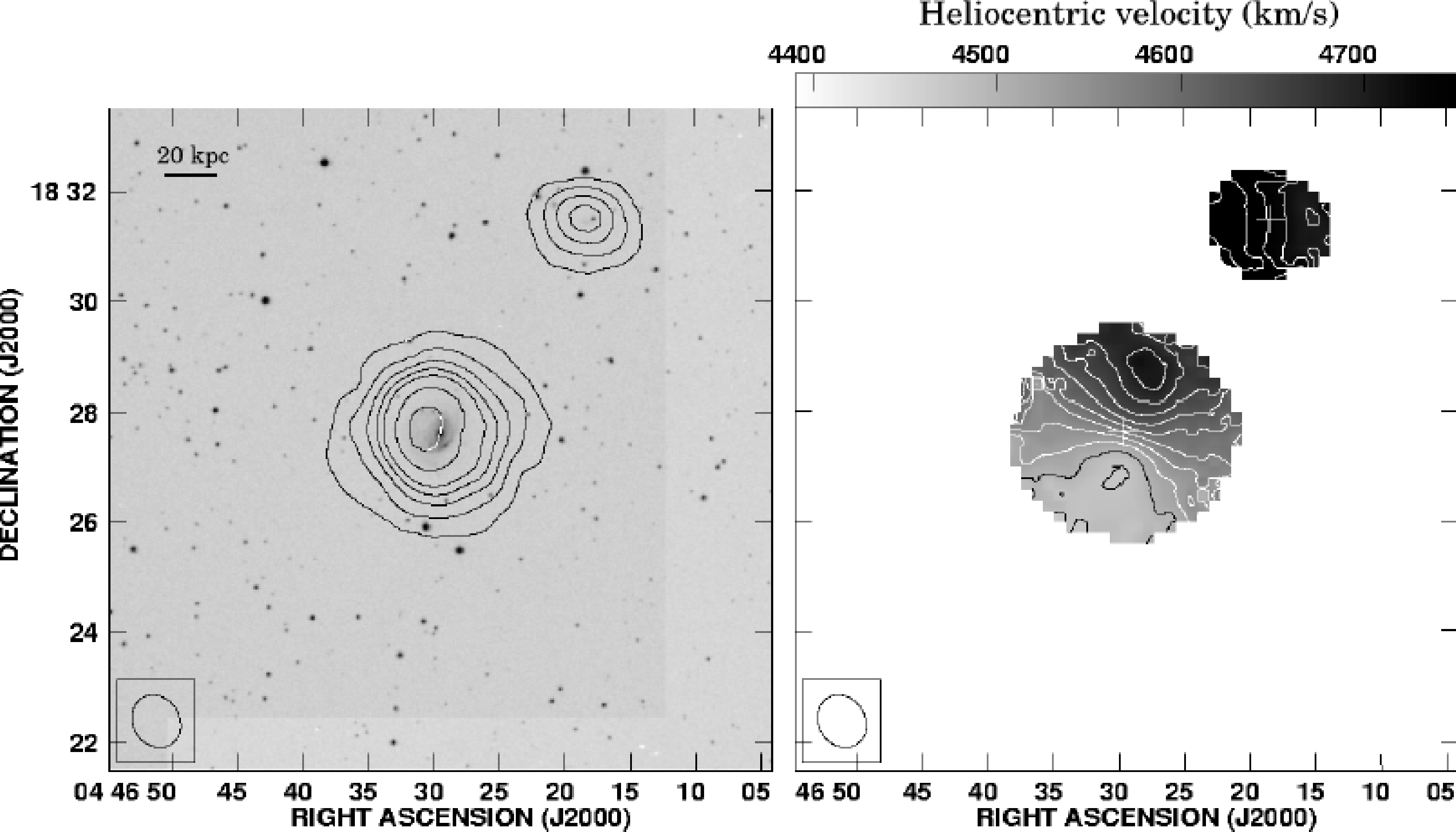} \\
\caption{
Upper panel: Optical image of Markarian~3157 (Mrk~3157) (active galaxy) and HI~0446$+$185  from the DSS2.  Lower panels: (\textsl{Left}) Contours of integrated HI intensity (zeroth moment) overlaid on the DSS2 image, and (\textsl{Right}) map of intensity-weighted HI mean velocity (first moment).  UGC~3157 lies at the
center of the map, whereas HI~0446$+$185 is located north-west of center. In the zeroth moment map, contours are plotted at 1, 5, 10, 15, 20, 30, $40 \times 30 {\rm \ mJy \ beam^{-1} \ km \ s^{-1}}$ ($1.7 \times 10^{19} {\rm \ cm^{-2}}$). In the first moment map, velocities are indicated by the scale wedge, and contours plotted at intervals of $25 {\rm \ km \ s^{-1}}$. The ellipse at the lower left corner of the lower panels indicates the half-power width of the synthesized beam, and has a size of $58\arcsec \times 50\arcsec$.}
\end{center}
\end{figure}

\begin{figure}
\begin{center}
\vspace*{-0.3 cm}
\hspace*{-3.0 cm}
\includegraphics[angle=0, scale=0.8]{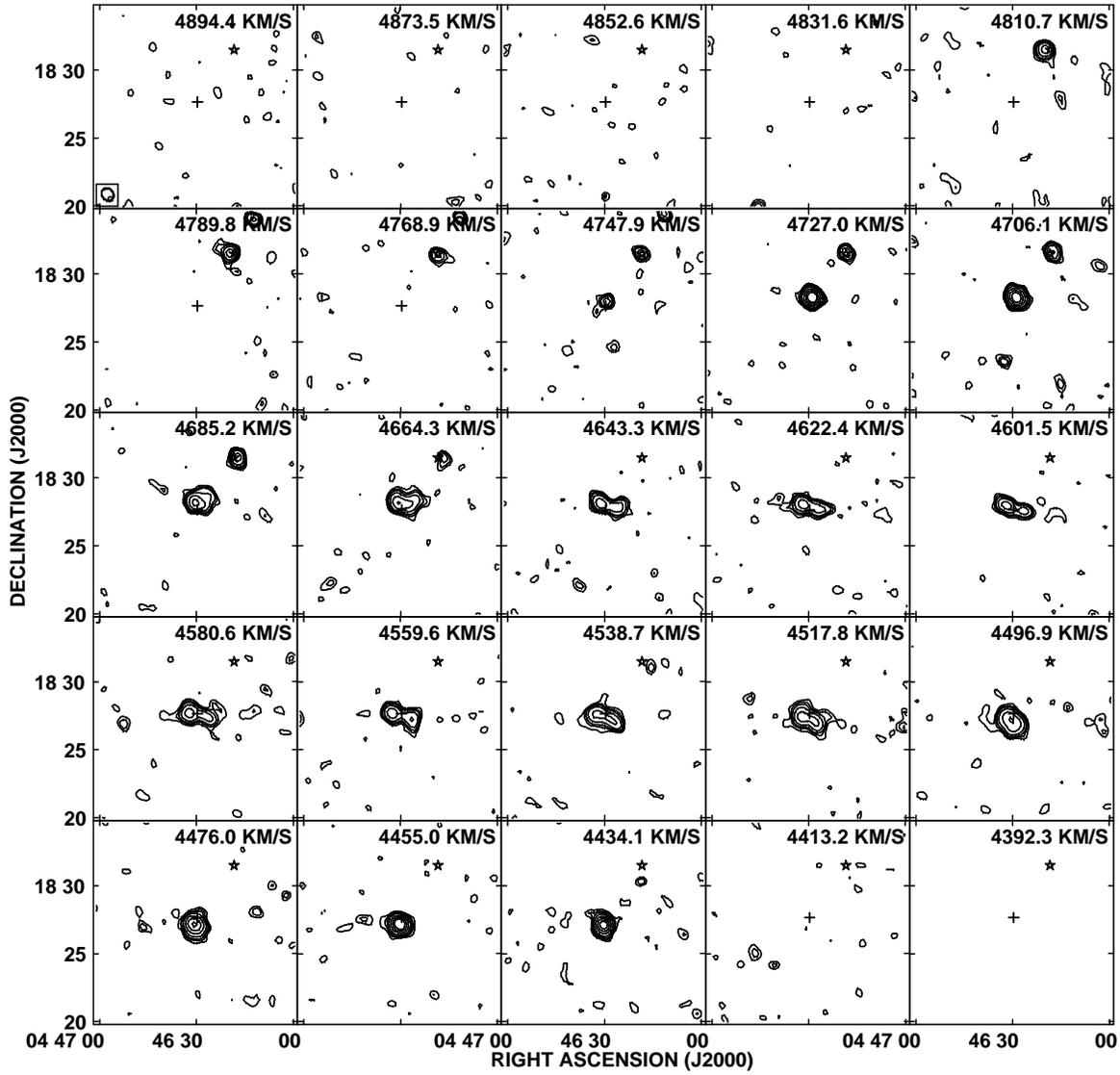}
\vspace*{0.0 cm}
\caption{HI channel maps of Mrk~3157 (active galaxy) and HI~0446$+$185. Contour levels are plotted at 2, 3, 4, 6, 10, 15, $20 \times 0.42 {\rm \ mJy \ beam^{-1}}$ ($1 \sigma$), which corresponds to a HI column density of $4.90 \times 10^{18} {\rm \ cm^{-2}}$. The central heliocentric velocity is shown for each channel. The cross marks the position of Mrk~3157, and star the position of HI~0446$+$185. The synthesized beam is shown by the ellipse at the lower left corner in the top left panel.}
\end{center}
\end{figure}

\clearpage
\begin{figure}
\begin{center}
\includegraphics[angle=0, scale=0.485]{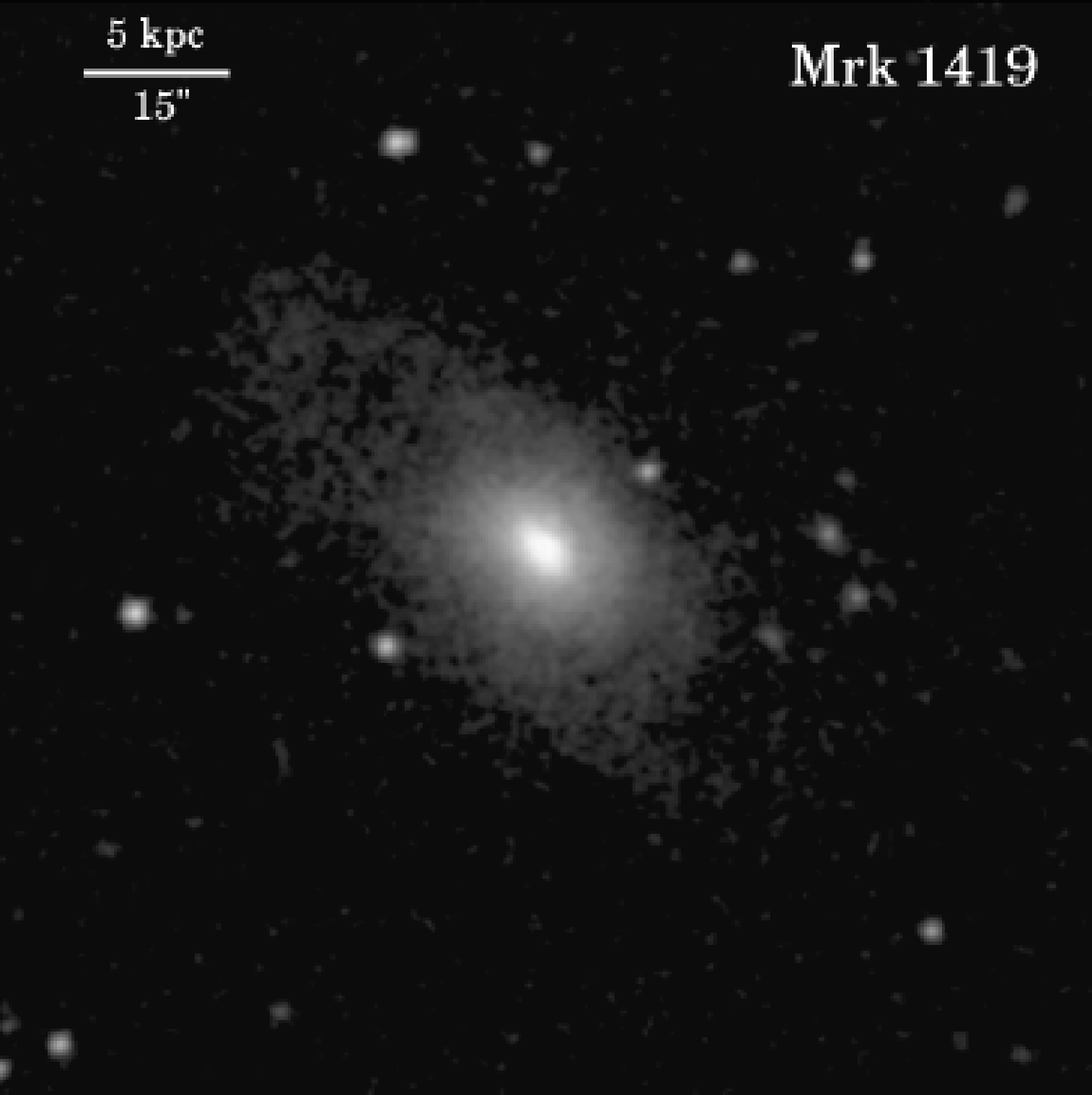}
\hspace*{-1.4 cm}
\includegraphics[angle=0, scale=0.482]{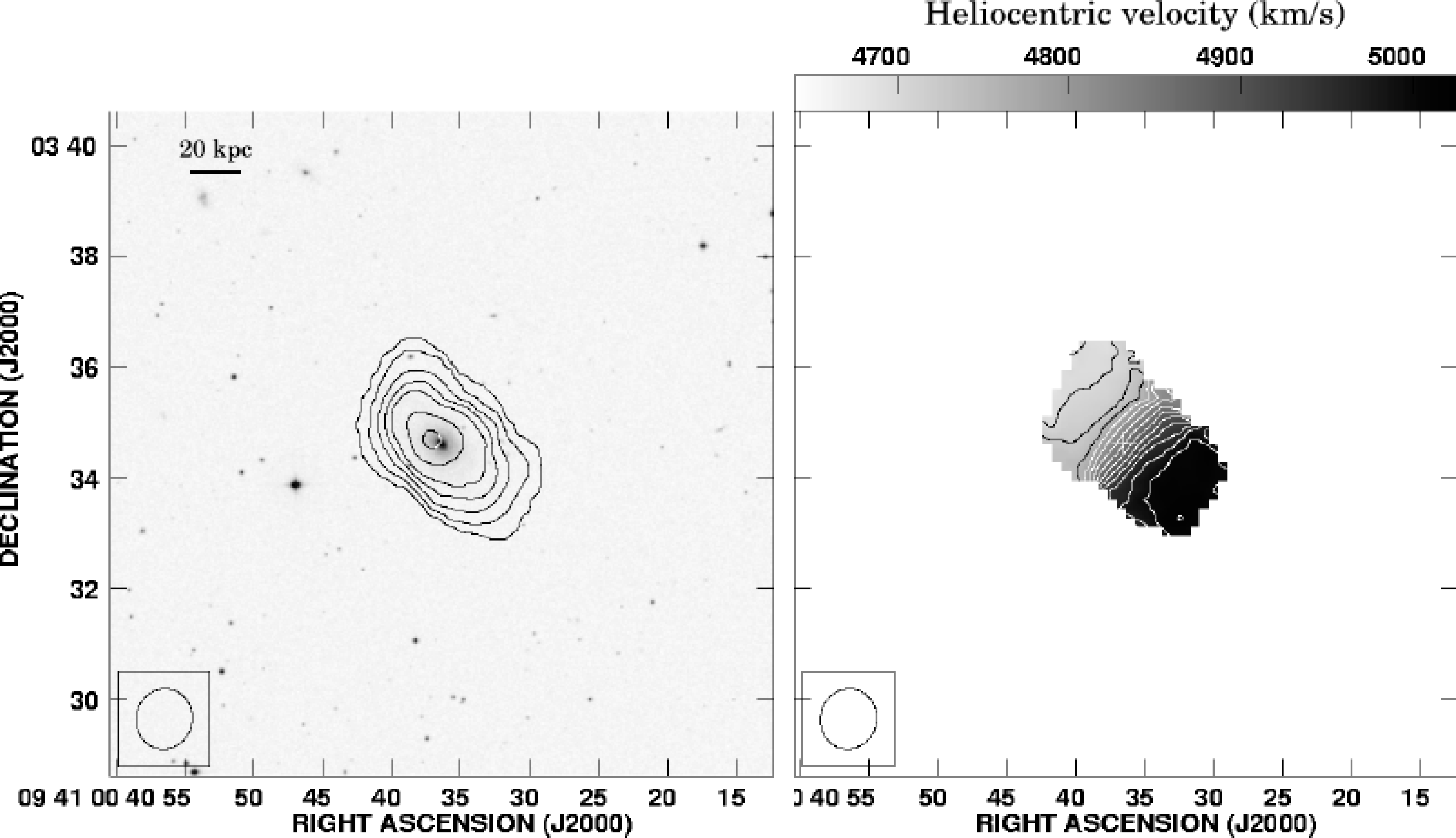} \\
\caption{
Upper panel: Optical image of Markarian~1419 (Mrk~1419) (active galaxy) from the DSS2.  Lower panels: (\textsl{Left}) Contours of integrated HI intensity (zeroth
moment) overlaid on the DSS2 image, and (\textsl{Right}) map of intensity-weighted HI mean velocity (first moment). In the zeroth moment map, contours are plotted at 1, 5, 10, 15, 20, 30, $40 \times 20 {\rm \ mJy \ beam^{-1} \ km \ s^{-1}}$ ($8.14 \times 10^{18} {\rm \ cm^{-2}}$).  In the first moment map, velocities are indicated by the scale wedge, and contours plotted at intervals of $25 {\rm \ km \ s^{-1}}$. The ellipse at the lower left corner of the lower panels indicates the half-power width of the synthesized beam, and has a size of $66\arcsec \times 61\arcsec$.}
\end{center}
\end{figure}

\begin{figure}
\begin{center}
\vspace*{-0.3 cm}
\hspace*{-1.8 cm}
\includegraphics[angle=0, scale=0.7]{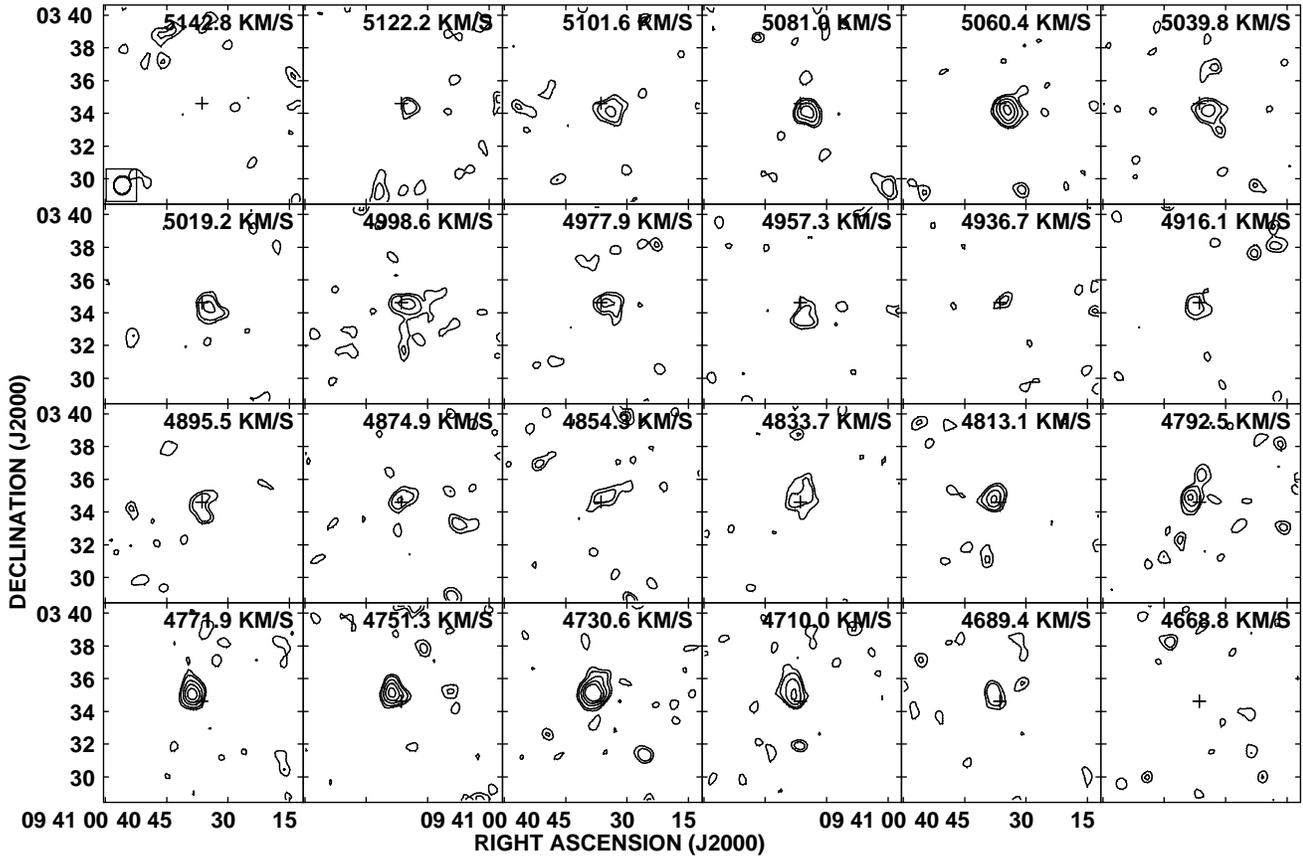}
\vspace*{0.0 cm}
\caption{
HI channel maps of Mrk~1419 (active galaxy). Contour levels are plotted at 2, 3, 5, 7, $9 \times 0.43 {\rm \ mJy \ beam^{-1}}$ ($1 \sigma$), which corresponds to a HI column density of $3.61 \times 10^{18} {\rm \ cm^{-2}}$. The central heliocentric velocity is shown for each channel. The cross marks the position of Mrk~1419. The synthesized beam is shown by the ellipse at the lower left corner in the top left panel.}
\end{center}
\end{figure}

\begin{figure}
\begin{center}
\includegraphics[angle=0, scale=0.485]{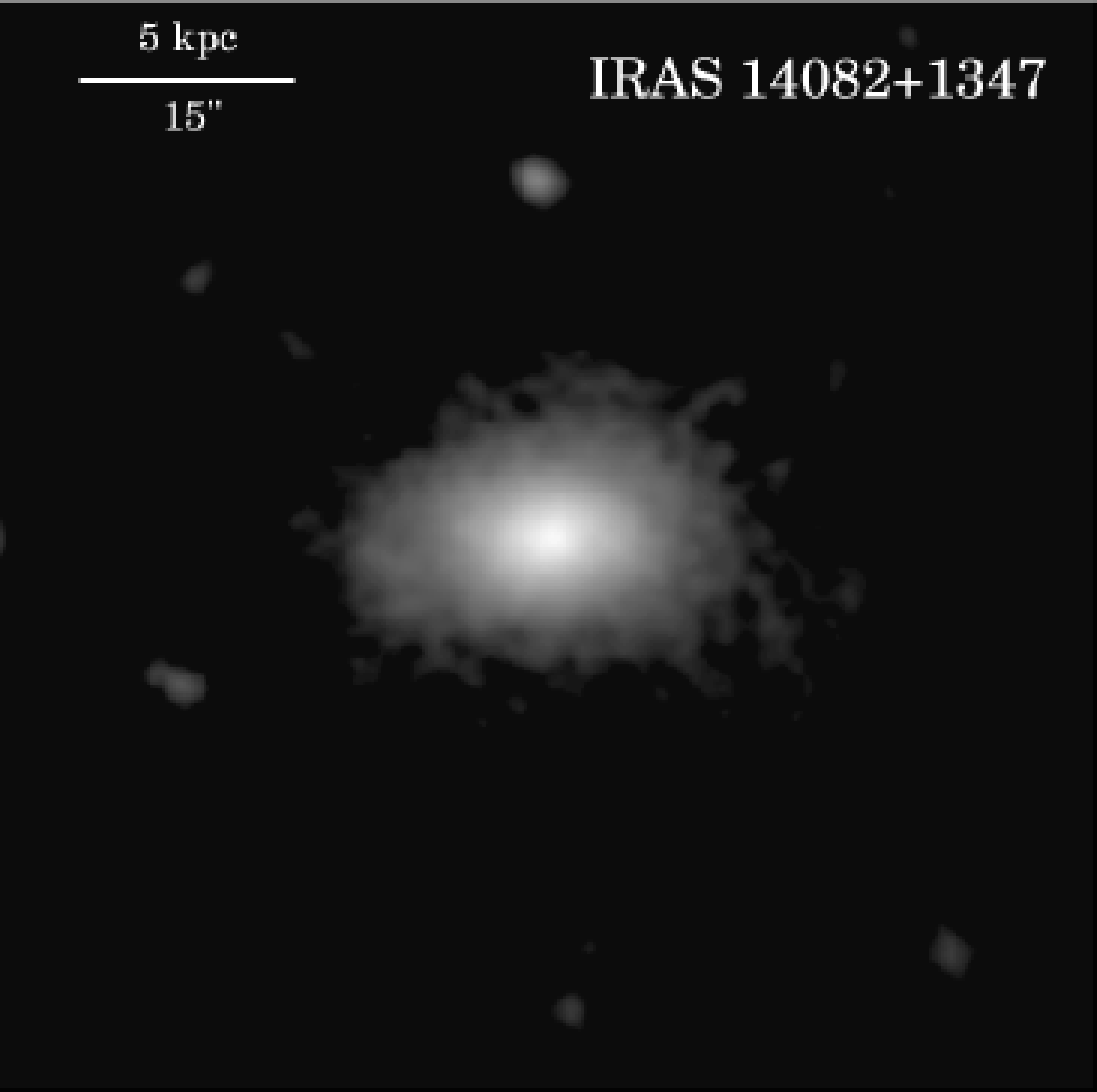}
\hspace*{-1.4 cm}
\includegraphics[angle=0, scale=0.482]{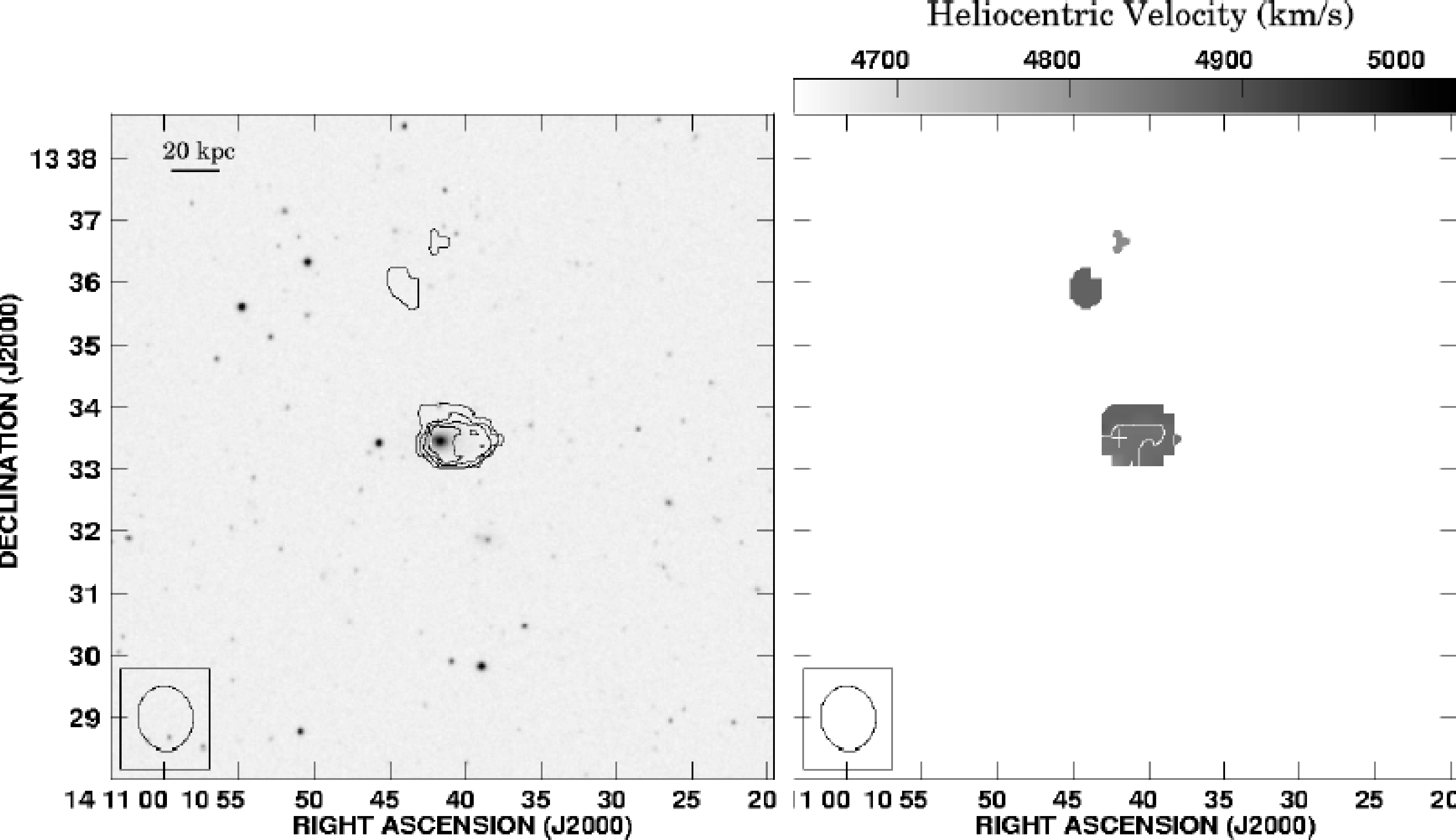} \\
\caption{
Upper panel: Optical image of IRAS~14082$+$1347 (active galaxy) from the DSS2. 
Lower panels: (\textsl{Left}) Contours of integrated HI intensity (zeroth
moment) overlaid on the DSS2 image, and (\textsl{Right}) map of intensity-weighted HI mean velocity (first moment). In the zeroth moment map, contours are plotted at 1, 3, 5, $7 \times 10 {\rm \ mJy \ beam^{-1} \ km \ s^{-1}}$ ($5.08 \times 10^{18} {\rm \ cm^{-2}}$).  In the first moment map, velocities are indicated by the scale wedge, and contours plotted at intervals of $25 {\rm \ km \ s^{-1}}$. The ellipse at the lower left corner of the lower panels indicates the half-power width of the synthesized beam, and has a size of $61\arcsec \times 53\arcsec$.}
\end{center}
\end{figure}

\begin{figure}
\begin{center}
\vspace*{-0.3 cm}
\hspace*{-2 cm}
\includegraphics[angle=0, scale=0.7]{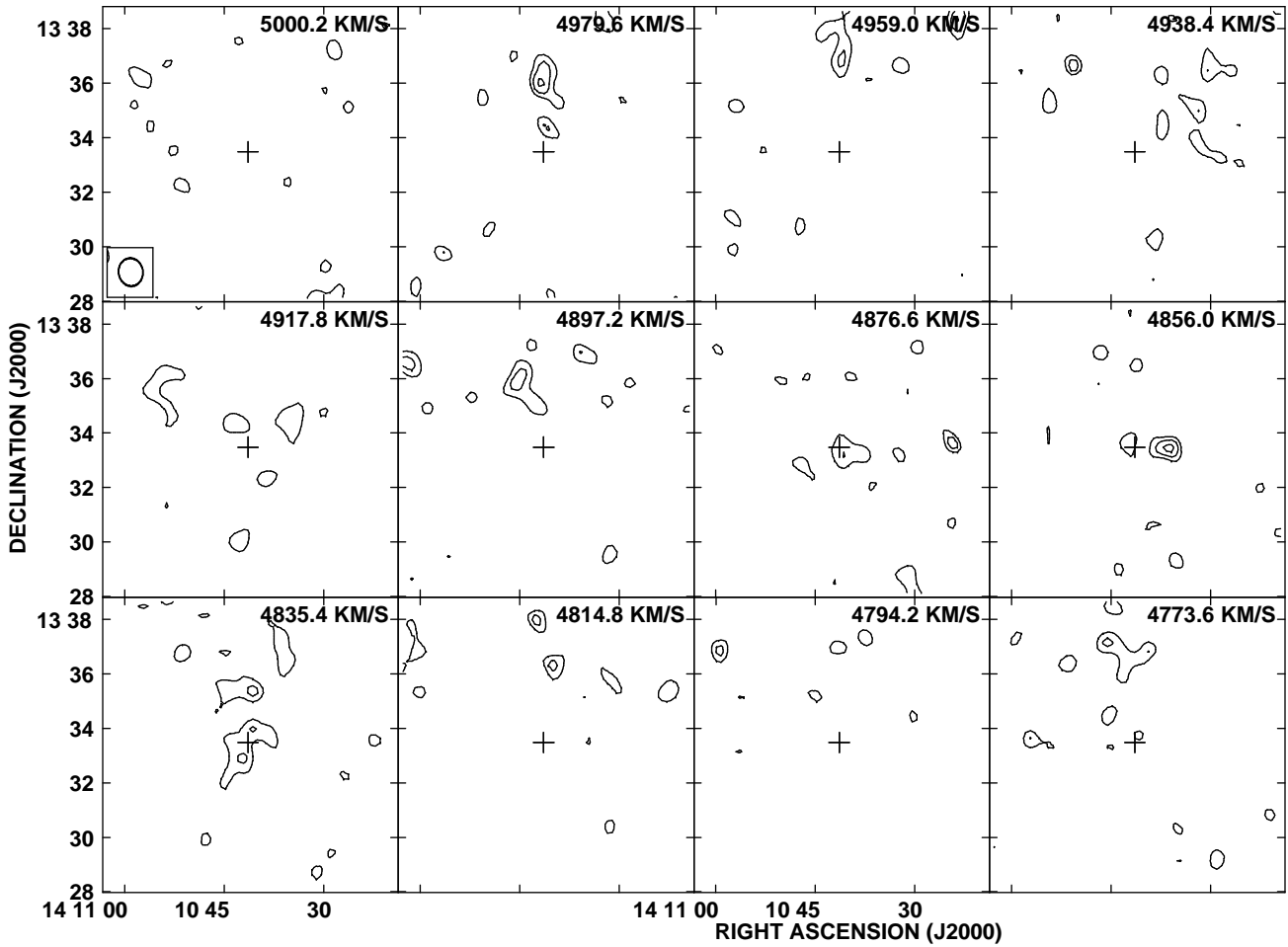}
\vspace*{0.0 cm}
\caption{
HI channel maps of IRAS~14082$+$1347. Contour levels are plotted at 2, 3, $4 \times 0.43 {\rm \ mJy \ beam^{-1}}$ ($1 \sigma$), which corresponds to a HI column density of $4.50 \times 10^{18} {\rm \ cm^{-2}}$. The central heliocentric velocity is shown for each channel. The cross marks the position of IRAS~14082$+$1347. The synthesized beam is shown by the ellipse at the lower left corner in the top left panel.}
\end{center}
\end{figure}

\begin{figure}
\begin{center}
\includegraphics[angle=0, scale=0.485]{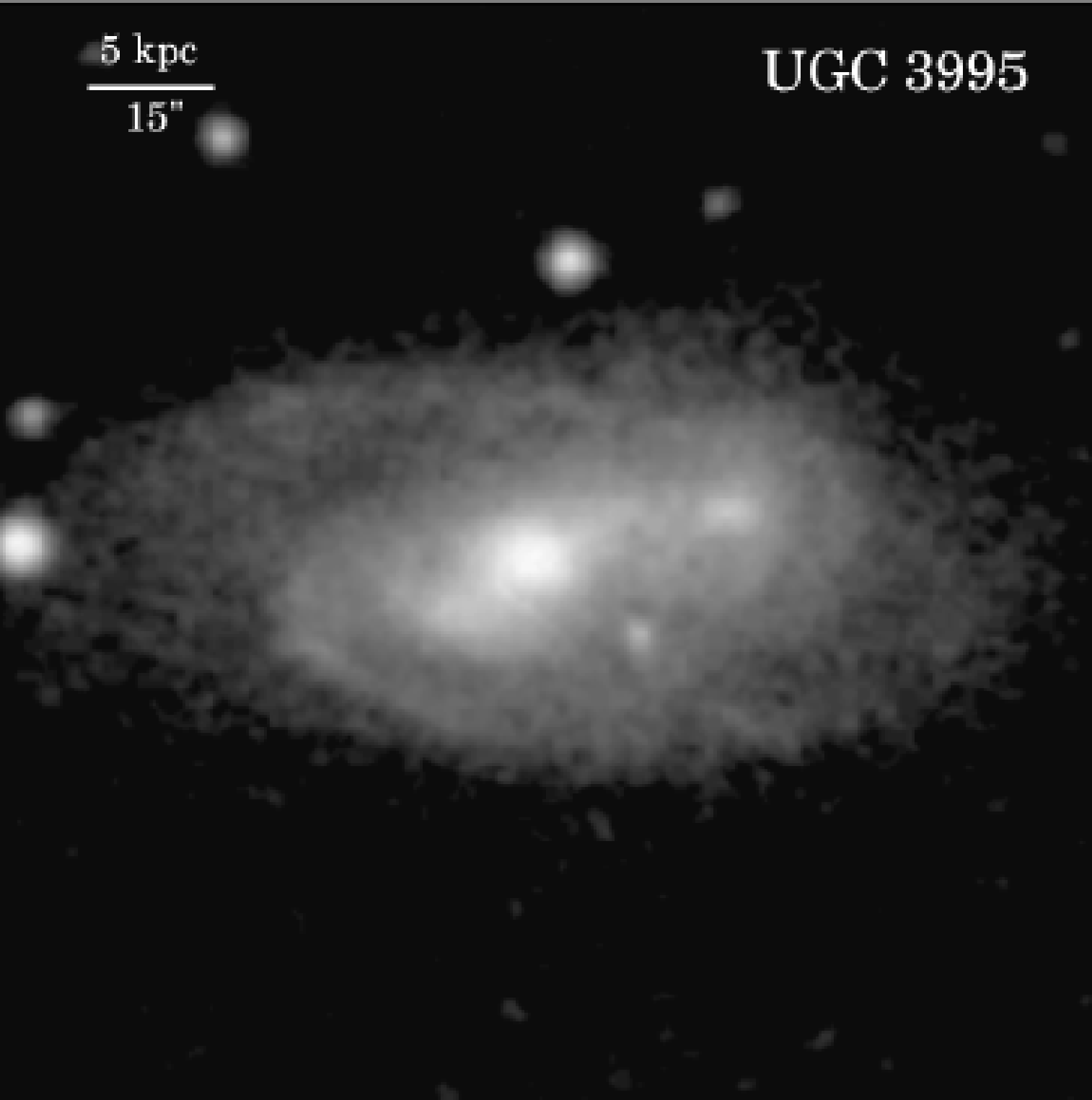}
\hspace*{-1.4 cm}
\includegraphics[angle=0, scale=0.482]{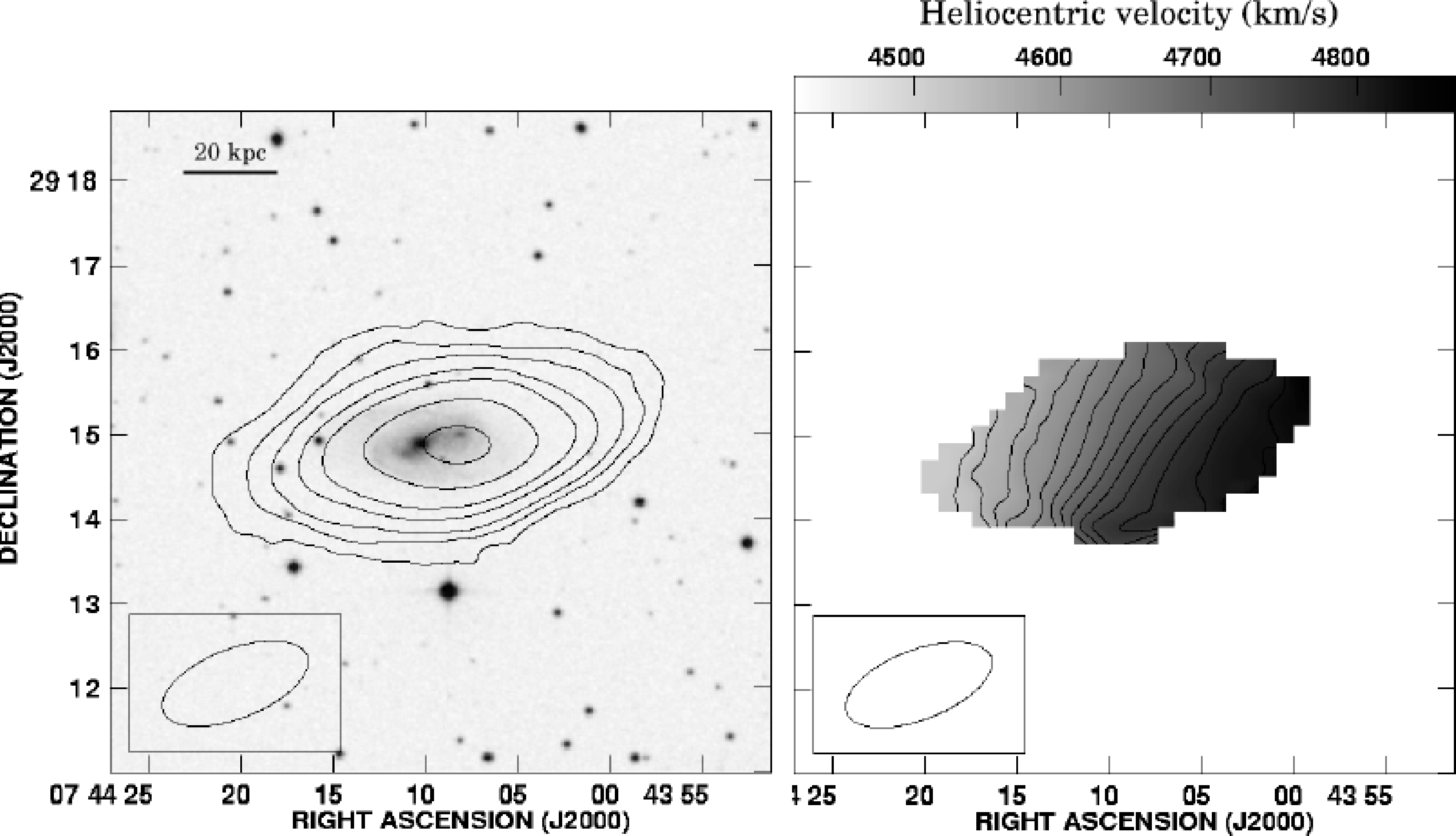} \\
\caption{
Upper panel: Optical image of UGC~3995 (active galaxy) and UGC 3995A from the DSS2.  Lower panels: (\textsl{Left}) Contours of integrated HI intensity (zeroth
moment) overlaid on the DSS2 image, and (\textsl{Right}) map of intensity-weighted HI mean velocity (first moment).  UGC~3995 lies to the east, whereas UGC~3995A to the west of the center. In the zeroth moment map, contours are plotted at 1, 5, 10, 20, 30, 50, $70 \times 40 {\rm \ mJy \ beam^{-1} \ km \ s^{-1}}$ ($1.22 \times 10^{19} {\rm \ cm^{-2}}$). In the first moment map, velocities are indicated by the scale wedge, and contours plotted at intervals of $25 {\rm \ km \ s^{-1}}$. The ellipse at the lower left corner of the lower panels indicates the half-power width of the synthesized beam, and has a size of $110\arcsec \times 49\arcsec$.}
\end{center}
\end{figure}

\begin{figure}
\begin{center}
\vspace*{-0.3 cm}
\hspace*{-3.0 cm}
\includegraphics[angle=0, scale=0.8]{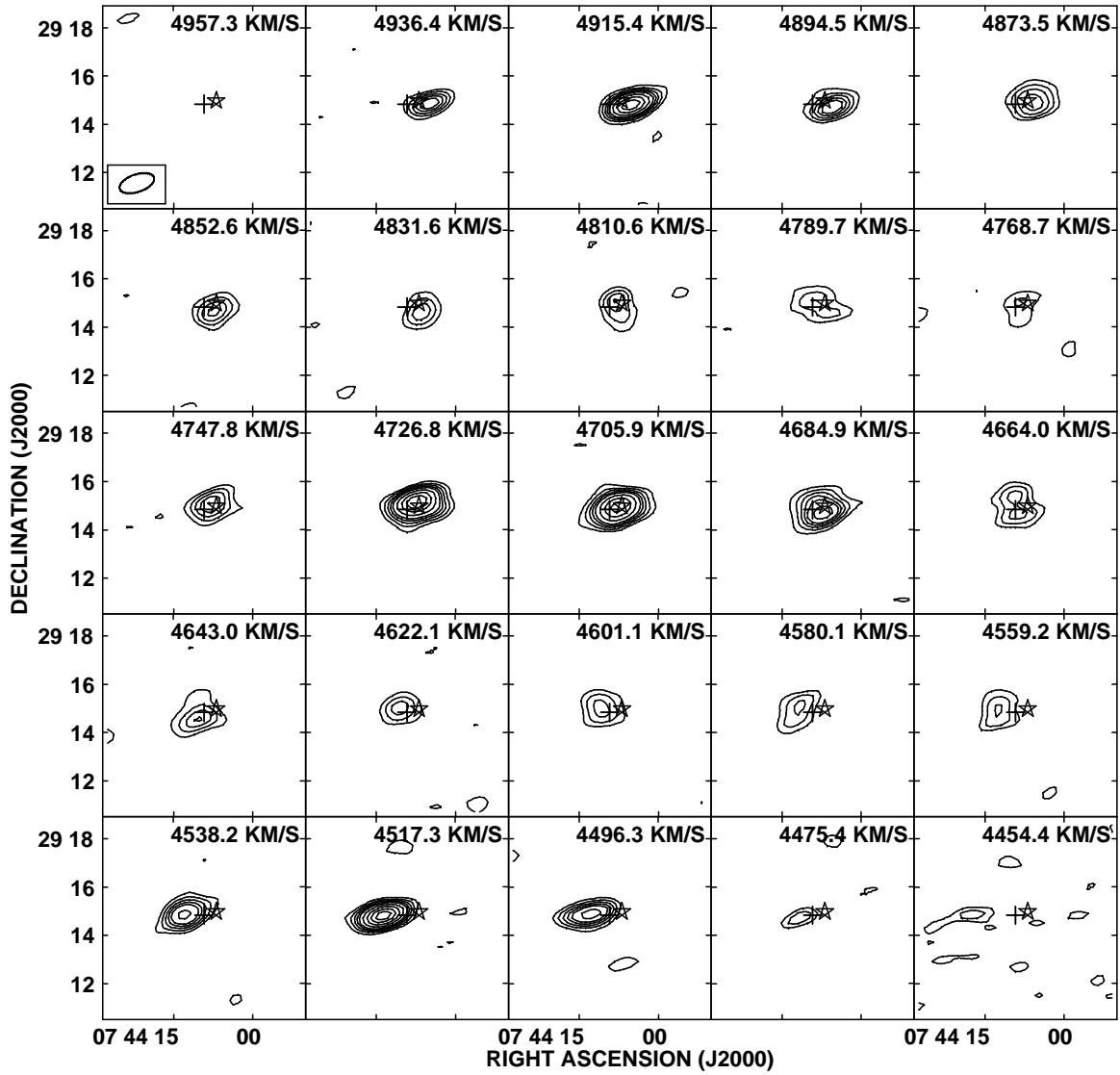}
\vspace*{0.0 cm}
\caption{
HI channel maps of UGC~3995 and UGC~3995A.  Contour levels are 
plotted at 3, 5, 7, 9, 11, 15, 20, 25, 30 $\times 0.62 {\rm \ mJy 
\ beam^{-1}}$ ($1 \sigma$), which corresponds to a HI column density of 3.90 $\times 10^{18} {\rm \ cm^{-2}}$. The central heliocentric velocity is shown for each channel. The cross marks the position of UGC~3995, and star the position of UGC~3995A. The synthesized beam is shown by the ellipse at the lower left corner in the top left panel.}
\end{center}
\end{figure}


\begin{figure}
\begin{center}
\includegraphics[angle=0, scale=0.485]{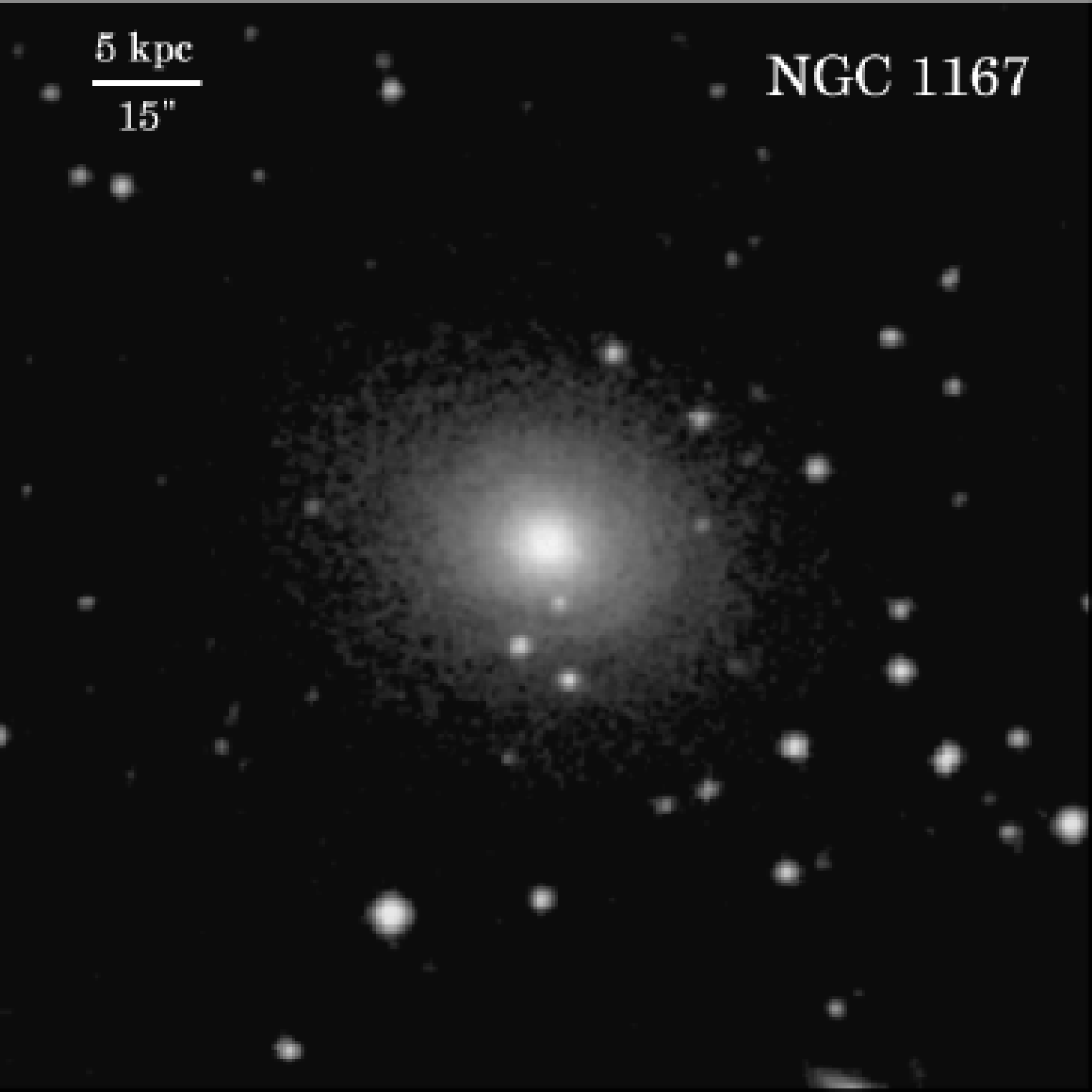}
\hspace*{-1.4 cm}
\includegraphics[angle=0, scale=0.482]{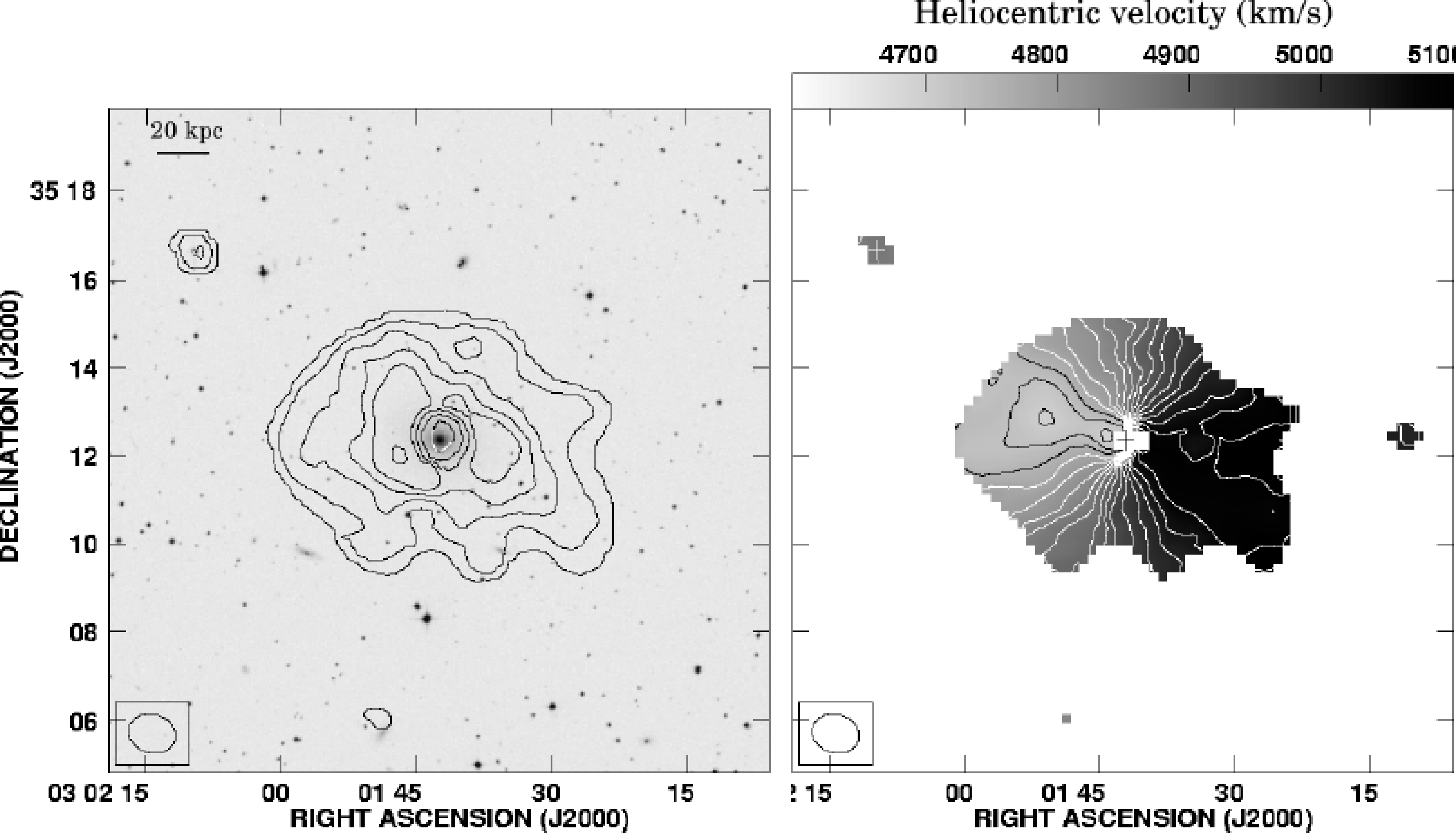} \\
\caption{
Upper panel: Optical image of NGC~1167 (active galaxy) from the DSS2.  Lower panels: (\textsl{Left}) Contours of integrated HI intensity (zeroth moment) overlaid on the DSS2 image, and (\textsl{Right}) map of intensity-weighted HI mean velocity (first moment).  NGC~1167 lies at the center of the map, whereas HI~0302$+$352 is located to the north-east. In the zeroth moment map, contours are plotted at 1, 5, 10, 15, 20, 30, 40, 50, $60 \times 30 {\rm \ mJy \ beam^{-1} \ km \ s^{-1}}$ ($1.45 \times 10^{19} {\rm \ cm^{-2}}$). In the first moment map, velocities are indicated by the scale wedge, and contours plotted at intervals of $25 {\rm \ km \ s^{-1}}$. The ellipse at the lower left corner of the lower panels indicates the half-power width of the synthesized beam, and has a size of $65\arcsec \times 52\arcsec$.}
\end{center}
\end{figure}

\begin{figure}
\begin{center}
\vspace*{-0.3 cm}
\hspace*{-0.5 cm}
\hspace*{-2.8 cm}
\includegraphics[angle=0, scale=0.80]{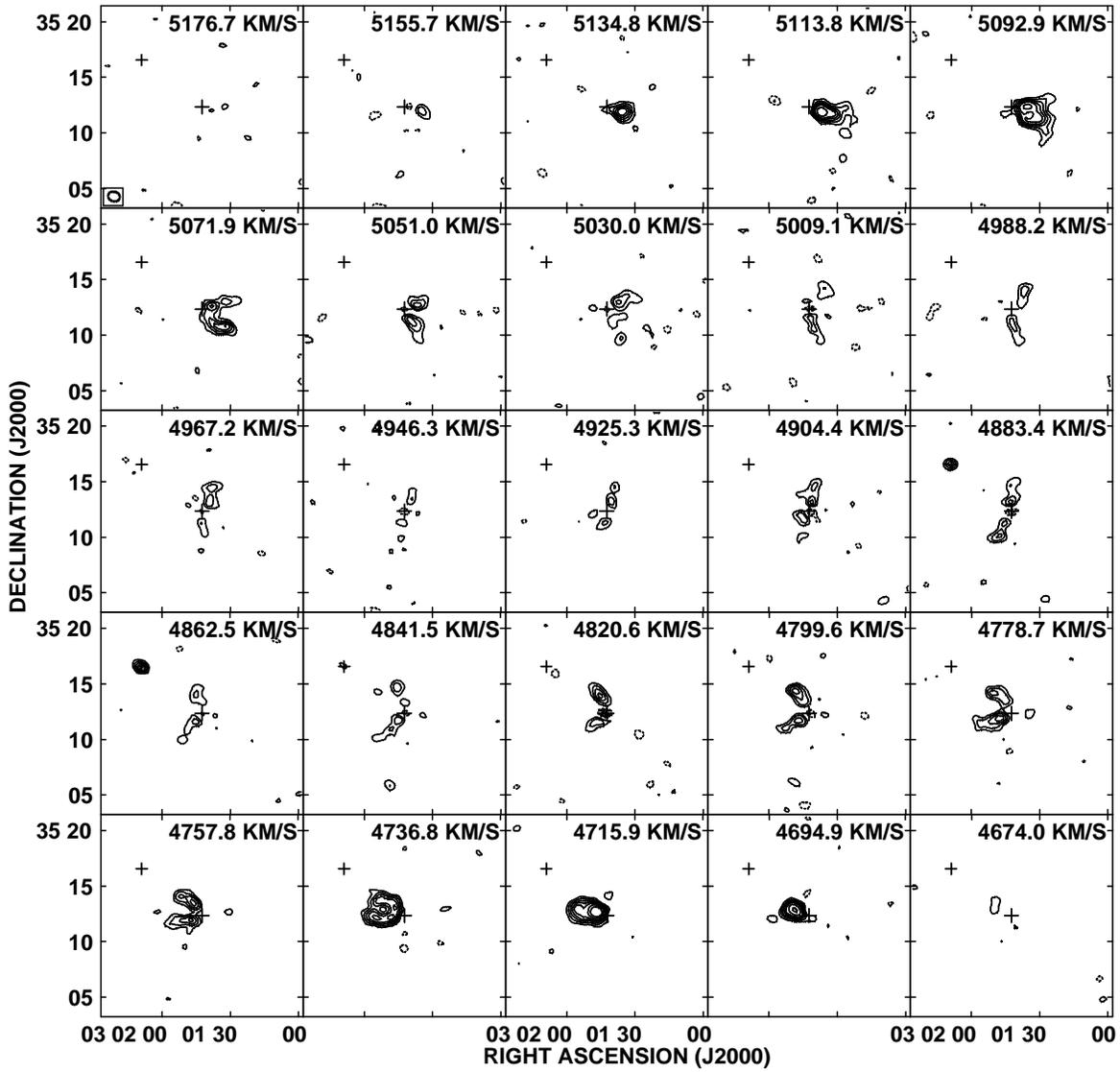}
\vspace*{0.0 cm}
\caption{
HI channel maps of NGC~1167 and HI~0302$+$352. Contour levels are 
plotted at $-$9, $-$7, $-$5, $-$3, 3, 5, 7, 9, 11 15, 20, $30 \times 0.6 {\rm \ mJy \ beam^{-1}}$ ($1 \sigma$), which corresponds to a HI column density of $5.97 \times 10^{18} {\rm \ cm^{-2}}$. The central heliocentric velocity is shown for each channel. The cross marks the position of NGC~1167, and star the position of HI~0302$+$352. The synthesized beam is shown by the ellipse at the lower left corner in the top left panel.}
\end{center}
\end{figure}

\begin{figure}
\begin{center}
\includegraphics[angle=0, scale=0.482]{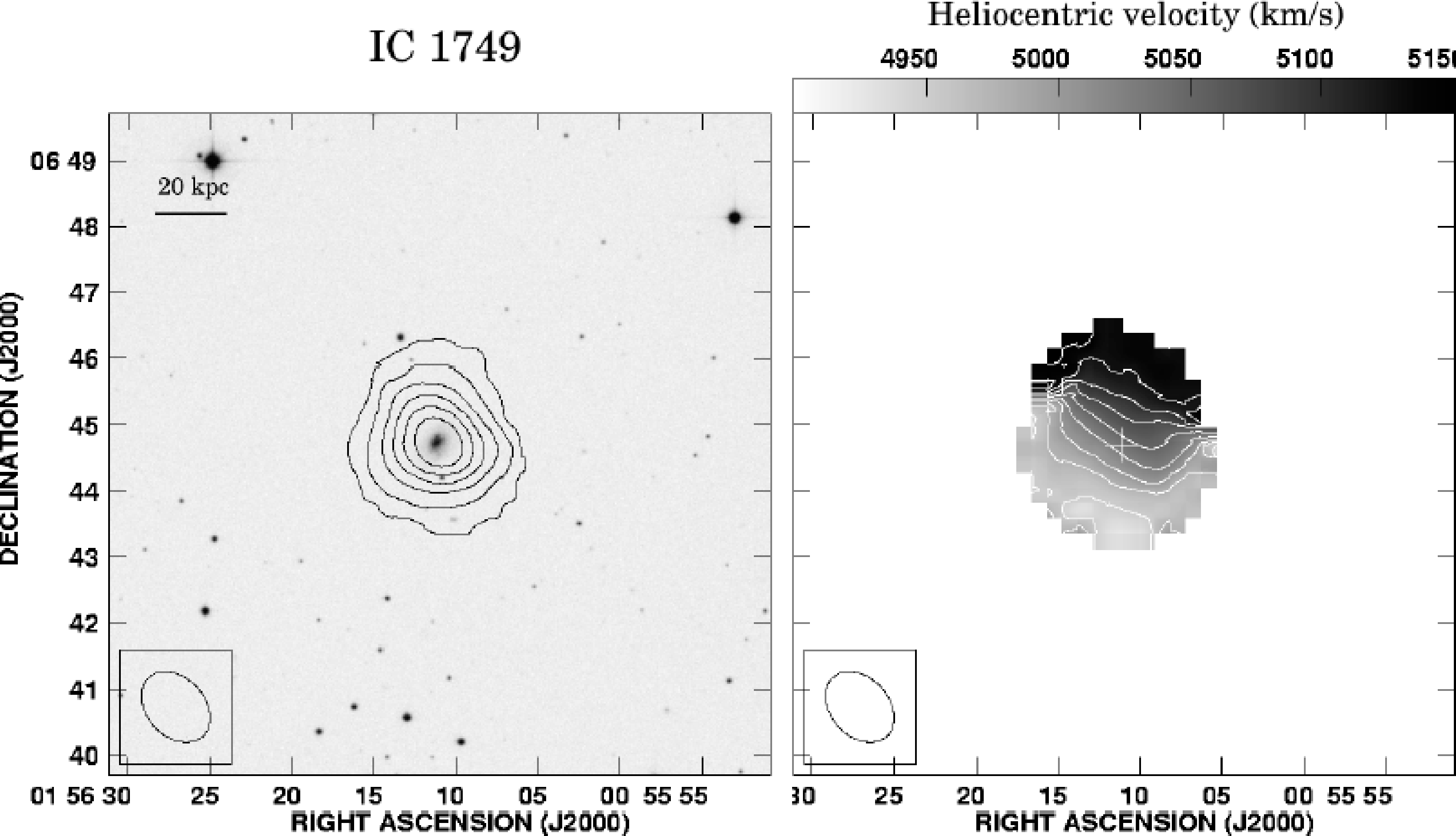} \\
\caption{Moment maps of IC~1749, detected in the same field as the target object UGC 1395. (\textsl{Left}) Contours of integrated HI intensity (zeroth moment) overlaid on the DSS2 image, and (\textsl{Right}) map of intensity-weighted HI mean velocity (first moment). In the zeroth moment map, contours are plotted at 1, 5, 10, 15, 20, $25 \times 35 {\rm \ mJy \ beam^{-1} \ km \ s^{-1}}$ ($1.49 \times 10^{19} {\rm \ cm^{-2}}$). In the first moment map, velocities are indicated by the scale wedge, and contours plotted at intervals of $25 {\rm \ km \ s^{-1}}$. The ellipse at the lower left corner of each panel indicates the half-power width of the synthesized beam, and has a size of $73\arcsec \times 53\arcsec$.}
\end{center}
\end{figure}

\begin{figure}
\begin{center}
\includegraphics[angle=0, scale=0.482]{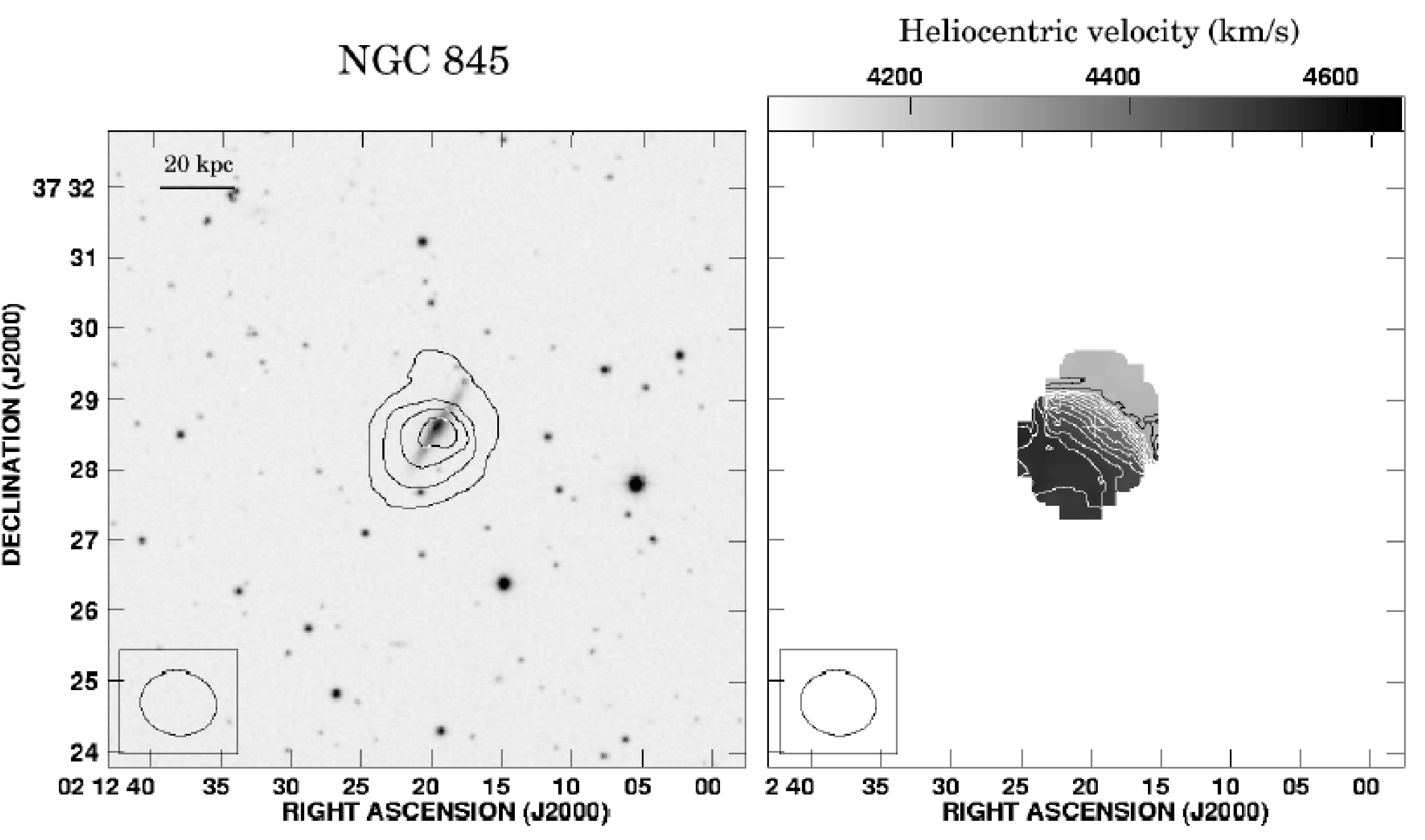} \\
\caption{Moment maps of NGC~845, detected in the same field as the target object NGC~841. (\textsl{Left})  Contours of integrated HI intensity (zeroth moment) overlaid on the DSS2 image, and (\textsl{Right}) map of intensity-weighted HI mean velocity (first moment).  In the zeroth moment map, contours are plotted at 1, 5, 10, $15 \times 30 {\rm \ mJy \ beam^{-1} \ km \ s^{-1}}$ ($1.39 \times 10^{19} {\rm \ cm^{-2}}$). In the first moment map, velocities are indicated by the scale wedge, and contours plotted at intervals of $25 {\rm \ km \ s^{-1}}$. The ellipse at the lower left corner of each panel indicates the half-power width of the synthesized beam, and has a size of $65\arcsec \times 54\arcsec$.}
\end{center}
\end{figure}

\begin{figure}
\begin{center}
\includegraphics[angle=0, scale=0.58]{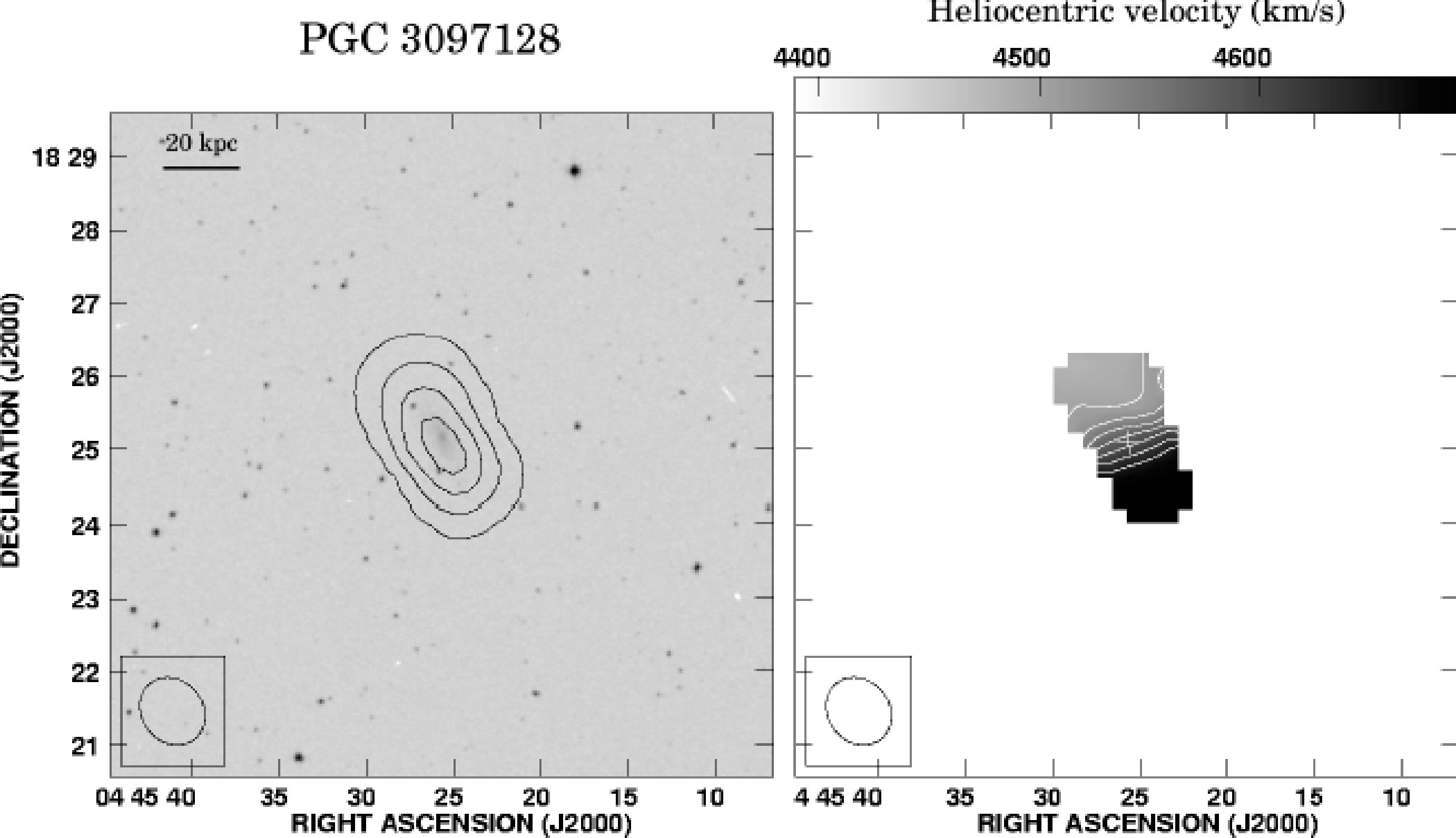} \\
\caption{Moment maps of PGC~3097128, detected in the same field as the target object UGC~3157. (\textsl{Left}) Contours of integrated HI intensity (zeroth moment) overlaid on the DSS2 image, and (\textsl{Right}) map of intensity-weighted HI mean velocity (first moment). In the zeroth moment map, contours are plotted at  1, 3, 5, 10, 15, $20 \times 35 {\rm \ mJy \ beam^{-1} \ km \ s^{-1}}$ ($1.83 \times 10^{19} {\rm \ cm^{-2}}$). In the first moment map, velocities are indicated by the scale wedge, and contours plotted at intervals of $25 {\rm \ km \ s^{-1}}$. The ellipse at the lower left corner of each panel indicates the half-power width of the synthesized beam, and has a size of $58\arcsec \times 54\arcsec$.}
\end{center}
\end{figure}

\begin{figure}
\begin{center}
\includegraphics[angle=0, scale=0.482]{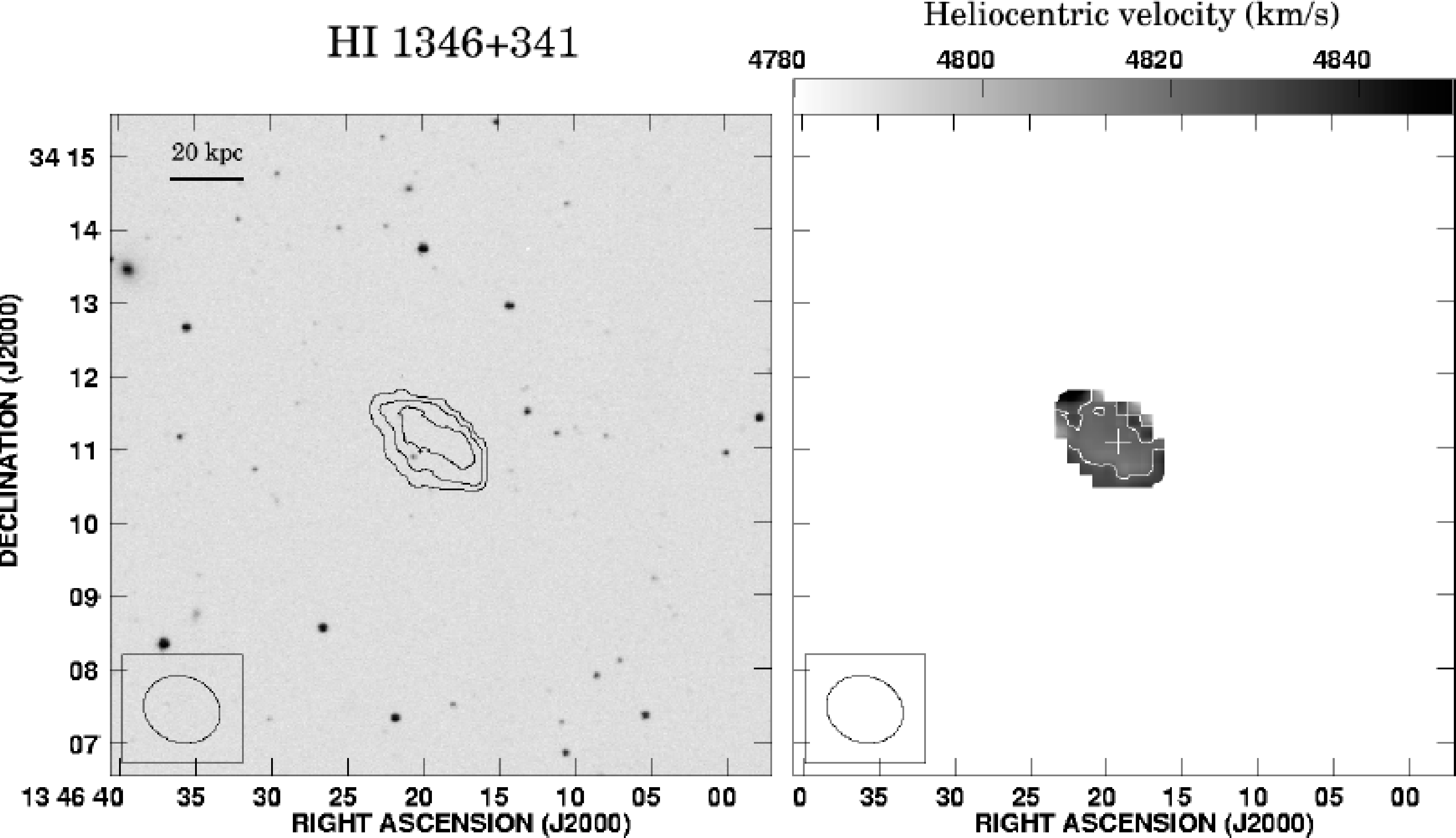} \\
\caption{
Moment maps of HI~1346$+$341, detected in the same field as the target object Markarian~461. (\textsl{Left}) Contours of integrated HI intensity (zeroth
moment) overlaid on the DSS2 image, and (\textsl{Right}) map of intensity-weighted HI mean velocity (first moment). In the zeroth moment map, contours are plotted at 1, 5, $10 \times 8 {\rm \ mJy \ beam^{-1} \ km \ s^{-1}}$ ($3.80 \times 10^{18} {\rm \ cm^{-2}}$). In the first moment map, velocities are indicated by the scale wedge, and contours plotted at intervals of $25 {\rm \ km \ s^{-1}}$. The ellipse at the lower left corner of each panel indicates the half-power width of the synthesized beam, and has a size of $64\arcsec \times 54\arcsec$.}
\end{center}
\end{figure}

\begin{figure}
\begin{center}
\includegraphics[angle=0, scale=0.482]{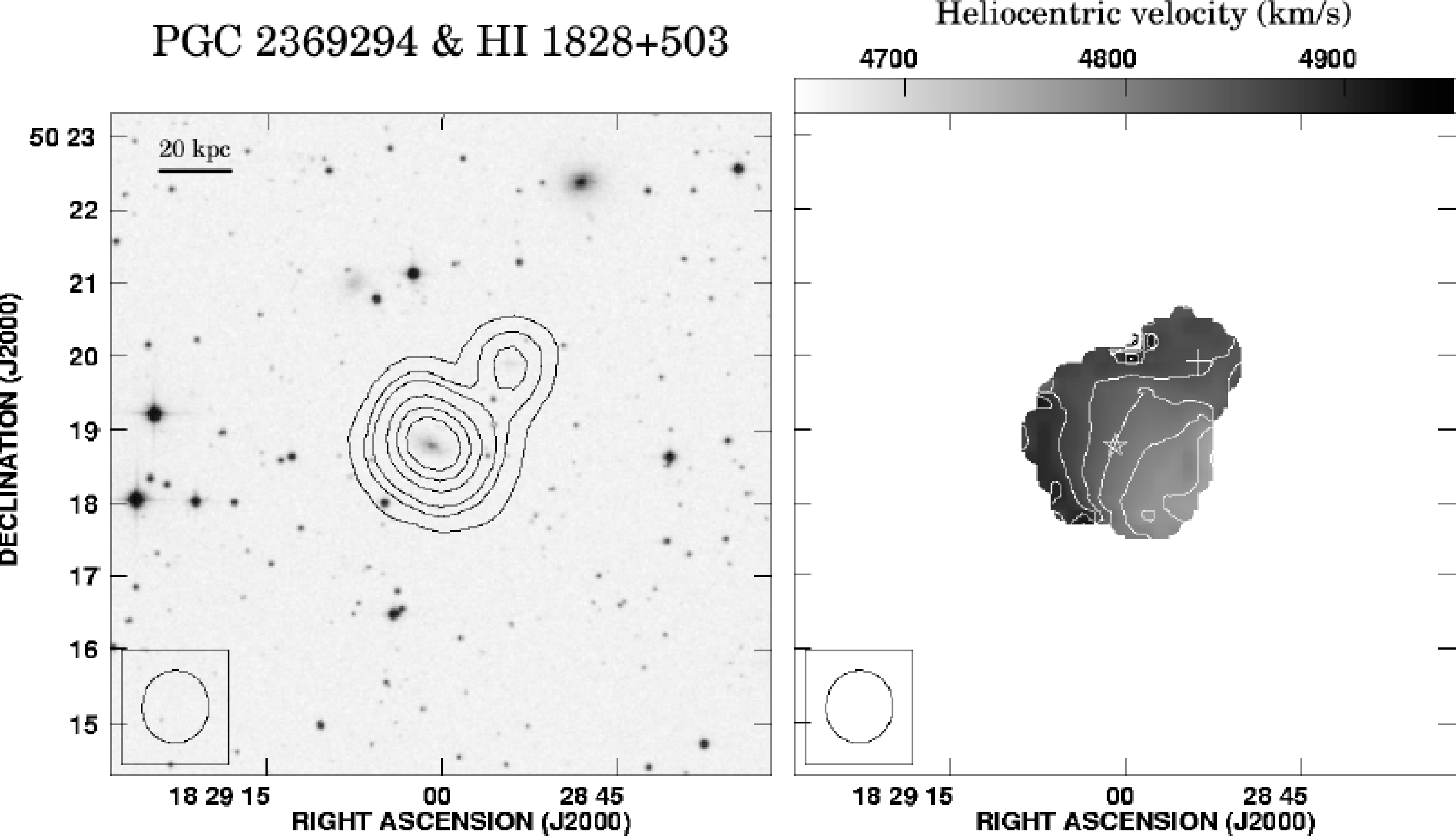} \\
\caption{Moment maps of PGC~2369294 and HI~1828$+$503, detected in the same field as the target object ARK~539. (\textsl{Left}) Contours of integrated HI intensity (zeroth moment) overlaid on the DSS2 image, and (\textsl{Right}) map of intensity-weighted HI mean velocity (first moment). PGC~2369294 lies at the center of the map, whereas HI~1828$+$503 is located north-west of center. In the zeroth moment map, contours are plotted at  1, 3, 5, 10, 15, 20 $\times 35 {\rm \ mJy \ beam^{-1} \ km \ s^{-1}}$ ($1.83 \times 10^{19} {\rm \ cm^{-2}}$). In the first moment map, velocities are indicated by the scale wedge, and contours plotted at intervals of $25 {\rm \ km \ s^{-1}}$. The ellipse at the lower left corner of each panel indicates the half-power width of the synthesized beam, and has a size of $58\arcsec \times 55\arcsec$.}
\end{center}
\end{figure}

\begin{figure}
\begin{center}
\includegraphics[angle=0, scale=0.482]{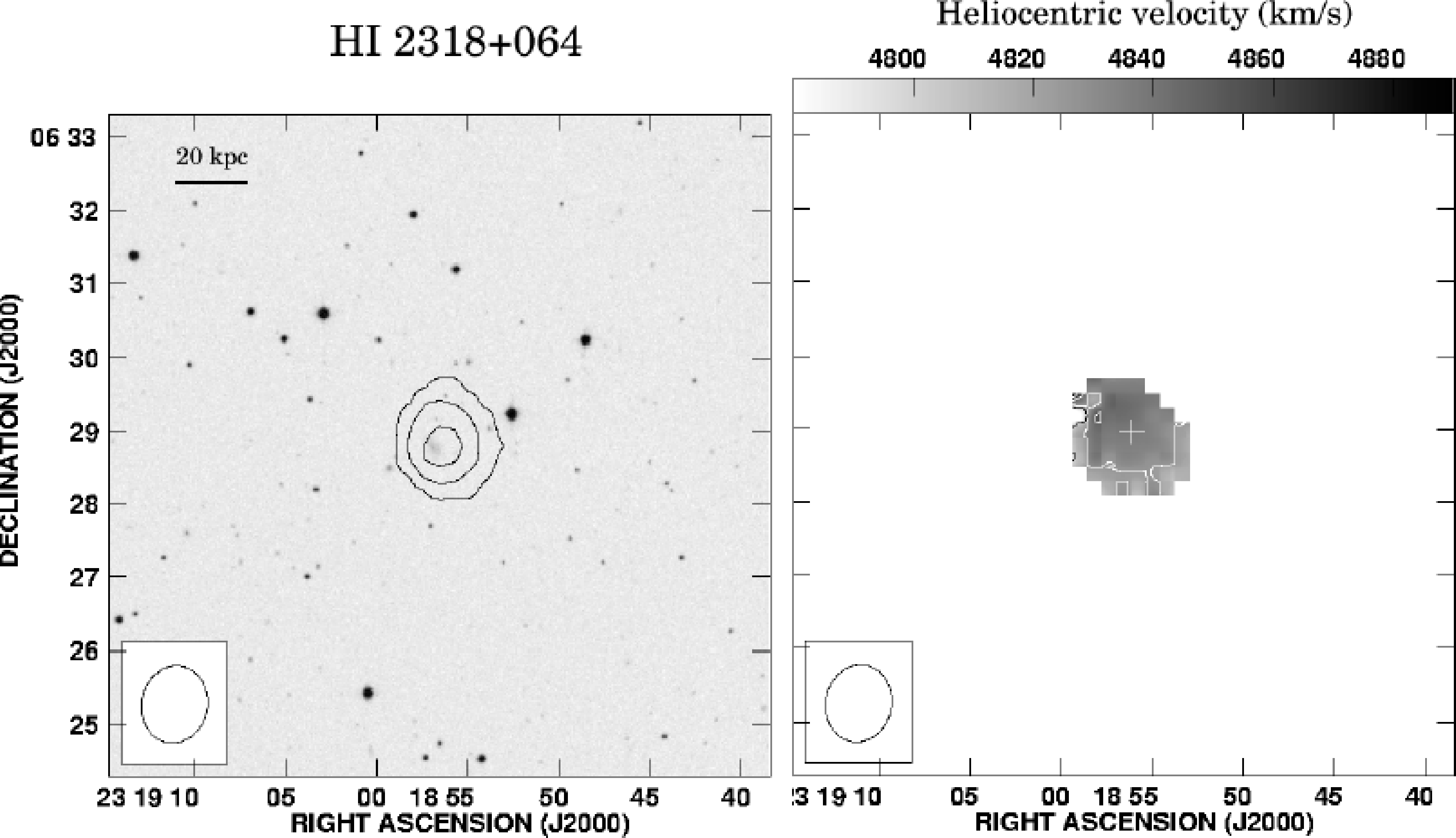} \\
\caption{Moment maps of HI~2318$+$064, detected in the same field as the target object NGC~7591. (\textsl{Left}) Contours of integrated HI intensity (zeroth moment) overlaid on the DSS2 image, and (\textsl{Right}) map of intensity-weighted HI mean velocity (first moment). In the zeroth moment map, contours are plotted at  1, 5, $10 \times 20 {\rm \ mJy \ beam^{-1} \ km \ s^{-1}}$ ($7.39 \times 10^{18} {\rm \ cm^{-2}}$). In the first moment map, velocities are indicated by the scale wedge, and contours plotted at intervals of $25 {\rm \ km \ s^{-1}}$. The ellipse at the lower left corner of each panel indicates the half-power width of the synthesized beam, and has a size of $63\arcsec \times 53\arcsec$.}
\end{center}
\end{figure}

\begin{figure}
\begin{center}
\includegraphics[angle=0, scale=0.482]{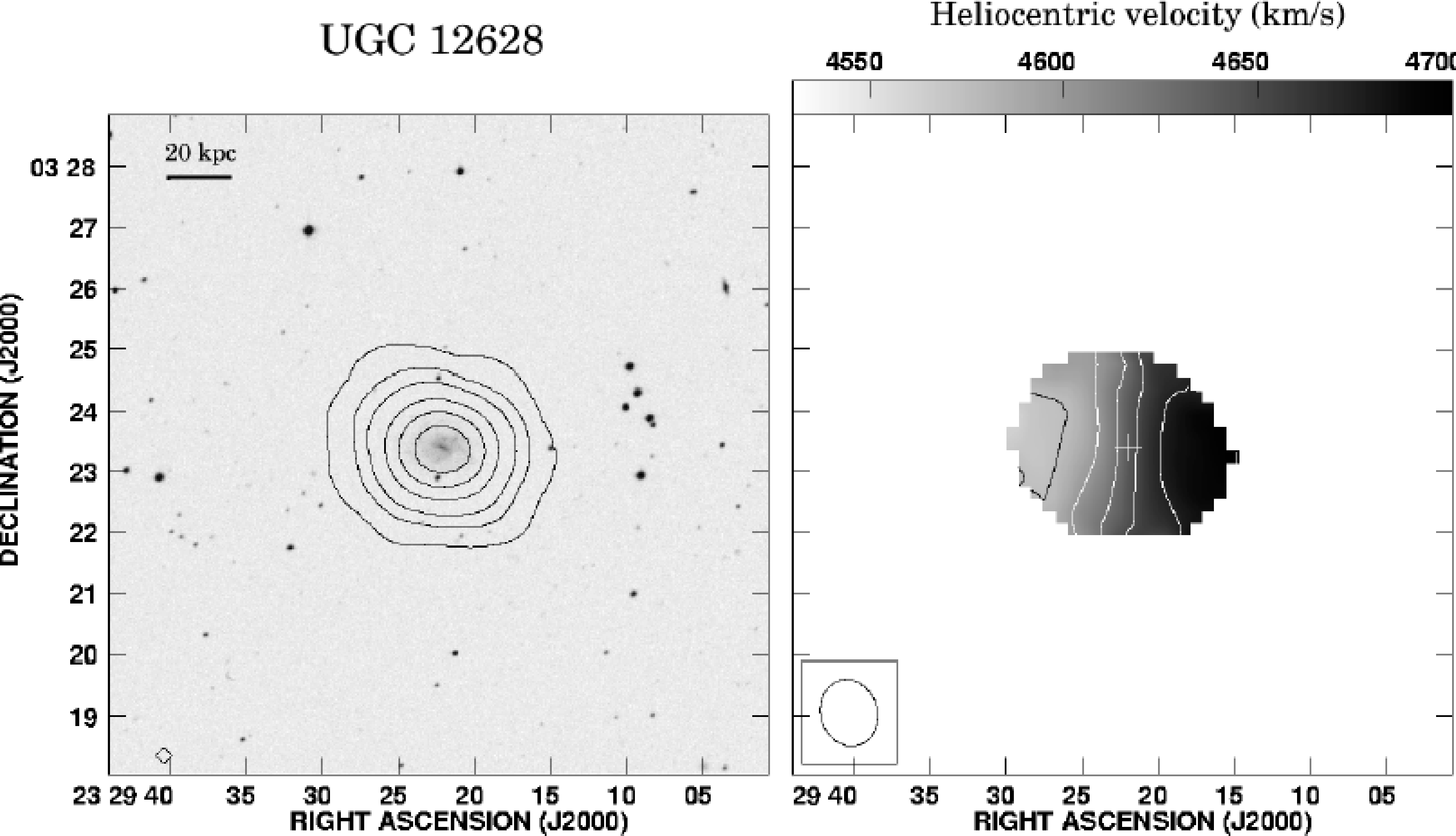} \\
\caption{Moment maps of UGC~12628, detected in the same field as the target objects NGC~7679 and NGC~7682 . (\textsl{Left}) Contours of integrated HI intensity (zeroth moment) overlaid on the DSS2 image, and (\textsl{Right}) map of intensity-weighted HI mean velocity (first moment). In the zeroth moment map, contours are plotted at 1, 5, 10, 20, 30, 40 $\times 40 {\rm \ mJy \ beam^{-1} \ km \ s^{-1}}$ ($1.82 \times 10^{19} {\rm \ cm^{-2}}$). In the first moment map, velocities are indicated by the scale wedge, and contours plotted at intervals of $25 {\rm \ km \ s^{-1}}$. The ellipse at the lower left corner of each panel indicates the half-power width of the synthesized beam, and has a size of $68\arcsec \times 53\arcsec$.}
\end{center}
\end{figure}

\begin{figure}
\begin{center}
\vspace{-11.5 cm}
\includegraphics[angle=0,scale=0.70]{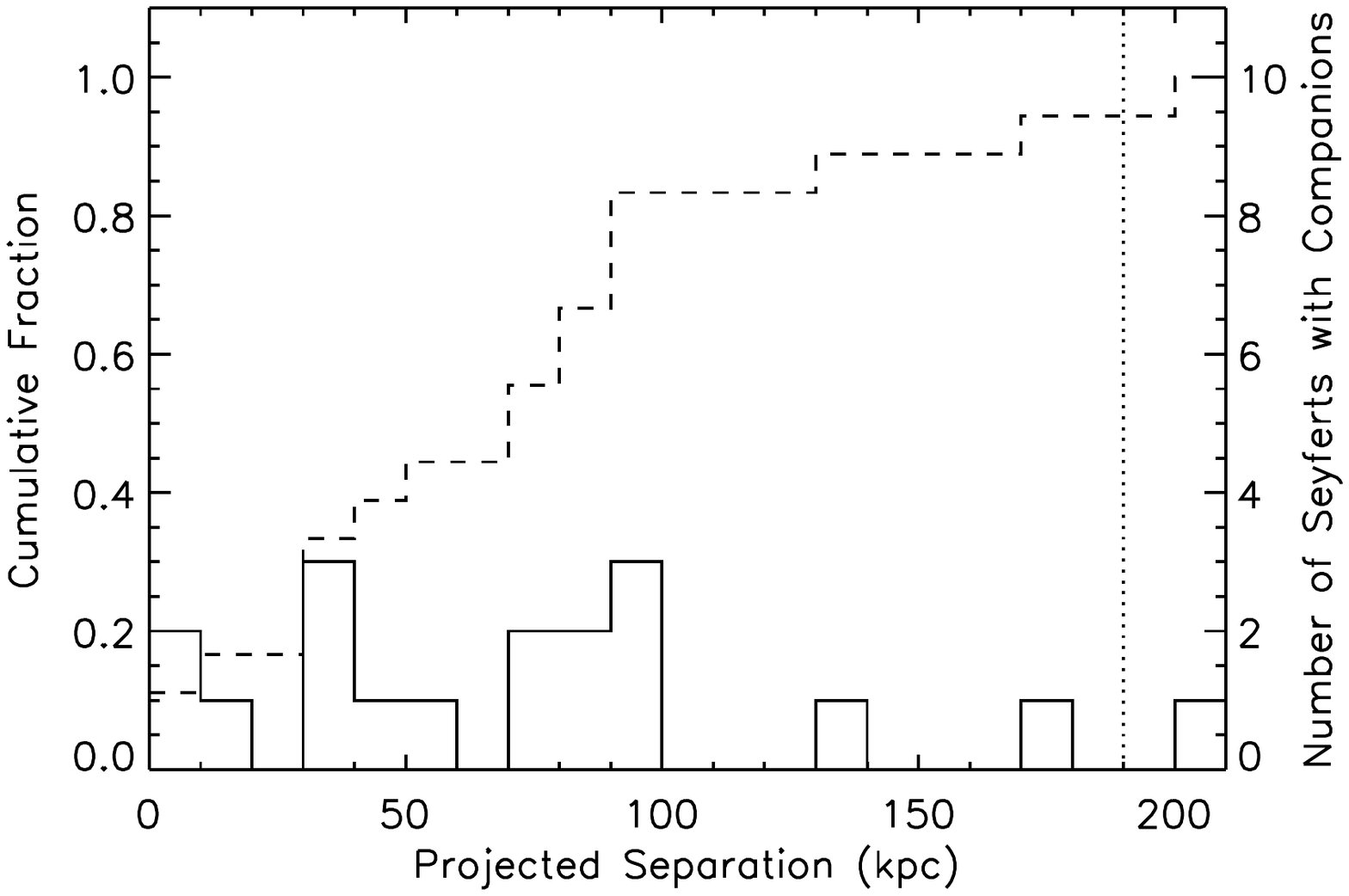}
\\
\vspace{-10.7 cm}
\hspace{0.0 cm}
\includegraphics[angle=0, scale=0.70]{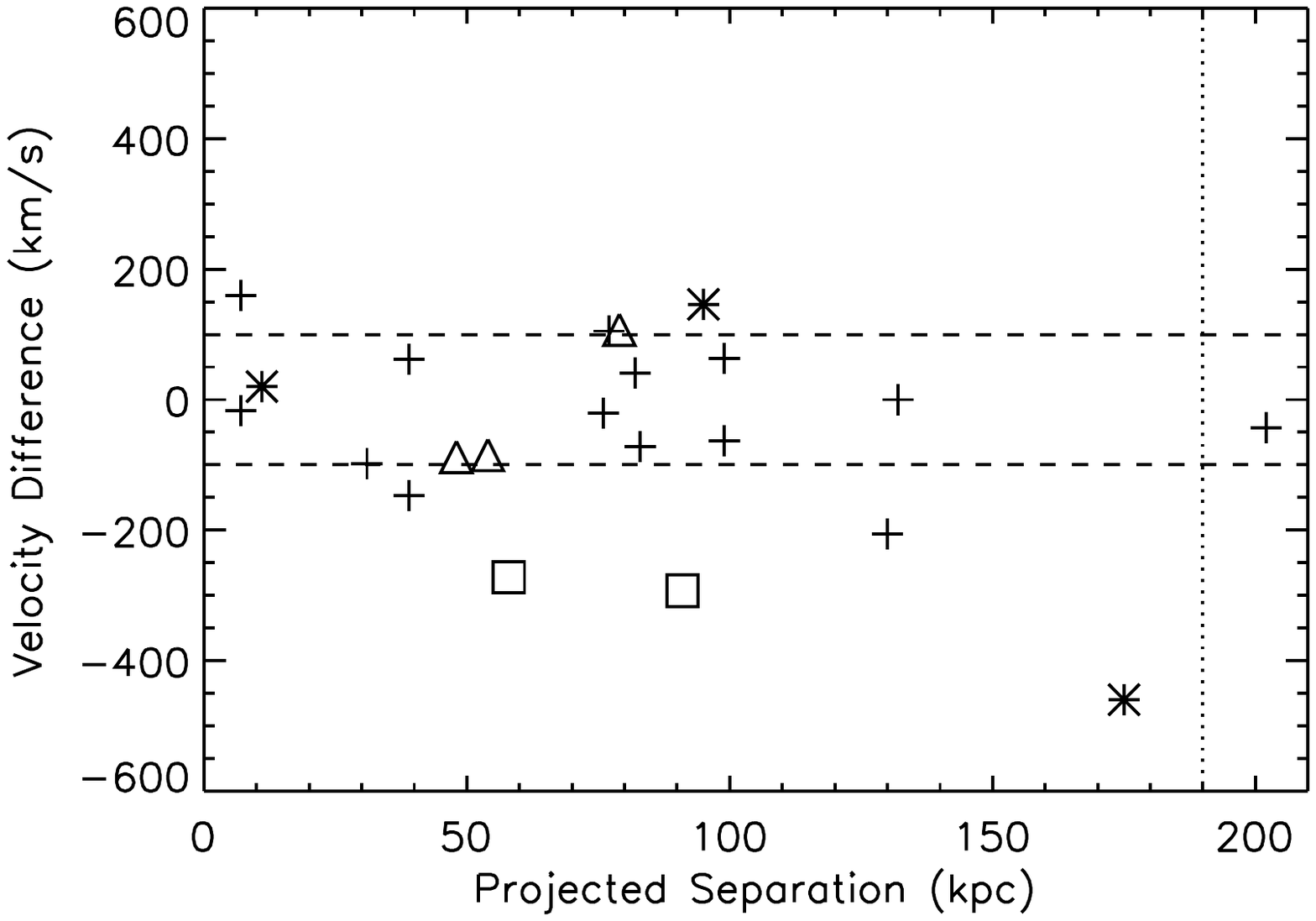}
\caption{(\textsl{Upper}) Plot of number (solid line) and cumulative fraction (dashed line) of Seyfert galaxies in our ensemble sample with (candidate) interacting neighboring galaxies as a function of their projected separation up to 200~kpc.  The data has been divided into bins of width 10~kpc.  A vertical dotted line is drawn at 190~kpc, within which the galaxies are all mapped with a nearly uniform sensitivity (at worst a difference of $\sim$$20\%$).  (\textsl{Lower}) Plot of the difference in radial velocity between the Sefyert and all detected neighboring galaxies within 210~kpc.  Crosses denote those in Group~I, triangles those in Group~II, asterisks those in Group~III, and boxes the remainder not identified to be interacting with their corresponding Seyfert galaxy.  Horizontal dashed lines are drawn at $\pm 100 {\rm \ km \ s^{-1}}$ to help guide the reader.} 
\end{center}
\end{figure}

\clearpage
\end{document}